\documentclass[twocolumn]{aastex7}
\usepackage{enumerate}
\usepackage{amssymb, amsmath}
\usepackage{natbib}
\usepackage{color}
\usepackage{ulem}
\usepackage{appendix}
\usepackage{hyperref}
\usepackage[stable]{footmisc}
\usepackage{relsize}
\usepackage{mathtools}

\definecolor{red}{rgb}{1.0,0.0,0.0}
\begin{document}
%\title{Astrometric and Precision Radial-Velocity Discovery of 12 Jovian Exoplanets Amenable to Future High-Contrast Imaging Observations}
\title{A New Population of Jovian Exoplanets Suitable For Direct Imaging with \\the \textit{Habitable Worlds Observatory}, the \textit{Roman Coronagraph}, and ELTs Identified from HARPS Radial-Velocity Data and Hipparcos-Gaia Astrometry}
%\title{A New Population of Jovian Exoplanets Amenable to Future High-Contrast Imaging Observations Identified from HARPS Radial-Velocity Data and Hipparcos-Gaia Astrometry}
%\title{Discovery of 15 Likely Jovian Exoplanets Amenable to Future High-Contrast Imaging Observations from HARPS Radial-Velocity Data and Hipparcos-Gaia Astrometry}
\author[0000-0003-2426-3852]{Jie Li}
\affiliation{Department of Physics and Astronomy, University of Texas at San Antonio, San Antonio, TX 78249, USA}
\email[show]{jie.lie@utsa.edu}
\author[0000-0002-7405-3119]{Thayne Currie}
%\altaffiliation{These authors contributed equally to this work as co-first authors.}
\affiliation{Department of Physics and Astronomy, University of Texas at San Antonio, San Antonio, TX 78249, USA}
 \affiliation{Subaru Telescope, National Astronomical Observatory of Japan, 
650 North A`oh$\bar{o}$k$\bar{u}$ Place, Hilo, HI  96720, USA}
\email[]{thayne.currie@utsa.edu}
\author[0000-0002-6845-9702]{Yiting Li}
%\altaffiliation{These authors contributed equally to this work as co-first authors.}
\affiliation{Department of Astronomy, University of Michigan, 1085 S. University, Ann Arbor, MI 48109, USA}
\email[]{lyiting@umich.edu}
\author[0000-0001-9441-7656]{Mona El Morsy}
\affiliation{Department of Physics and Astronomy, University of Texas at San Antonio, San Antonio, TX 78249, USA}
\email[]{mona.elmorsy@utsa.edu}
\author[0000-0001-8342-7736]{Jack Lubin}
\affiliation{Department of Physics and Astronomy, University of California, Los Angeles, Los Angeles, CA 90095, USA}
\email[]{jblubin@ucla.edu}
\author[0000-0001-7967-9922]{Danielle Bovie}
\affiliation{Department of Physics and Astronomy, University of Texas at San Antonio, San Antonio, TX 78249, USA}
\email[]{danielle.bovie@my.utsa.edu}
\author[0009-0007-4163-9336]{Nathaniel Lujan}
\affiliation{Department of Physics and Astronomy, University of Texas at San Antonio, San Antonio, TX 78249, USA}
\email[]{nate.lujan@my.utsa.edu}
\author[0009-0002-8645-1736]{Erica Dykes}
\affiliation{Department of Physics and Astronomy, University of Texas at San Antonio, San Antonio, TX 78249, USA}
\email[]{erica.dykes@my.utsa.edu}
%Should also consider adding the following people

\shortauthors{Li et al. 2026}
\begin{abstract}
%To identify new along-period giant planets,
We jointly analyze archival radial-velocity data from the HARPS spectrograph and astrometry from Hipparcos and Gaia to identify and constrain the dynamical properties of new and known long-period jovian exoplanets.   Our search yields 14 new confirmed planets with masses of 0.04--6.1 $M_{\rm Jup}$ and semimajor axes of 0.1--12.6 au.  Planets include a super-Jupiter in a near-Saturn-like orbit producing a strong astrometric acceleration (HIP 669 b), an eccentric Jupiter around a K star with a modest acceleration (HIP 39039 b), a 74-day-orbit warm Neptune (HIP 64550 b), and a giant planet around a Tier A Habitable Worlds Observatory target that may endanger any habitable zone Earth (HIP 17378 b). Three systems show evidence for candidate planets requiring additional data for confirmation.   Our modeling also constrains the dynamical masses and orbits for 32 previously known radial-velocity-discovered planets.  
Orbits, masses, and ages for our sample show that five targets are potentially detectable in reflected light with the Roman Coronagraph, 28 should be imageable in reflected light with the Habitable Worlds Observatory, and six may be detectable in thermal emission with E-ELT/METIS.  These results demonstrate the power of combining archival radial-velocity data with astrometry to discover and characterize long-period giant planets, a capability that will only grow with Gaia DR4. It provides a new sample of targets for future direct-imaging observations, and shows a path forward for timely vetting of targets for exo-Earth imaging with the Habitable Worlds Observatory.
\end{abstract}

\section{Introduction}
Radial velocity (RV) observations have played a central role in exoplanet science since the discovery of 51 Pegasi b, the first exoplanet discovered around a Sun-like star \citep{Mayor_1995}.  Over the past three decades, advances in instrumental precision and long-term monitoring (e.g., \citealp{Mayor_2003, Wittenmyer_2006, Cumming_2008, Vogt_2014, Fischer_2014}) have led to the discovery of a large population of exoplanets with RV surveys. Of the nearly 6000 confirmed exoplanets, approximately 19\% were initially discovered using the RV technique \citep{Howe_2025}. Beyond planet discovery, RV measurements provide important constraints on orbital properties, including orbital period, eccentricity, and velocity semi-amplitude, enabling the dynamical characterization of planetary systems \citep{Fischer_2014, Hara_2023}.

Despite these remarkable successes, RV measurements alone suffer from an inherent degeneracy between the companion mass $m$ and orbital inclination $i$. Because RV observations measure only the line-of-sight component of the host star’s reflex motion, they constrain the companion mass as m~$\sin i$ rather than its true mass.
%, where $i$ is the orbital inclination. 
In addition, RV surveys have limited sensitivity to planets with long periods, for which the observational baseline may cover only a fraction of the orbit \citep[e.g.][]{Rosenthal_2021}. Such incomplete orbital coverage can lead to poorly constrained companion masses and orbital parameters.
%orbital periods, companion masses, and other orbital parameters. 
%These limitations make it difficult to determine the true masses and three-dimensional orbital properties of long-period companions from RV observations alone.

Astrometry provides a complementary approach to detecting long-period companions by measuring the motion of the host star in the plane of the sky. The combination of \textit{Hipparcos} and \textit{Gaia}, separated by a time baseline of more than two decades, enables the identification of proper-motion anomalies (PMA) induced by unseen companions \citep{Kervella2022,Brandt2021b}. While a significant PMA can indicate the presence of a companion and place constraints on its mass as a function of orbital separation, it generally cannot determine the companion mass and orbital properties without additional information \citep{Kervella2022}. 

Compilations like the \textit{Hipparcos-Gaia Catalog of Accelerations} (HGCA; \citealp{Brandt2018, Brandt2021b}) and \textit{G23H} \citep{Thompson_2026} further improves the astrometric information by providing cross-calibrated proper motions and precise measurements of stellar accelerations. Subsequent studies have used these astrometric measurements together with RV observations to break the degeneracies inherent to RV measurements, yielding true dynamical masses and improved orbital constraints for planetary and substellar companions (e.g., \citealp{Li_2021, Feng_2022, Feng_2024, An_2025}). Other work has directly imaged planets around stars showing PMAs -- i.e. \textit{accelerating stars} --  and combined the resulting relative astrometry with absolute astrometric measurements to constrain the orbits and dynamical masses of their companions (e.g., \citealp{Currie2023b, Currie2023a, Mesa2023, Currie_2026}). More recently, joint analyses incorporating RV, absolute astrometry, and direct imaging have demonstrated the power of combining all three techniques to achieve a more complete dynamical characterization of planetary systems \citep{Li_2024, ElMorsy2024b,Bovie_2025}.

Precise orbital solutions obtained by combining astrometry with other observational techniques can predict the projected separation and position angle of a companion at future epochs. This ability to determine where and when a companion will be most favorable for detection is particularly valuable for efficiently planning and targeting direct imaging observations. For example, orbital constraints have been used to guide the direct imaging of long-period, cold giant planets such as 14 Her c \citep{Gagliuffi_2021, Gagliuffi_2025}, and $\epsilon$ Ind Ab \citep{Matthews_2024, Sanghi_2026}. Similar joint analyses have been used to predict the future location and observability of $\epsilon$ Eridani b \citep{Thompson_2025}. More broadly, this approach is increasingly being applied to the planning of future reflected-light observations with the \textit{Nancy Grace Roman Space Telescope} Coronagraph Instrument (Roman/CGI), including systems such as HIP 54515 and HIP 71618 (OASIS; \citealt{ElMorsy2025b, Currie_2026}).

The availability of extensive archival RV data allows us to apply this approach systematically to large samples of nearby stars by combining long-baseline RV observations with astrometric measurements. Such systematic searches can identify new companions, refine the orbital properties of known companions, and provide well-characterized targets for direct imaging observations across a broad range of wavelengths. These targets can be followed up with current ground-based instruments, such as Keck/NIRC2 \citep{Xuan_2018} and Subaru/SCExAO \citep{Jovanovic_2015}, while also preparing for the next generation of direct imaging instruments, including the Mid-infrared ELT Imager and Spectrograph on the Extremely Large Telescope (E-ELT/METIS; \citealt{Brandl_2021}), the Roman Coronagraph \citep{Kasdin_2020}, and the Habitable Worlds Observatory (HWO) \citep{Kane_2024}. 

Building on these advances, in this paper, we systematically analyze the archival RV data from High Accuracy Radial velocity Planet Searcher (HARPS) \citep{Mayor_2003}, together with absolute astrometry from the HGCA to identify the systems exhibiting acceleration signatures. Using the orbit-fitting codes \texttt{orvara} \citep{Brandt2021} and \texttt{Octofitter} \citep{Thompson_2023}, we jointly model these data to constrain the masses and orbital parameters of companions. Based on these dynamical constraints, we identify the highest-priority targets for subsequent ground-based and space-based direct imaging observations. 

This paper is organized as follows. In Section \ref{sec: data}, we describe our data from which we identify vetted exoplanet candidates through an initial triage process in \ref{sec: initial_ID} based on the star's acceleration signatures, non-stellar RV signal, and a preliminary dynamical modeling.  In Section \ref{sec: discovery}, we derive dynamical masses and orbital properties for these companions -- both newly-identified exoplanets and previously-known candidates that only had mass lower limits. In Section \ref{sec: DI}, we investigate their prospects for direct imaging in reflected light and thermal emission with future instruments. Finally, in Section \ref{sec: summary}, we provide a brief summary of our results.

\section{Data}
%and Joint Dynamical Analysis}
\label{sec: data}

\subsection{Hipparcos and Gaia eDR3 Astrometry}
\label{sub: HGCA}
%\TC{\textit{I think we can substantially condense: see below}}\\
%An unseen companion can induce reflex motion on its host star, causing the star to exhibit a small astrometric ``wobble'' on the plane of the sky. By precisely tracking changes in the stellar position and proper motion over long temporal baselines, these subtle acceleration signals can be detected and used as evidence for the presence of companions. Astrometric acceleration measurements are therefore particularly powerful for identifying long-period companions that may only produce incomplete orbital coverage in radial velocity observations.

%In this work, we use the Hipparcos–Gaia Catalog of Accelerations (HGCA; \citealt{Brandt2018, Brandt2021b}) to obtain precise astrometric measurements of accelerating stars. The HGCA combines data from the Hipparcos and Gaia EDR3, and provides a cross-calibrated astrometric catalog on the Gaia EDR3 reference frame, covering just over 115000 stars with a temporal baseline of nearly 30 years. For each star, it includes three proper motion measurements with calibrated uncertainties: the Hipparcos proper motion, the Gaia EDR3 proper motion, and a long-term proper motion derived from the positional difference between Hipparcos and Gaia EDR3. In addition, it reports a $\chi^2$ statistic for a model assuming constant proper motion with two degrees of freedom. 

We incorporate astrometric data from the most recent edition of the Hipparcos–Gaia Catalog of Accelerations (HGCA; \citealt{Brandt2018, Brandt2021b}), which combines precision astrometry of host stars from Hipparcos and Gaia EDR3 to produce a cross-calibrated astrometric catalog on the Gaia EDR3 reference frame covering 115,000 stars with a temporal baseline of $\sim$ 25 years.   HGCA includes parallax measurements and three proper motion measurements -- instantaneously-measured Hipparcos and Gaia eDR3 proper motions as well as a long-term motion derived from the positional difference between Hipparcos and Gaia.  It reports a $\chi^{2}$ statistic for a model assuming constant proper motion with two degrees of freedom (e.g. $\chi^{2}$ $\sim$ 6.2, 11.8, and 28.8 correspond to 2-$\sigma$, 3-$\sigma$, and 5-$\sigma$ deviations from linear motion).

%Previous studies \citep{An_2025} typically focused on stars with very clear accelerations (i.e., selecting targets with $\chi^2 > 25$, corresponding approximately to a significance above the 5 $\sigma$ level). However, since weak signals may arise from either low-mass companions, which produce intrinsically small astrometric accelerations, or long-period companions whose orbital timescales exceed the Hipparcos–Gaia baseline, resulting in nearly linear motion and suppressed measurable curvature, we adopt a less restrictive threshold and consider systems with detectable astrometric accelerations (i.e., $\chi^2 > 7$, corresponding approximately to a significance above the $2\sigma$ level) as candidate acceleration systems for further analysis. For the remaining lower-significance systems ($\chi^2 < 7$), the astrometric evidence alone is generally insufficient to establish the presence of companions reliably. Therefore, these systems require a more careful examination of their RV signals in the subsequent analysis.

%\textcolor{blue}{A paragraph on the G23H catalog incorporating Gaia DR2 data in addition to Gaia DR3.   Important topics include: i) this is a second Gaia measurement -- over a shorter time window and less precise than DR3 but more than Hipparcos, ii) useful for constraining systems with a planet on a $\sim$ many months to few years orbit.   Use Thompson et al recalibration of Gaia DR2 to correct for rotation}

\subsection{The G23H Catalog}
\label{sub: g23h}

Incorporating Gaia DR2 astrometry obtained over the first 22 months of the 34-month Gaia DR3 data collection into dynamical modeling may in some cases improve orbital constraints \citep{Feng_2024, Thompson_2026}, especially for systems whose companions have orbital periods comparable to or slightly greater than difference between the Gaia DR2 and DR3 mean epochs ($\approx$0.5 years).    The \textit{Renormalized Unit Weight Error} (RUWE) -- a measurement of astrometric excess noise -- is also well suited for identifying massive companions on short orbital periods.  For companions identified from data listed in \S\ref{sub: HGCA} and \ref{subsec: rv}, we therefore considered the G23H catalog \citep{Thompson_2026}, which adds Gaia DR2 astrometry, the Gaia DR3 RUWE, and Gaia RV variability constraints (also sensitive to close-in binaries).   Because we have precision RV measurements, we disregard the RV variability parameter.  The DR2 measurements were recalibrated to the DR3 reference frame to correct for rotation. 
%Although Gaia DR2 has a shorter observational baseline and lower astrometric precision than DR3, it remains more precise than Hipparcos and provides an additional astrometric measurements. This additional information may help improve orbital constraints, particluarly for companions with periods ranging from several months to a few years.

%TC{one other question: you used these complementary data in Sect 3.1.1. to identify periodic signals, correct? If so, the text there needs to be adjusted}
%\textcolor{red}{Jie: When I initially used RVSearch to select targets based on whether the FAP was below 0.1\%, I did not include any additional RV data beyond the HARPS measurements. After Sect. 3.2, once I had obtained the preliminary sample, I searched for additional RV measurements before running orvara. I then reran RVSearch using the expanded RV datasets and used these data for the subsequent orvara fitting. I think maybe we can mention this in the first paragraph of Sect. 3.3. After explaining why we need additional RV data, we can state that we reran RVSearch with the expanded datasets and performed the orvara analysis.}

\subsection{HARPS and Other Radial Velocity Data}
\label{subsec: rv}

We downloaded the HARPS target catalog \citep{Barbieri_2023}\footnote{\url{https://www.eso.org/qi/catalog/show/416}} and then cross-matched it with the HGCA, identifying 3805 targets with both HARPS RV measurements and astrometry. 
Then, we acquired HARPS RV data for matched sources from the Data \& Analysis Center for Exoplanets (DACE),\footnote{\url{https://dace.unige.ch}} a web platform hosted at the University of Geneva that provides access to published exoplanet-related observational data and analysis tools.  We further downselected our sample to focus only on those targets with at least 10 HARPS measurements.

%\TCtext{We downloaded the HARPS target catalog \citep{Barbieri_2023} and then cross-matched it with the HGCA, identifying 3805 targets with both HARPS RV measurements and astrometry. 
%Then, we downloaded HARPS RV data for matched sources from the Data \& Analysis Center for Exoplanets (DACE),\footnote{\url{https://dace.unige.ch}} a web platform hosted at the University of Geneva that provides access to published exoplanet-related observational data and analysis tools.  We further downselected our sample to focus only on those targets with at least 10 HARPS measurements.}

While HARPS data contained in DACE lack a nightly zero-point (NZP) RV correction computed in earlier HARPS archive compilations \citep{Trifonov_2020,Perdelwitz_2024}, the NZP offsets typically have little night-to-night jitter and DACE includes a larger sample of key stellar activity measurements than in prior catalog sources.  Furthermore, HARPS underwent instrumental upgrades in 2015 \citep{Curto_2015}, which can introduce RV zero-point offsets between pre- and post-upgrade observations.   Thus, we opted to use the DACE RV measurements.  In our later analysis, we treat the data taken before and after each upgrade as independent instruments, fitting separate RV offsets and jitter terms for each subset.

%\TCtext{While HARPS data contained in DACE lack a nightly zero-point (NZP) RV correction computed in earlier HARPS archive compilations \citep{Trifonov_2020,Perdelwitz_2024}, the NZP offsets typically have little night-to-night jitter\TC{(verify)} and DACE includes key stellar activity measurements we found were sometimes malformed in prior catalog sources.  Furthermore, HARPS underwent instrumental upgrades in 2015 \citep{Curto_2015}, which can introduce RV zero-point offsets between pre- and post-upgrade observations. In our later analysis, we treat the data taken before and after each upgrade as independent instruments, fitting separate RV offsets and jitter terms for each subset.} 

Finally, we correct the HARPS RV measurements for secular acceleration, an apparent long-term RV trend caused by the changing line-of-sight projection of a star’s space velocity due to its proper motion. This effect is particularly important for high-proper-motion targets. For example, HIP 63833 has a secular acceleration of $0.4689~{\rm m\,s^{-1}\,yr^{-1}}$, corresponding to an RV change of approximately $\sim4.7~{\rm m\,s^{-1}}$ over a 10 yr baseline.

%\TC{My understanding here is that you used logRHK -- and if that isn't available -- the S index.  I.e. you didn't actually use BIS.  Could you clarify?}
Along with the HARPS RV measurements, we retrieved the corresponding time-series stellar activity indicators from DACE, including the bisector inverse span (BIS; \citealt{Queloz_2001}) and the Mount Wilson $S$-index ($S_{\rm MW}$; \citealt{Wilson_1978, Baliunas_1995}). We use BIS primarily to trace short-timescale activity-induced variations, which can arise from spots. In contrast, $S_{\rm MW}$, which trace chromospheric activity through the Ca II H \& K lines, are used to characterize activity variations over longer timescales. Together, these indicators allow us to assess stellar activity over a range of timescales and evaluate whether the observed RV variability may be associated with stellar activity rather than orbital motion.   While the $\log R'_{\rm HK}$ index is often more standard, we found that many HARPS-observed stars in DACE lack such measurements.  Where both $\log R'_{\rm HK}$ and $S_{\rm MW}$ were available, they yielded almost identical correlations (or lack thereof) with RV measurements.

To supplement HARPS data, we downloaded RV measurements obtained with other spectrographs from DACE.
%For the targets in our final sample, we will search DACE for all available RV measurements from additional spectrographs to supplement the HARPS observations. 
Incorporating these measurements increases the number of RV epochs and, in many cases, extends the temporal baseline, thereby providing stronger constraints on the orbital parameters of candidate companions. Depending on data availability for each target, these additional measurements include RVs obtained with high-resolution spectrographs such as HIRES \citep{Vogt_1994}, APF \citep{Vogt_2014}, UCLES \citep{Diego_1990}, CORALIE \citep{Queloz_2000, Segransan_2010} HDS \citep{Noguchi_2002}, MIKE \citep{Bernstein_2003} and ESPRESSO \citep{Pepe_2021}, etc. For several of these spectrographs, instrumental upgrades introduced changes in the RV zero point and/or instrumental response. For example, CORALIE underwent instrumental upgrades in 2007, 2014 \citep{Barbato_2023}, and 2024 \citep{Apergis_2026}, resulting in four distinct RV datasets, CORALIE98, CORALIE07, CORALIE14, and CORALIE24. We therefore adopt the same treatment as for HARPS and model measurements obtained before and after upgrades as separate instruments, with independent RV zero-point offsets and jitter terms for each configuration.
Again, we also correct these additional RV measurements for secular acceleration when this correction has not already been applied.

\section{Identification of Candidate Companions}\label{sec: initial_ID}
\subsection{Candidate Selection from Radial Velocities and Astrometry}
\subsubsection{RV Selection}
We first performed an initial search for periodic signals in the RV data for the 3805 targets in our sample. As described in the previous section, we treated the pre- and post-2015 HARPS observations as two independent instruments and fitted a separate RV zero-point offset for each data set. We also removed $5\sigma$ outliers from the RV time series before performing the period search.
%\TC{could you show me an example of what you did?   if the signal has a la}

Given the large size of the initial sample and the computational cost of performing a full Keplerian search for every target, we first carried out a preliminary period search using the Lomb–Scargle periodogram (LSP; \citealt{VanderPlas_2018}). At this stage, we required the strongest periodic signal to have a false-alarm probability (FAP) below 1\%. This initial selection reduced the sample from 3805 to 1015 targets. 

We then performed a more detailed Keplerian search on these 1015 targets using \texttt{RVSearch} \citep{Rosenthal_2021}. The RV offsets determined during the preliminary LSP analysis were not carried over to this stage. Instead, the pre- and post-2015 HARPS observations were again treated as two independent instruments, with separate RV zero-point offsets and jitter terms fitted by \texttt{RVSearch} as part of the Keplerian modeling. We adopted a blind, iterative search strategy similar to that described in \citet{Rosenthal_2021}. Starting from a zero- or single-companion model, \texttt{RVSearch} constructs a Bayesian Inference Criterion-based periodogram ($\Delta$BIC) by fitting Keplerian signals over a grid of trial orbital periods and comparing models with and without an additional companion. 

\begin{figure}[!tp]
\centering
\includegraphics[width=\columnwidth,height=\textheight,keepaspectratio]{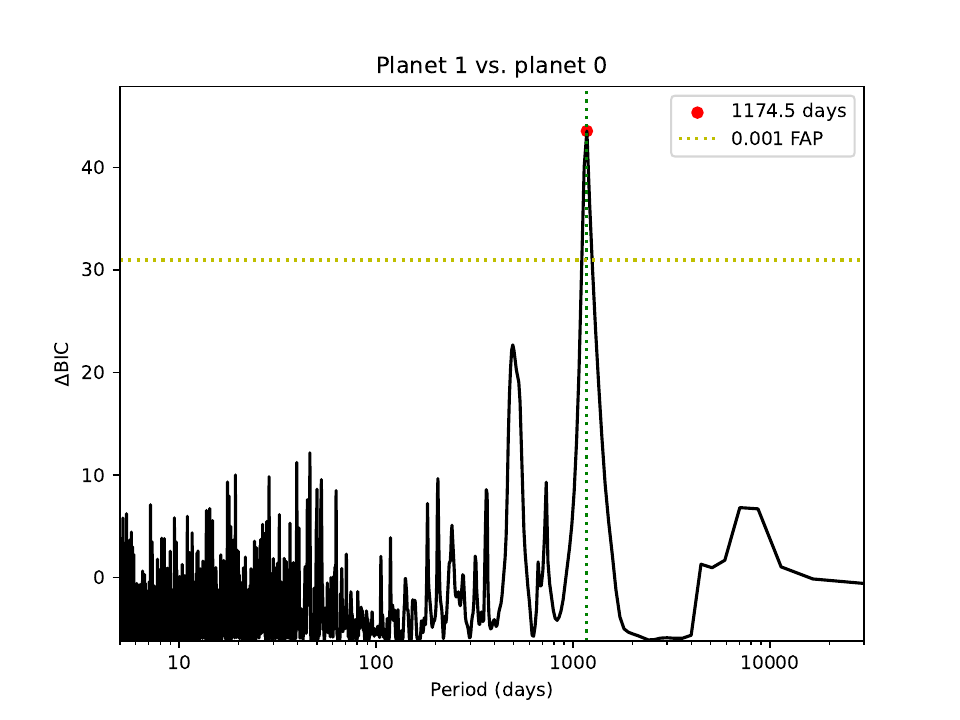}
\caption{\texttt{RVSearch} results for HIP 118009. The periodogram reveals a highly significant periodic signal in the radial velocity time series, with a strong detection and a false alarm probability below 0.001.
\label{fig: RVSearch}}
\end{figure}

Following the signal-detection criteria adopted by \citet{Howard_2016} and \citet{Rosenthal_2021}, we required detected periodic signals to have an FAP below 0.1\%: for most of our sample, this threshold covered $\Delta$BIC $\sim$ 30--50. When a signal satisfied this criterion, the corresponding Keplerian model was refined and the search was repeated for additional signals until no further significant periodicity was detected. For systems with multiple candidate companions, we adopted the model favored by the $\Delta$BIC comparison in \texttt{RVSearch} as the preferred RV solution. Following this selection, 803 targets remained for the subsequent analysis. Figure \ref{fig: RVSearch} shows a representative example of the analysis, revealing a previously unreported 1175 day-periodic signal around the star HIP 118009.

%Most of the RV signals retained in our sample have orbital periods of hundreds of days or longer. This is expected given the design of our survey, which focuses on companions that can be constrained using both RV and astrometric measurements. A small number of systems in the retained sample, however, have RV periods shorter than 100 days. For these short-period systems, the available astrometric constraints may not provide reliable measurements on the companion orbits. We therefore retained these systems based on their significant RV detections and performed a more detailed RV-only MCMC analysis with \texttt{RVSearch} to constrain their orbital parameters. Because the orbital inclinations cannot be constrained from the RV data alone, we derived and report the minimum companion masses, $M\sin i$, rather than the true dynamical masses.

%The relevant \texttt{RVSearch} results are listed in Table \ref{tab: new_planetnew_harpshgca} and Table \ref{tab: recovered_planetnew_harpshgca}, while 

\subsubsection{Astrometric-Acceleration Selection}
\begin{figure}[!tp]
\centering
\includegraphics[width=\columnwidth,height=\textheight,keepaspectratio]{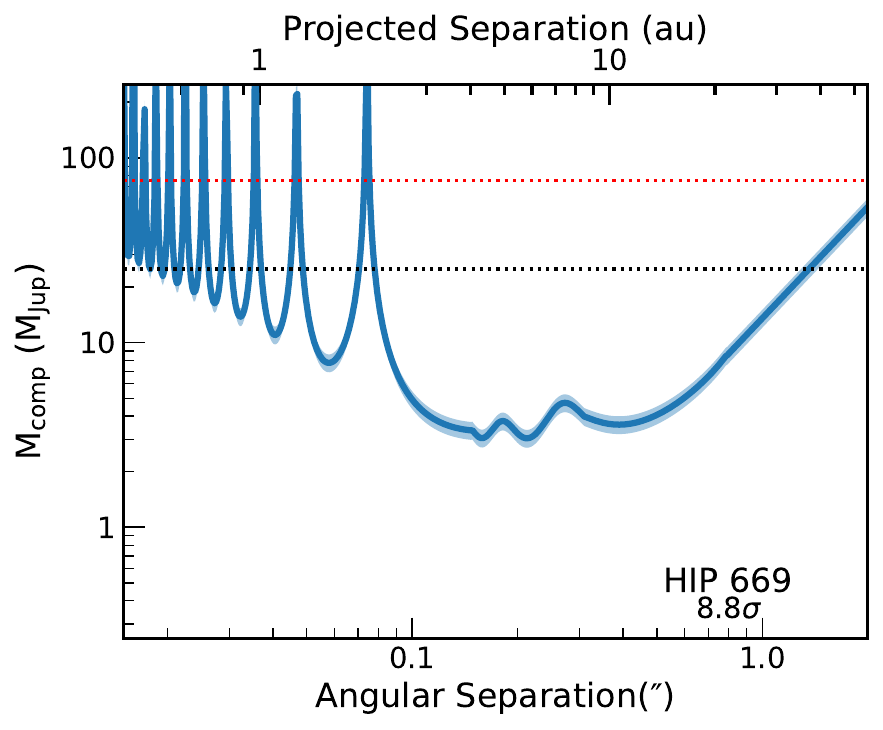}
\caption{Astrometric sensitivity curve for HIP 669. The blue solid line shows the companion mass inferred from the measured astrometric acceleration as a function of projected separation, while the shaded region denotes the corresponding uncertainty. The host star exhibits a significant astrometric acceleration with a detection significance of approximately 8.8 $\sigma$. The black dashed line marks a companion mass of 25 $M_{\rm Jup}$, approximately corresponding to the upper-mass boundary of giant planets, while the red dashed line indicates 75 $M_{\rm Jup}$, near the conventional upper-mass limit for brown dwarfs.
\label{fig: astrometric}}
\end{figure}

In parallel with the RV-based selection described above, we independently searched for evidence of companions using the astrometric accelerations from the HGCA. We applied this selection to the same initial sample of 3805 targets and evaluated the significance of the proper-motion anomaly (PMA) using the corresponding $\chi^2$ statistic. We required $\chi^2_{\rm HGCA}>7$, approximately corresponding to a $\sim~2\sigma$ detection of astrometric acceleration. Targets satisfying this criterion were retained as astrometrically selected companion candidates, independent of whether a significant periodic signal was identified in the RV data at the FAP $\le$ 0.1\% level. Applying this criterion resulted in 1146 targets for subsequent analysis. Figure \ref{fig: astrometric} shows a representative example of the PMA for a target (HIP 669) satisfying our astrometric-acceleration selection criterion.

%The relevant astrometric information for these targets is also provided in Table \ref{tab: new_planetnew_harpshgca} and Table \ref{tab: recovered_planetnew_harpshgca}. 

\subsection{Assessment of Stellar Activity}
Stellar activity can induce RV variations that mimic or obscure Doppler signals produced by orbiting companions. We therefore assessed the potential contribution of stellar activity to the RV signals identified in the preceding selection before carrying the candidates forward for further analysis. We used two activity indicators, the Mount Wilson $S$-index ($S_{\rm MW}$; \citealt{Wilson_1978, Baliunas_1995}) and the bisector inverse slope (BIS; \citealt{Queloz_2001}), depending on the characteristic timescale of the RV signal.
 
Because most of the systems in our sample exhibit astrometric signatures suggestive of companions with orbital periods on the order of hundreds of days or longer, we first adopted $S_{\rm MW}$ as the primary activity indicator to investigate the impact of long-term stellar activity on the observed RV variations. The $S_{\rm MW}$ index traces chromospheric emission in the Ca II H\&K lines and is sensitive to long-term variations associated with stellar magnetic activity, which can produce RV variations that resemble the signatures of long-period companions. However, our sample also contains a small number of targets with RV periods shorter than 180 days. For these systems, we additionally used BIS to investigate the effects of stellar activity on shorter timescales, particularly RV variations associated with rotationally modulated surface features such as spots and plages. BIS traces asymmetries in spectral line profiles and therefore provides a complementary indicator of such activity-induced RV signals. 

The $S_{\rm MW}$ and BIS measurements used in this analysis were derived from the HARPS spectra and are available through DACE for the targets in our sample. Again, we treated the pre- and post-2015 HARPS observations separately and corrected the corresponding offsets in both the RV measurements and the adopted activity indicator before evaluating the RV–activity correlations.

\begin{figure}[!tp]
\centering
\includegraphics[width=\columnwidth,height=\textheight,keepaspectratio]{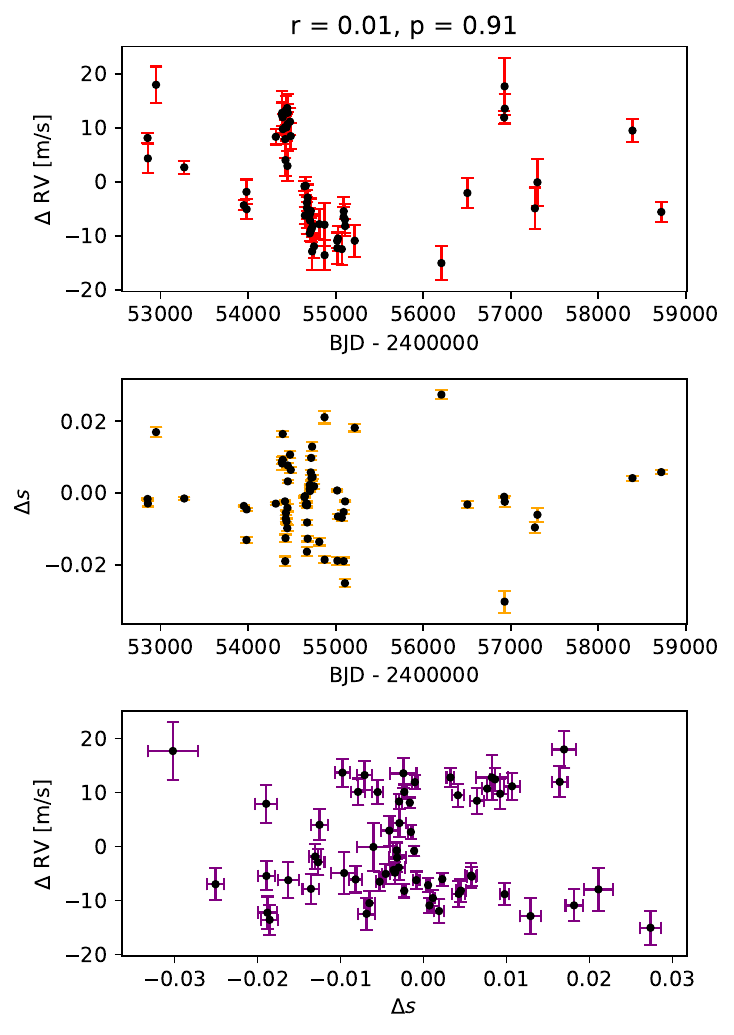}
\caption{Example of the stellar activity assessment for HIP 6511. The top and middle panels show the offset-corrected RV measurements ($\Delta$RV) and S-index measurements ($\Delta S$) as a function of time, respectively, where $\Delta$ denotes the measurements after subtracting the fitted instrument-dependent zero-point offsets. The bottom panel shows $\Delta$RV as a function of $\Delta S$. The Pearson correlation coefficient and corresponding $p$-value are $r=0.01$ and $p=0.91$, respectively, indicating no significant correlation between the RVs and the stellar activity.
\label{fig: stellar}}
\end{figure}

We quantified the relationship between the RV measurements and the selected activity indicator using the Pearson correlation coefficient, $r$, and its associated $p$-value. We retained targets satisfying $|r|<0.2$ or $p>0.05$ to minimize the potential contribution of stellar activity to our candidate sample. Figure \ref{fig: stellar} shows an example of a target (HIP 6511) for which stellar activity was ruled out as the likely origin of the RV signal. Following this assessment, 462 of the 803 RV-selected targets and 252 of the 1146 astrometrically selected targets remained for subsequent analysis.

As an additional check, we compared the three strongest peaks in the activity-indicator periodograms with the periods identified by \texttt{RVSearch}. We treat the RV signal as suspect – warranting further follow up – if the difference in signal peaks is less than one periodogram resolution element (1/T).  No clear evidence of matching periodicities between the RV signals and stellar activity indicators was found.

\subsection{Joint RV--Astrometry Modeling}
\subsubsection{Orbital Fitting with \texttt{orvara}}
Following the candidate selection and stellar-activity assessment described above, we proceeded to jointly model the RV and astrometric data for the remaining systems. We combined the RV-selected and astrometrically selected samples, which contain 462 and 252 targets, respectively. Because some targets satisfy both selection criteria, the combined sample consists of 534 unique systems. Before performing the joint analysis, we searched DACE for additional archival RV measurements beyond the HARPS data in order to improve the orbital constraints and extend the temporal baseline where possible. The additional RV data were obtained from HIRES, APF, UCLES, ESPRESSO, HDS, CORALIE, ELODIE, Hamilton, and MIKE.

We then performed joint RV–astrometry orbital modeling for these systems using \texttt{orvara}. Given the size of the sample and the associated computational cost, we first carried out a preliminary fit for each target. For systems in which \texttt{RVSearch} identified one or more companions, we adopted the number identified by \texttt{RVSearch} in the \texttt{orvara} model. For the remaining systems, we assumed a single-companion model. Stellar-mass priors were primarily determined from the spectral types of the host stars, using the stellar masses corresponding to each spectral type provided by \citet{Pecaut_2013}. When a spectral type was not covered by \citet{Pecaut_2013}, we adopted the stellar mass reported by \citet{Anders_2019}. For a small number of stars with unusually high metallicities, we instead adopted the stellar masses reported by \citet{Pinamonti_2023, Schorr_2025}. All stellar-mass priors were assigned a standard deviation of $0.25 \rm M_{\odot}$. The remaining prior settings are summarized in Table~\ref{tab: prior}.

For this preliminary stage, our Markov Chain Monte Carlo (MCMC) setup employed 20 temperatures, 100 walkers per temperature, and 100000 steps per walker. The first 500 steps were discarded as burn-in, and the chains were thinned by retaining every 50th step. Several of the RV instruments included in our analysis underwent one or more instrumental upgrades. As above, observations obtained before and after such upgrades were treated as separate instrumental data sets. \texttt{Orvara} simultaneously fitted all available RV data, assigning an independent RV zero-point offset and jitter term to each instrumental configuration as part of the joint orbital fit. 

Based on these preliminary fits, we excluded systems with inferred companion masses exceeding 100 $\rm M_{\rm Jup}$, unless their uncertainties allowed for masses below 80 $\rm M_{\rm Jup}$. We then visually re-examined the orbital solutions and posterior distributions for the remaining systems, particularly those with significant RV detections for which the joint RV–astrometry fits failed to converge to reliable orbital solutions. 
In some of these cases, the available astrometric measurements were insufficient to constrain the companion orbit, often because of their relatively short orbital periods or a combination of noisy RV data and weak astrometric signals.
%or bad astrometric constraints. 
Given the significance of their RV detections, we retained these systems and performed a more detailed RV-only MCMC analysis with \texttt{RVSearch} to constrain their orbital parameters. Because the orbital inclinations cannot be constrained from the RV data alone, we derived and report the minimum companion masses, m$\sin i$, for these systems rather than their true dynamical masses.

For the remaining candidates, we performed a second, more detailed \texttt{orvara} analysis. Compared with the preliminary fitting stage, we increased the number of temperatures to at least 30 and the number of steps per walker to at least 200,000 to improve the convergence and sampling of the posterior distributions. 
%We used the preliminary \texttt{orvara} solutions as prior information to initialize the companion parameters for the refined fits , 
We used the preliminary \texttt{orvara} solutions as initial guesses for the companion parameters in the refined fits, thereby improving sampling efficiency and reducing convergence toward poor local solutions. After the refined orbital analysis, we retained systems that either represent previously unreported companion candidates or correspond to known companions for which only the minimum mass, m$\sin i$, had previously been reported. 

\subsubsection{Complementary Dynamical Analysis with \texttt{Octofitter}}
%\textcolor{red}{Following the refined \texttt{orvara} analysis, we found that several companions exhibited inclination posterior distributions centered near 90 deg, with uncertainties ranging from 30 to 50 deg. These broad distributions suggest that the inclination constraints may be strongly influenced by the isotropic orientation prior. In addition, some systems with significant RV detections yielded only minimum companion masses due to sifufficient astrometric constraints, as discussed above. To further assess the consistency of our orbital solutions and investigate whether additional astrometric information could provide meaningful constraints on true masses of these companions, we performed joint RV-astrometry analyses using \texttt{Octofitter} \citep{Thompson_2023} with the G23H catalog.}

The refined \texttt{orvara} analysis left several companions with inclination posterior distributions centered near 90$^{\circ}$, with uncertainties of $\sim$30--50$^{\circ}$, strongly resembling the input (isotropic) system orientation prior and only ruling out nearly face-on solutions implying massive companions that can be ruled out by HGCA astrometry.  Additionally, \texttt{orvara} failed to yield solutions with many RV detections having periods significantly shorter than the HGCA temporal baseline but comparable to the Gaia DR2--DR3 baseline (e.g. P $\lesssim$ 2 years). To further assess the consistency of our orbital solutions and investigate whether additional astrometric information could provide meaningful constraints on true masses of these companions, we performed joint RV-astrometry analyses using \texttt{Octofitter} \citep{Thompson_2023} with the G23H catalog.

We adopted prior distributions broadly consistent with those used in the \texttt{orvara} analysis, with additional bounds on several parameters. Specifically, the host-star mass was assigned a Gaussian prior truncated at 0, while the companion mass and semimajor axis followed log-flat priors bounded by $0.01\, M_{\rm Jup} < M_{\rm sec} < M_{pri,\rm prior}$ and $0.001 < a < 1024$ au, respectively. The Gaia parallax prior was modeled as a Gaussian distribution truncated below half the catalog parallax. These prior bounds are not expected to significantly affect the inferred orbital parameters, provided that the posterior distributions remain well within the adopted limits. However, for multi-companion systems, the \texttt{Octofitter} model assumed coplanar orbits, with all companions sharing a common inclination and longitude of ascending node.

The host star's mass posterior is prior dominated, and \texttt{Octofitter} largely served as a check on the \texttt{orvara} results.  Thus, when the \texttt{Octofitter} stellar mass posterior is multimodal under the initial $\pm$ 0.25 $M_{\rm \odot}$ prior, we reran the fits using a narrower Gaussian prior of width 0.05 $M_{\rm odot}$.  We adopt the \texttt{orvara} results by default but use \texttt{Octofitter} when the \texttt{orvara} did not converge to a solution, or when \texttt{Octofitter} yielded tighter constraints.

\begin{deluxetable}{lc}\label{tab: prior}
\tablecaption{Priors Adopted for \texttt{orvara}}
\tablewidth{\textwidth}
\tablehead{
\colhead{Parameter} & \colhead{Prior}
} 
\startdata
RV jitter, $\sigma_{\rm jit}$ ($\mathrm{m\,s^{-1}}$) & $1/\sigma_{\rm jit}$ (log-flat) \\
$M_{pri}$ ($M_\odot$) & $\mathcal{N}(M_{\rm prior},\,0.25 M_\odot)$\\
%$M_{pri}$ ($M_\odot$) & $\mathcal{N}(M_{\rm prior},\,0.25^2 M_\odot^2)$\\
$M_{sec}$ ($M_{\rm Jup}$) & $1/M_{\rm sec}$ (log-flat)\\
$a$ (au) & $1/a$ (log-flat) \\
$\sqrt{e}\sin\omega$ & uniform \\ 
$\sqrt{e}\cos\omega$ & uniform \\
Inclination $i$ & $\sin i$ $(0^\circ < i < 180^\circ$) \\
Mean longitude at 2010.0, $\lambda_{\rm ref}$ (deg) & uniform \\
Ascending node, $\Omega$ (deg) & uniform \\
Parallax $\varpi$ & $\mathcal{N}(\varpi_{\rm Gaia},\,\sigma_{\varpi,\rm Gaia}^2)$ \\
\enddata
\end{deluxetable}

\section{New Discoveries and Dynamical Masses}\label{sec: discovery}

Our final sample contains 45 systems. The relevant \texttt{RVSearch} results, astrometric information and Pearson correlation coefficients with their associated $p$-values are listed in Appendix A in Tables~\ref{tab: new_planetnew_harpshgca} and \ref{tab: recovered_planetnew_harpshgca} for new companions and known ones, respectively.  

%Analysis results for all targets are included in a Zenodo link\footnote{\url{doi:10.5281/zenodo.24455353}}.   
Analysis results for all targets are archived on Zenodo\footnote{\url{https://doi.org/10.5281/zenodo.24455353}}.

\begin{figure*}
\setlength{\unitlength}{\textwidth}
\begin{picture}(1,0.95)   % height = corner-plot aspect ratio
  \put(0,0){\includegraphics[width=\textwidth]{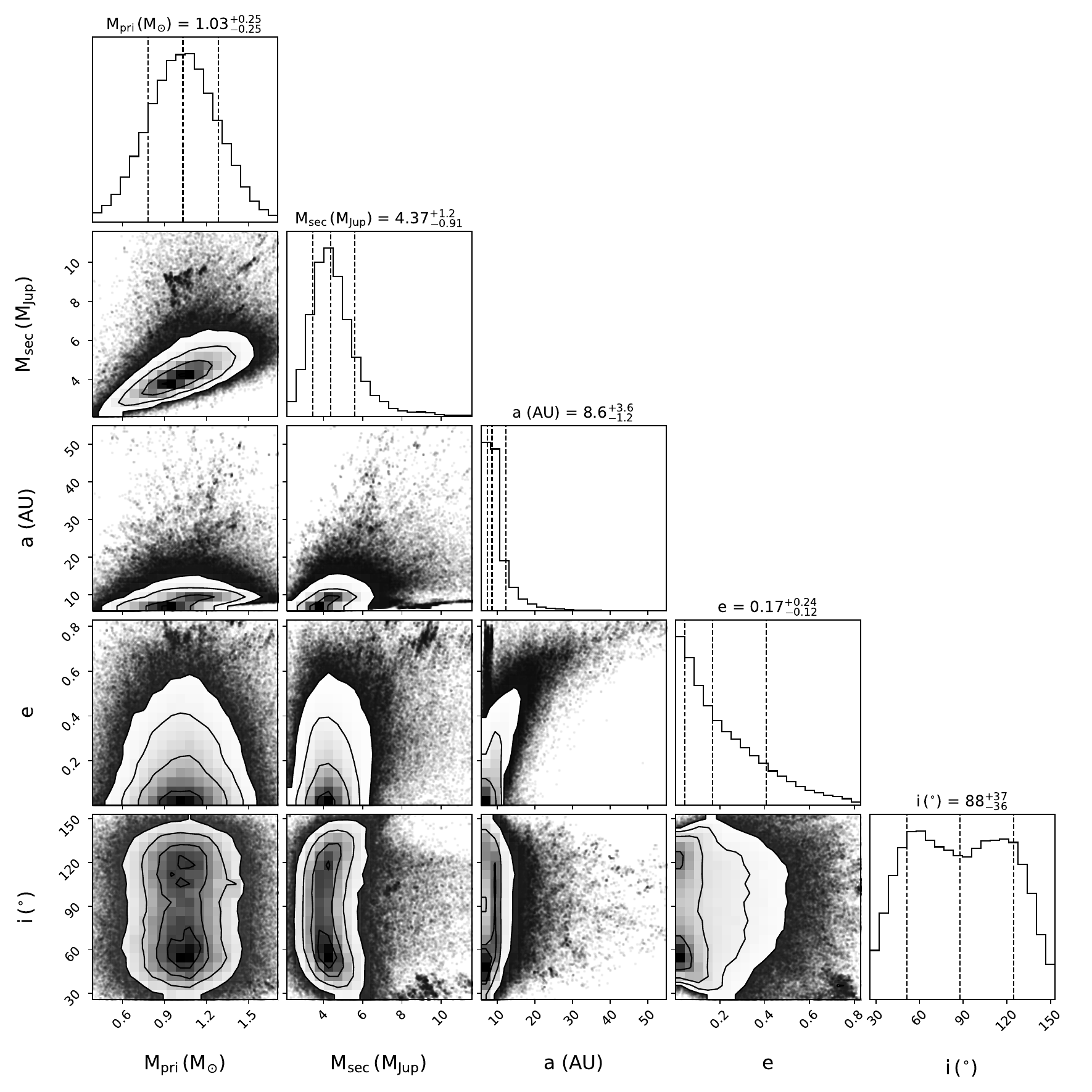}}
  \put(0.725,0.7){\includegraphics[width=0.30\textwidth]{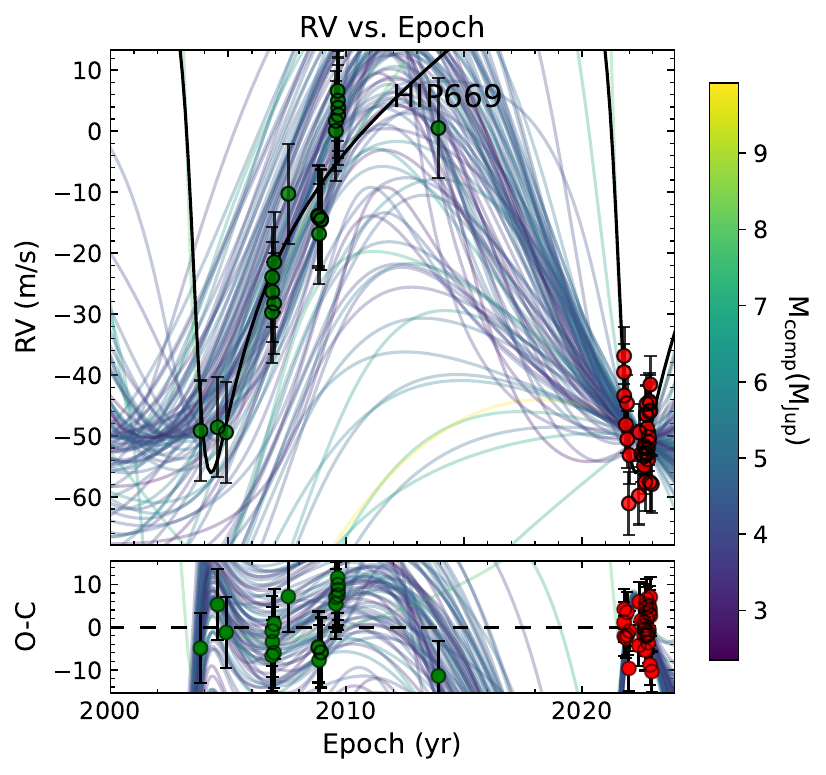}}
  \put(0.70,0.41){\includegraphics[width=0.30\textwidth]{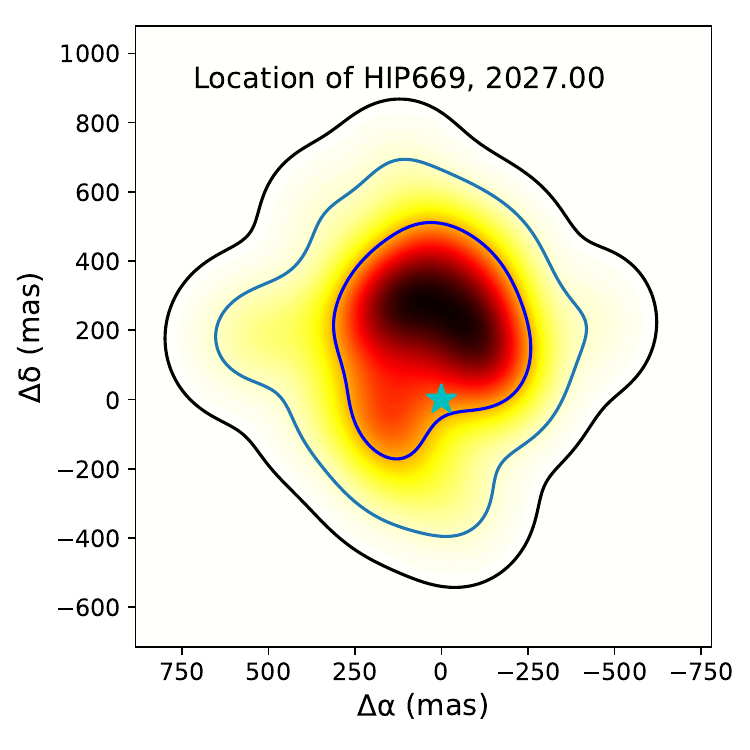}}
\end{picture}
    \caption{Corner plot showing posterior distributions of selected orbital parameters for HIP 669 b from jointly modeling RV and astrometric data with \texttt{orvara}.  The contour lines delineate the regions encompassing 68, 95, and 99\% of the posteriors. The posterior distribution of the inclination, $i$, is bimodal, with peaks at approximately 61 deg and 116 deg, corresponding to prograde and retrograde orbits, respectively.  The insets show the \texttt{orvara} fits to RV data and the posterior distribution of the predicted location for HIP 669 b on January 1, 2027. The inset values give the median and 68\% interval of the full, bimodal inclination posterior; separating the two modes gives $i$ = 61 $\pm$ 20 deg and 116$^{+20}_{-18}$ deg.}
    \vspace{-0.in}
    \label{fig:corner_plotgauss_hip669}
\end{figure*}

We identify a total of 14 new planets -- 9 at ``long" periods fit with \texttt{orvara}, typically greater than 2 years, and 5 at shorter periods -- and two long-period binaries with dynamical masses.  An additional three signals are candidate planets requiring new data for confirmation.  Table~\ref{tab: new_planet} in Appendix B lists the dynamical properties of the long-period planets and binaries, which we discuss in Section 4.1: Figure \ref{fig:corner_plotgauss_hip669} illustrates dynamical analysis of one detection.  
%A total of 11 newly-identified companions have dynamical masses determined from \texttt{orvara}, including 10 exoplanets and one likely brown dwarf.  Tables~\ref{tab: new_planet} in Appendix B lists the dynamical properties of these companions, which we discuss in Section 4.1: Figure \ref{fig:corner_plotgauss_hip669} illustrates dynamical analysis of one detection.  With few exceptions (noted below), complementary modeling from \texttt{Octofitter} finds identical orbital parameters.

For the short-period planets, \texttt{RVSearch} provides evidence for a planet with a period typically $\lesssim$ 1 year that is not well explained by stellar activity (Table \ref{tab:new_planet_low_mass}).  While these planets cannot be well modeled by \texttt{orvara} since the orbital period is significantly shorter than the Hipparcos-Gaia astrometric baseline, \texttt{Octofitter} successfully yields constrained orbital solutions (Table \ref{tab: new_planet_octofitter} in Appendix B, Section \ref{sub: newshort}): Figure \ref{fig: HIP64550} illustrates analysis demonstrating one detection.   The three candidate planets are described in Section \ref{sub: candidates}.

Finally, we update the dynamical masses and orbits of 32 previously known companions (Table \ref{tab: update} in Appendix B, Section \ref{sub: updated_mass}): Figures \ref{fig:corner_plotgauss_hip15527} and \ref{fig: HIP109388} illustrate dynamical analysis of a single-planet system and multi-planet system, respectively.  
%Section 4.4 discusses how orbital solutions calculated from \texttt{orvara} compare to those from \texttt{Octofitter}. 

With few exceptions (noted below), complementary modeling from \texttt{Octofitter} finds consistent orbital parameters.  \texttt{Octofitter} failed to obtain reliable solutions for HIP 23575 and HIP 52409, and recovered only one companion in HIP 30579.  For nine systems (HIP 15527, HIP 17378, HIP 30579, HIP 31540, HIP 83941, HIP 96078, HIP 109378, HIP 112414, and HIP 116906), \texttt{Octofitter} showed multimodal stellar mass posteriors even though the companion masses and most orbital parameters remained relatively well constrained.  We refit these systems using the narrower 0.05 $M_{\rm \odot}$ prior.  For HIP 15527, HIP 31540, and HIP 116906, the host-star mass posteriors remained multimodal even after adopting a narrower Gaussian prior, albeit concentrated over a much narrower mass range. Nevertheless, the companion-mass and eccentricity posteriors were approximately Gaussian and consistent with the corresponding \texttt{orvara} results, which we adopt for these three systems.  

In addition, \texttt{Octofitter} provided orbital solutions for five systems that had not been successfully fitted with \texttt{orvara}: HIP 17378, HIP 18918, HIP 47513, HIP 51317, and HIP 64550.  \texttt{Octofitter} provided tighter inclination constraints for some planets 
%HIP 53437, HIP 63833 b, HIP 83547, HIP 84787, HIP 109378, and HIP 112414 
compared with the corresponding \texttt{orvara} solutions (e.g. HIP 63833 b). For these systems, we adopted the \texttt{Octofitter} orbital solutions for subsequent analyses, while retaining the \texttt{orvara} solutions for the remaining systems.

\subsection{New Long-Period Planets Fit with \texttt{orvara}}
\label{sub: new}

\subsubsection{HIP 669 b}
\label{subsub: HIP 669}
HIP 669 (HD 361) is a G1 V star located at a distance of 27.31 pc (Figure \ref{fig:corner_plotgauss_hip669}). The system exhibits an 8.8 $\sigma$ significant astrometric acceleration; our RV data reveal a periodic signal with a false-alarm probability below 0.1\%, which is well described by a Keplerian model. We find no significant correlation between the RV variations and stellar activity indicators, supporting an orbital origin of the signal rather than stellar activity. The \texttt{orvara} fit reveals a planet with a true mass of ${4.37}_{-0.91}^{+1.2}\,M_{\rm Jup}$, a semi-major axis of ${8.6}_{-1.2}^{+3.6}$ au, and an eccentricity of ${0.17}_{-0.12}^{+0.24}$. The inclination posterior is bimodal, with two degenerate solutions at ${61}_{-20}^{+20}$ and ${116}_{-18}^{+20}$ deg, corresponding to prograde and retrograde orbits, respectively. The orbital period is ${24.4}_{-3.7}^{+18}$ yr, and the maximum projected angular separation is estimated to be $0.35^{+0.19}_{-0.06}\arcsec$. \texttt{Octofitter} finds nearly identical parameters: $4.50^{+1.01}_{-0.88}$ $M_{\rm Jup}$,  $9.04^{+4.15}_{-1.45}$ au, $e$ = $0.19^{+0.21}_{-0.14}$, and $i$ = $62^{+20}_{-17}$ or $116^{+18}_{-17}$$^{\circ}$.
%The maximum a posteriori (MAP) orbit further implies a maximum projected angular separation of approximately 0\farcs{}38. 
We identify the new companion as HIP 669 b (HD 361 b) and classify it as a giant planet based on its dynamical true mass.

%\subsubsection{HIP 5750 B}
%\label{subsub: HIP 5750}
%HIP 5750 (HD 7399) is a F5 V star located at a distance of 111.82 pc. The system exhibits a very strong astrometric acceleration (10.3 $\sigma$) and RV signal.   We find no significant correlation between the RV variations and stellar activity indicators. \textcolor{red}{The \texttt{orvara} fit reveals a low-mass companion with a true mass of ${101}_{-23}^{+30}\,M_{\rm Jup}$, a semi-major axis of ${2.55}_{-0.49}^{+0.62}$ au, an eccentricity of ${0.813}_{-0.16}^{+0.078}$, and an inclination of ${41}_{-17}^{+55}$ deg. The orbital period is ${3.44}_{-0.88}^{+1.4}$ yr, and the maximum projected angular separation is estimated to be $0.04^{+0.01}_{-0.01}\arcsec$}.
%The orbit inferred from the MAP solution further implies a maximum projected angular separation of approximately 0\farcs{}05. 
%\textcolor{red}{\texttt{Octofitter} dynamical modeling corrobrates these solutions, although the inclination posterior distribution is much more tightly constrained.}
%We identify the new companion as HIP 5750 B (HD 7399 B) and classify it as a low-mass binary near the hydrogen burning limit: a candidate brown dwarf.  Follow up data may clarify whether it likely resides above or below the hydrogen burning limit.

\subsubsection{HIP 25421 b}
\label{subsub: HIP 25421}
HIP 25421 (HD 35854, GJ 1079) is a K2/3 V star located at a distance of 17.86 pc. While it displays a tentative astrometric acceleration ($\sim$ 2-$\sigma$), RV data reveal a clear periodic signal at $\sim$ 1871 days with a false-alarm probability below 0.1\%, which is well described by a Keplerian model. We find no significant correlation between the RV variations and stellar activity indicators, supporting an orbital origin of the signal rather than stellar activity. The \texttt{orvara} fit reveals a planet with a true mass of ${0.102}_{-0.037}^{+0.088}\,M_{\rm Jup}$, an eccentricity of ${0.18}_{-0.13}^{+0.22}$, and an inclination of ${92}_{-50}^{+50}$ deg. The semi-major axis posterior exhibits two distinct peaks, at ${1.36}_{-0.15}^{+0.12}$ au and ${2.81}_{-0.37}^{+0.34}$ au, corresponding to orbital periods of approximately ${1.78}_{-0.48}^{+0.64}$ yr and ${5.30}_{-1.56}^{+2.24}$ yr, respectively: \texttt{Octofitter} arrives at consistent conclusions. Additional RV and astrometric observations will be required to distinguish between these orbital solutions and better constrain the orbit. For the subsequent analysis, we adopted the solution with $a = 2.81$ au, as the posterior probability is predominantly concentrated around this solution.
The maximum projected angular separation is estimated to be $0.17^{+0.03}_{-0.04}\arcsec$.
%The orbit inferred from the MAP solution further implies a maximum projected angular separation of approximately 0\farcs{}11. 
We identify the new companion as HIP 25421 b (HD 35854 b, GJ 1079 b) and classify it as a giant planet based on its dynamical true mass.

\subsubsection{HIP 31540 c}
\label{subsub: HIP 31540}
HIP 31540 (HD 47186) is a G6 V star located at a distance of 37.39 pc.  This system was previously reported to host two planets with orbital periods of $4.0845\pm0.0002$ days and $1353.6\pm57.1$ days, identified as HD 47186 b and c \citep{Bouchy_2009}, respectively. However, their data covered only one RV minimum for the outer claimed signal; \citet{Lagrange2023} showed that a $\approx$1350 day orbit is not confirmed.  Neither previous study combined the RVs with astrometry to constrain the true mass: the Hipparcos-only analysis of S.~Reffert \& A.~Quirrenbach (2011), which assumed the 1354-d orbit, allowed masses up to $66\,M_{\rm Jup}$ at $3\sigma$, and A.-M.~Lagrange et al.\ (2023) used RVs alone. Thus, we argue that this companion at its 46-year period was not established until now.  

Our RV data reveal two periodic signals with a false-alarm probability below 0.1\%, which are well described by  Keplerian models.   We find no significant correlation between the RV variations and stellar activity indicators, supporting an orbital origin of the signals rather than stellar activity.  The shorter-period signal is consistent with the previously reported 4.08-day orbit of the b companion.   However, for the outer companion, our \texttt{RVSearch} analysis identifies a substantially longer period of 6252.95 days. Although the \texttt{RVSearch} and \texttt{orvara} solutions differ in their preferred periods, both indicate an orbit substantially longer (5--12 times) than the previously reported $1353.6$-day period corresponding to a second planet.  The \texttt{orvara} fit reveals a planet with a true mass of ${0.75}_{-0.17}^{+0.30}\, M_{\rm Jup}$, a semi-major axis of ${12.6}_{-4.1}^{+10}$ au, an eccentricity of ${0.78}_{-0.13}^{+0.10}$, and an inclination of ${89}_{-39}^{+39}$ deg. The orbital period is ${46}_{-20}^{+61}$ yr, and the maximum projected angular separation is estimated to be $0.39^{+0.30}_{-0.13}\arcsec$. 
%The orbit inferred from the MAP solution further implies a maximum projected angular separation of approximately 0\farcs{}57.  
Octofitter's parameter posteriors agree with \texttt{orvara}'s, albeit with some evidence for multimodality (e.g. a peak for semi-major axis at $\sim$10 au and $e$ $\sim$ 0.6 with a much smaller second peak at 25 au and $e$ $\sim$ 0.85).

The inclination is not well constrained, while our dynamical mass ($\sim$ 0.75 $M_{\rm Jup}$) is $\sim$2 times higher than the $m~sin(i)$ reported from \citealt{Bouchy_2009}.
%: thus our detection of a c planet appears to be a substantially different kind of object.
%Because our analysis incorporates a substantially larger RV data set and a longer observational baseline, these results suggest that the previously reported orbital period of HD 47186 c may need to be revised. 
Additional long-baseline RV and astrometric observations will be important for further constraining the orbit of this outer companion.

\subsubsection{HIP 39039 b}
\label{subsub: HIP 39039}
HIP 39039 (HD 66040) is a K0 V star located at a distance of 54.20 pc. The system exhibits a 2.3-$\sigma$ significant astrometric acceleration, while \texttt{RVSearch} identifies a potential periodic signal at 3486 days.  The RV signal's FAP value is high due to multiple peaks at smaller periods and the fact that \texttt{RVSearch} considers only circular orbits.  However, a full Keplerian model is strongly preferred over a no-planet model ($\Delta{\rm BIC}\approx39$), and periods shorter than $\sim$8\,yr are excluded ($\Delta{\rm BIC}\geq10$ relative to the best fit). The RV evidence derives primarily from a single periastron passage sampled by HARPS between 2016 and 2021.   We find no significant correlation between the RV variations and stellar activity indicators. 

The \texttt{orvara} fit reveals a planet with a true mass of ${1.37}_{-0.36}^{+0.52}\, M_{\rm Jup}$, a semi-major axis of ${4.51}_{-0.52}^{+0.54}$ au, and an eccentricity of ${0.66}_{-0.10}^{+0.11}$. The inclination posterior is bimodal, with two degenerate solutions at ${59}_{-21}^{+22}$ and ${120}_{-22}^{+22}$ deg, corresponding to prograde and retrograde orbits, respectively. The orbital period is ${10.27}_{-0.82}^{+0.53}$ yr, and the maximum projected angular separation is estimated to be $0.12^{+0.02}_{-0.02}\arcsec$.
Aside from a long tail of large semimajor axis solutions, \texttt{Octofitter} arrives at consistent answers. We identify the new companion as HIP 39039 b (HD 66040 b) and classify it as a giant planet based on its dynamical mass, albeit one with more poorly constrained parameters due to its multimodality.

\subsubsection{HIP 66487 b}
\label{subsub: HIP 66487}
HIP 66487 (HD 118466) is a K2 V star located at a distance of 47.87 pc. While the star lacks a significant astrometric acceleration, our RV data reveal a periodic signal with a false-alarm probability below 0.1\%, which is well described by a Keplerian model.  Although the period is close to one year, the signal is not an annual systematic: a period of exactly 365.25\,days is disfavored ($\Delta{\rm BIC}\approx23$); the pre- and post-2015 HARPS data independently recover the same amplitude and phase; the signal persists when only within-season RV variations are fitted.  We find no significant correlation between the RV variations and stellar activity indicators, supporting an orbital origin of the signal rather than stellar activity.  The S-index shows a magnetic cycle at $\sim$4200\,days but no power near the RV period.

The \texttt{orvara} fit reveals a planet with a true mass of ${0.66}_{-0.19}^{+0.40}\,M_{\rm Jup}$, a semi-major axis of ${0.945}_{-0.11}^{+0.087}$ au, an eccentricity of ${0.14}_{-0.10}^{+0.14}$, and an inclination of ${89}_{-43}^{+43}$ deg. The orbital period is ${1.0129}_{-0.0034}^{+0.0045}$ yr, and the maximum projected angular separation is estimated to be $0.021^{+0.003}_{-0.003}\arcsec$.
%The orbit inferred from the MAP solution further implies a maximum projected angular separation of approximately 0\farcs{}02. 
We identify the new companion as HIP 66487 b (HD 118466 b) and classify it as a giant planet based on its dynamical mass.

\subsubsection{HIP 71639 b}
\label{subsub: HIP 71639}
HIP 71639 (HD 128582) is a F8 V star located at a distance of 34.00 pc.  The star lacks a detectable astrometric acceleration, but our RV data reveal a periodic signal with a false-alarm probability below 0.1\%, which is well described by a Keplerian model. We find no significant correlation between the RV variations and stellar activity indicators. The \texttt{orvara} fit reveals a planet with a true mass of ${6.1}_{-1.3}^{+2.1}\,M_{\rm Jup}$, a semi-major axis of ${3.15}_{-0.87}^{+1.3}$ au, an eccentricity of ${0.34}_{-0.22}^{+0.23}$, and an inclination of ${89}_{-12}^{+12}$ deg. The orbital period is ${5.6}_{-2.2}^{+2.9}$ yr, and the maximum projected angular separation is estimated to be $0.10^{+0.08}_{-0.03}\arcsec$.  Octofitter yields nearly identical results.   For both dynamical codes, the inclination posterior is significantly narrower than an isotropic prior, indicating that HGCA and G23H astrometry are ruling out long-period, high-mass solutions.
%The orbit inferred from the MAP solution further implies a maximum projected angular separation of approximately 0\farcs{}19. 
We identify the new companion as HIP 71639 b (HD 128582 b) and classify it as a giant planet based on its dynamical mass.

\subsubsection{HIP 103854 b}
\label{subsub: HIP 103854}
HIP 103854 (HD 199847) is a G2 V star located at a distance of 89.52 pc. Although the star lacks a detectable acceleration, our RV data reveal a periodic signal with a false-alarm probability below 0.1\%, which is well described by a Keplerian model. We find no significant correlation between the RV variations and stellar activity indicators. \texttt{RVSearch} finds an orbital period of $\sim$ 6.5 years.  Additional analysis suggests that sparse HARPS data between 2003 and 2008 are almost solely responsible for this constraint.  Without these data, the post-2015 data still require a long-period signal but allow a wide range of periods ($\sim$4--25 yr).  The \texttt{orvara} fit reveals a planet with a true mass of ${0.271}_{-0.073}^{+0.17}\,M_{\rm Jup}$, a semi-major axis of ${3.62}_{-0.36}^{+0.44}$ au, and an eccentricity of ${0.127}_{-0.089}^{+0.15}$. The orbital period is ${6.76}_{-0.24}^{+0.48}$ yr, and the maximum projected angular separation is estimated to be $0.044^{+0.008}_{-0.005}\arcsec$. The inclination posterior mirrors the isotropic prior.  Octofitter corroborates these solutions.
%The orbit inferred from the MAP solution further implies a maximum projected angular separation of approximately 0\farcs{}04. 
We identify the new companion as HIP 103854 b (HD 199847 b) and classify it as a giant, Saturn-mass planet on a multi-year orbit whose period requires additional RV monitoring or astrometry to more precisely estimate.
%based on its inferred true mass whose .

\subsubsection{HIP 104045 b}
\label{subsub: HIP 104045}
HIP 104045 (HD 200633) is a G5 V star located at a distance of 53.79 pc. While the star lacks a detectable acceleration, our RV data reveal a periodic signal at 2462 days with a false-alarm probability below 0.1\%.  The S-index shows a magnetic-cycle signal at $\sim$3300\,days, which is not clearly resolved from the RV period given our baseline.  However, we attribute the RV signal to a planet.  We find no significant correlation between the RV variations and stellar activity indicators, and between 2011 and 2014 the RVs vary by $\sim$ 20 m s$^{-1}$, while the S-index remains constant, supporting an orbital origin of the signal rather than stellar activity.  The pre- and post-2015 HARPS data independently recover the RV signal with a consistent phase, whereas the S-index's phase changes.

The \texttt{orvara} fit reveals a planet with a true mass of ${0.62}_{-0.14}^{+0.21}\,M_{\rm Jup}$, a semi-major axis of ${3.52}_{-0.33}^{+0.29}$ au, an eccentricity of ${0.142}_{-0.082}^{+0.079}$, and an inclination of ${91}_{-35}^{+35}$ deg. The orbital period is ${6.69}_{-0.18}^{+0.18}$ yr, and the maximum projected angular separation is estimated to be $0.07^{+0.01}_{-0.01}\arcsec$. Octofitter finds consistent solutions.
%The orbit inferred from the MAP solution further implies a maximum projected angular separation of approximately 0\farcs{}08. 
We identify the new companion as HIP 104045 b (HD 200633 b) and classify it as a giant planet based on its dynamical mass.

A prior arxiv submission identified two planets around this star -- a cold Neptune at 316 days and a jovian planet at 2315 days -- but this submission was withdrawn \citep{Ferreira_2023}, and to our knowledge no subsequent claim of a planet has been presented in an arxiv posting or in published peer-reviewed literature.   Our analysis does not support the existence of a 316 day planetary signal: this period, though, is close to a peak in the S-index periodogram.  The longer-period signal is very similar to that we find.  Our data include several additional HARPS data points beyond those considered in that arxiv submission that appear to confirm a periodicity at $\sim$2462 days.

\subsubsection{HIP 118009 b}
\label{subsub: HIP 118009}
HIP 118009 (HD 224230) is a K3 V star located at a distance of 49.63 pc. The star lacks a detectable acceleration, but our RV data reveal a periodic signal with a false-alarm probability below 0.1\%, which is well described by a Keplerian model with an MCMC period of $\sim$1166 days ($\sim$3.2 years).  While the \texttt{RVSearch} periodogram shows a second peak at $\sim$500 days, this appears to be a yearly alias and is suppressed once the longer-period signal is removed. We find no significant correlation between the RV variations and stellar activity indicators, supporting an orbital origin of the signal rather than stellar activity.

The \texttt{orvara} fit reveals a planet with a true mass of ${0.59}_{-0.15}^{+0.24}\,M_{\rm Jup}$, a semi-major axis of ${2.00}_{-0.24}^{+0.20}$ au, an eccentricity of ${0.096}_{-0.066}^{+0.093}$, and an inclination of ${91}_{-38}^{+37}$ deg. The orbital period is ${3.192}_{-0.043}^{+0.045}$ yr, and the maximum projected angular separation is estimated to be $0.04^{+0.01}_{-0.01}\arcsec$.  Octofitter yields nearly identical posteriors.
%The orbit inferred from the MAP solution further implies a maximum projected angular separation of approximately 0\farcs{}04. 
We identify the new companion as HIP 118009 b (HD 224230 b) and classify it as a giant planet based on its dynamical mass.

\begin{figure*}[!tp]
%\begin{figure}[!tp]
\centering
\includegraphics[width=0.32\textwidth,trim=8mm 0mm 15mm 0mm,clip]{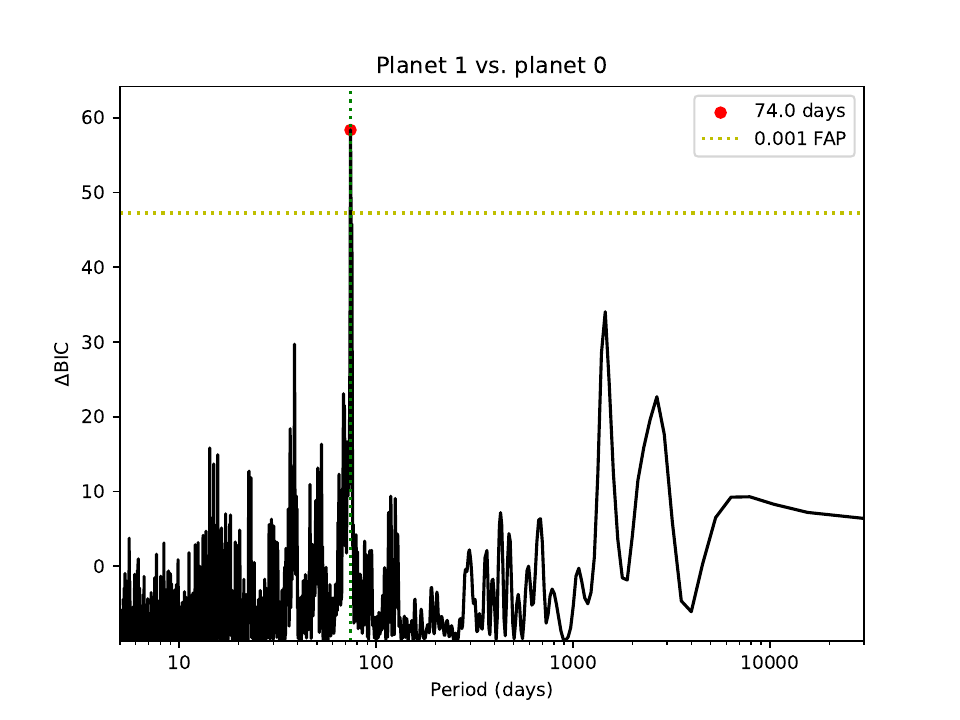}
\includegraphics[width=0.39\textwidth,trim=2mm 0mm 5mm 0mm,clip]{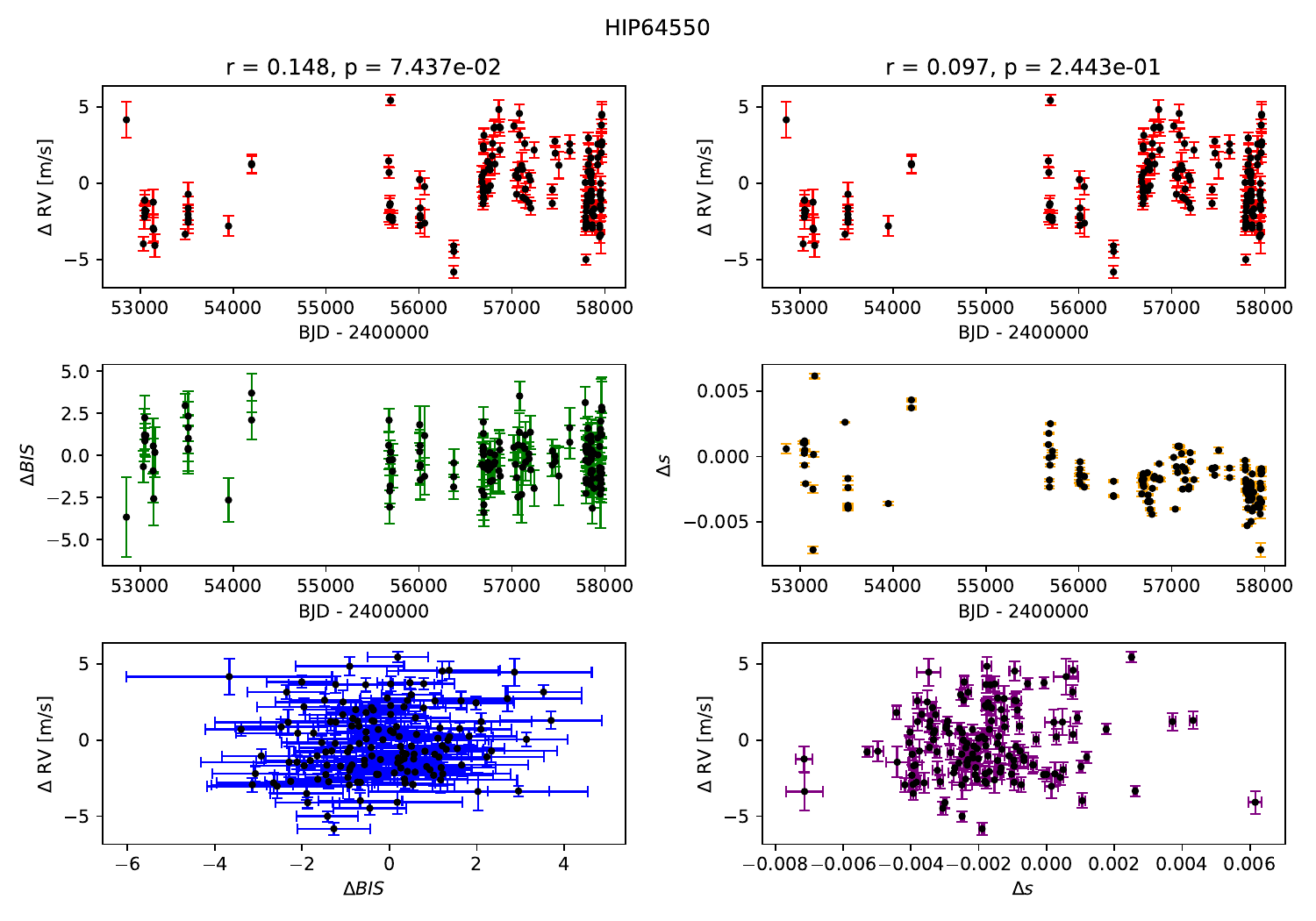}
\includegraphics[width=0.26\textwidth,trim=2mm 0mm 5mm 0mm,clip]{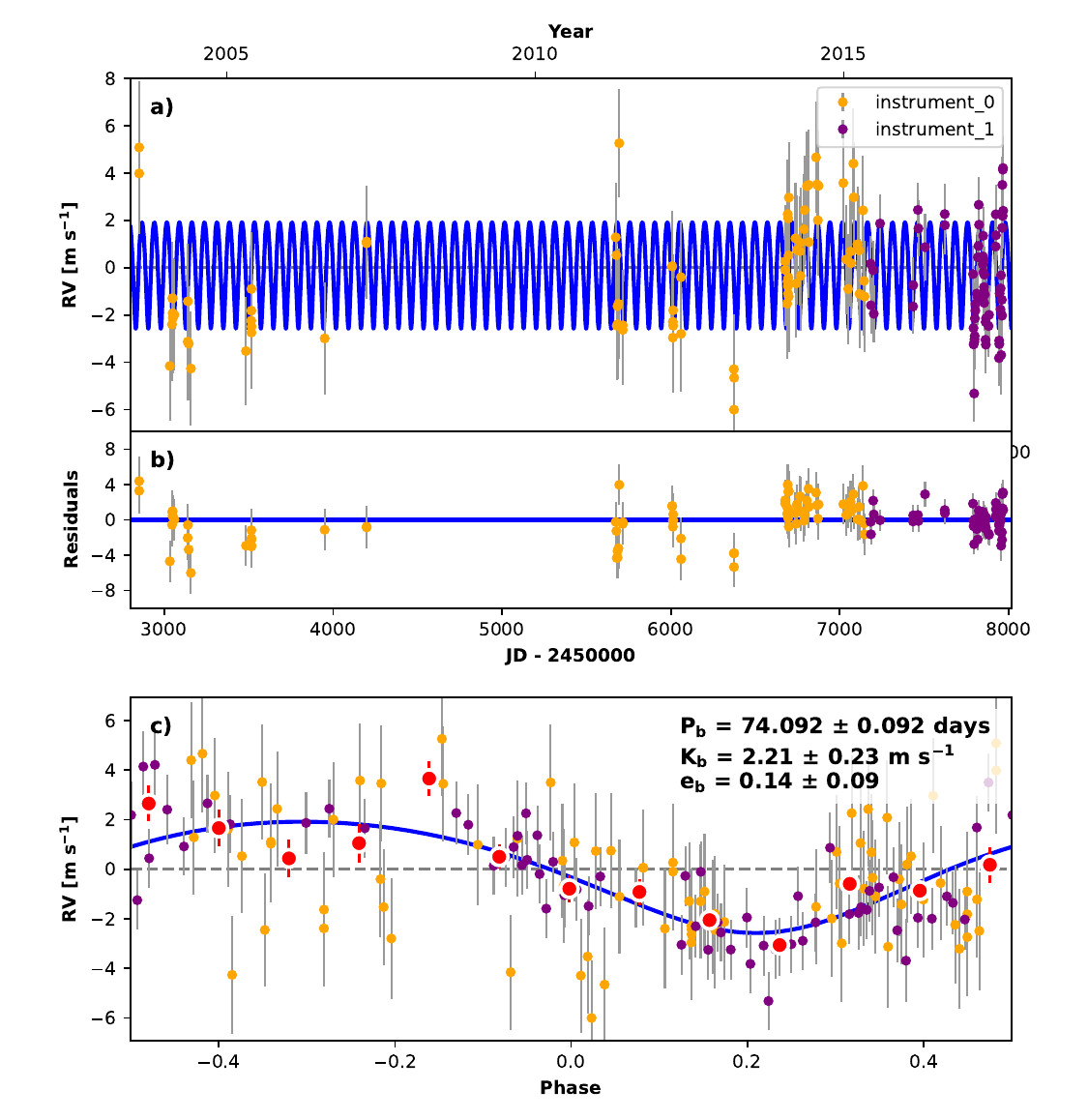}
\caption{Analysis for HIP 64550, which shows evidence for a 16 $M_{\rm \oplus}$ planet. (left) The \texttt{RVSearch} periodogram identifies a significant 74-day peak corresponding to a periodic signal.  (middle) This signal is uncorrelated with stellar activity.  (right) \texttt{RVSearch} Monte Carlo orbit fitting converges on the same 74-day period, an RV semi-amplitude of 2.21 $\pm$ 0.23 m s$^{-1}$, and a low eccentricity ($e$ = 0.14 $\pm$ 0.09). Dynamical modeling with \texttt{Octofitter} finds a mass posterior centered on 0.05 $M_{\rm Jup}$ or $\sim$16 $M_{\rm \oplus}$.}
\label{fig: HIP64550}
\end{figure*}

\subsection{Representative Newly Identified Short-Period Planets with Masses Solely Fit by \texttt{Octofitter}}
\label{sub: newshort}
%\TC{add a few cases here.  Some options: HIP 17378 b and HIP 64550 b}
\subsubsection{HIP 17378 b}
\label{subsub: HIP 17378}
HIP 17378 (HD 23249, GJ 150) is a K0 IV star located at a distance of 9.08 pc and Tier A member of the HWO Provisional Target List \citep{Mamajek2024}.  Previously, \citet{Laliotis2023} identified a Keplerian signal at $\sim$596 days, classifying it as a ``source requiring confirmation" due to the influence of several UCLES measurements pulling the \texttt{RVSearch} solution towards a high eccentricity.  While \citet{Harada2025} recover this signal, they argue that its high eccentricity and the lack of a well-defined peak in the RV periodogram argues in favor of the signal being a false positive.

Our RV data reveal a periodic signal at 601.5 days with a false-alarm probability of 6.45 $\times$10$^{-5}$, well below the 0.1\% threshold. 
%We find no significant correlation between the RV variations and stellar activity indicators, 
The RV-activity correlation is negligible, supporting an orbital origin of the signal rather than stellar activity. Further analysis with the \texttt{RVSearch} MCMC analysis yields a minimum planet mass of $0.103 \pm 0.006\,M_{\rm Jup}$, a semi-major axis of $1.542 \pm 0.004$ au, an eccentricity of $0.185 \pm 0.092$, a radial-velocity semi-amplitude of $2.07 \pm 0.11$ m/s, and an orbital period of $601.6 \pm 2.6$ days.  However, we do see a large secondary peak just below the 0.1\% FAP threshold at $\approx$2300 days whose existence is consistent with \citeauthor{Harada2025}'s qualitative interpretation of the star's RV periodogram lacking a well-defined peak.

Due to its contentious history, we performed additional analysis to assess binning effects and the uniqueness of the 600-day signal in the RV power spectrum.  Our full RV data set for HIP 17378 consists of over 1800 RV measurements, including many HARPS measurements obtained on the same night.  Because the densely-sampled HARPS measurements might not be fully independent on sub-night timescales, we analyzed nightly-binned RVs.   \texttt{RVSearch} analysis on the binned data corroborates our unbinned data result: a $\sim$600-day period with a false alarm probability of 4.6$\times$10$^{-5}$ ($\Delta$BIC $\approx$38 relative to the no-planet hypothesis).  MCMC modeling finds $K=1.86$\,m\,s$^{-1}$ and $e=0.14$, corresponding to $m\sin i\approx0.094\,M_{\rm Jup}$: also consistent with analysis on the unbinned data.  While the $\sim$600 day signal remains significant, the 2300 day signal is highly suppressed and indistinguishable from numerous other peaks once the 600-day signal is removed: ambiguity about the peak periodicity in the unbinned data may be an artifact of the dense HARPS sampling.  For both the unbinned and binned data, any peak remaining after removing the 600 day periodicity fails the FAP $\le$ 0.1\% threshold.

While the existence of a $\approx$600-day planetary signal is decisive, details about its orbit are less certain.  \texttt{RVSearch} favors a low-eccentricity orbit ($e$ $\sim$ 0.185 unbinned; $e\sim0.14$ binned), but \texttt{Octofitter} analysis incorporating astrometry converges to a far higher eccentricity ($e$ $\sim$ 0.9), more massive ($\approx$ 5 $M_{\rm Jup}$), and nearly face-on solution. HIP 17378 lacks a detectable astrometric acceleration, so the independent support for a face-on solution is weak.  Thus we conclude that HIP 17378 b is a bona fide giant planet at $\approx$ 1.5 au with a poorly constrained eccentricity.

\subsubsection{HIP 64550 b}
\label{subsub: HIP 64550}
HIP 64550 (HD 114853) is a G1.5 V star located at a distance of 25.01 pc (Figure \ref{fig: HIP64550}). Our RV data reveal a periodic signal at $\sim$ 74 days with a false-alarm probability below 0.1\%, which is well described by a Keplerian model. The second most significant peak is at $\sim$38 days, which is suppressed once the primary signal is removed and thus is likely an alias.  We find no significant correlation between the RV variations and stellar activity indicators and no power in the activity signals' periodograms at 50-100 days, supporting an orbital origin of the signal rather than stellar activity. 
The \texttt{orvara} fit did not converge and appears to favor a longer-period solution ($\sim$7 yr). Because the HGCA shows no significant acceleration
($\chi^2_{\rm HGCA} = 0.41$), this preference is driven not by astrometry but
by sub-threshold long-period RV power near $\sim$1500\,days, which coincides with
a peak in the $S_{\rm MW}$ periodogram and is likely due to a magnetic activity
cycle.  
%This may indicate that the astrometric signal is not dominated by the companion detected by \texttt{RVSearch}. 
Nevertheless, the \texttt{RVSearch} MCMC converged successfully, yielding a minimum planet mass of $0.043 \pm 0.005\,M_{\rm Jup}$ (13.7 $M_{\rm \oplus})$, a semi-major axis of $0.3382 \pm 0.0003$ au, an eccentricity of $0.14 \pm 0.09$, a radial-velocity semi-amplitude of $2.21 \pm 0.23$ m/s, and an orbital period of $74.092 \pm 0.092$ days.   Dynamical modeling with \texttt{Octofitter} yields similar parameters, favoring a dynamical mass of 15.9 $M_{\rm \oplus}$ -- mainly constrained by the RV signal and inclination prior -- and a semimajor axis of 0.33 au.   We identify the new companion as HIP 64550 b (HD 114853 b) and classify it as a planet, plausibly a warm Neptune, based on its mass.

\subsection{Candidate Planets Requiring New Data For Confirmation}
\label{sub: candidates}

 \subsubsection{HIP 47513 b}
 HIP 47513 (GJ 361) is an M2.5\,V star at a distance of 11.54\,pc. In a submitted preprint that has not been published in a peer-reviewed journal, \citet{Tuomi_2019} identify a candidate around this star with a period of $\sim$29 days.   Our HARPS and HIRES RVs
reveal the same signal -- 28.93\,days --  with a low false-alarm probability (FAP $=2.2\times10^{-4}$) with a lower mass limit of 0.0364 $M_{\rm Jup}$ ($\sim$ 12 $M_{\rm \oplus}$). The S-index activity indicator shows no power at this period and the RV measurements are not correlated with this indicator's values.  \texttt{Octofitter} dynamical modeling fits favor a mass of 0.06 $M_{\rm Jup}$ but otherwise the same parameters.  However, the period lies extremely close to the 29.5-day lunar month: aliasing with the lunar month cannot be firmly excluded with the present data.  Thus, we classify HIP 47513 b as a compelling candidate planet at this point.

\subsubsection{HIP 56572 c}
HIP 56572 hosts a planet at $\approx$384\,days. A second RV signal at $\approx$192\,days identified by \texttt{RVSearch} is most likely a harmonic or alias of planet b. However, the star shows a highly significant HGCA proper-motion anomaly ($|\Delta\mu|\approx0.61$\,mas\,yr$^{-1}$, $\chi^2=173$) that the known planet b cannot produce. Gaia DR3 shows no sources within 15$''$ and a RUWE of 1.10, giving no evidence of a stellar companion or blend. 
The joint fit with \texttt{Octofitter} favors a planet (10--25 $M_{\rm Jup}$) or brown dwarf (25--35 $M_{\rm Jup}$) at 6--25 au.
We therefore designate the source of the acceleration HIP 56572 c, a candidate outer companion whose mass and orbit should be constrained by Gaia DR4 epoch astrometry and direct imaging.  Because the joint fit is dominated by the astrometric signal of candidate c, which the RVs do not constrain, we do not report updated parameters for confirmed planet b.

\subsubsection{HIP 3559 b}
HIP 3559 shows a significant HGCA acceleration 
and a long-term RV trend. Both are consistent with the known comoving M dwarf HIP 3559 B (separation 4.09$''$, $\approx$133\,au projected; $M_G=11.6$, $\approx$0.2--0.3\,$M_\odot$), which lies at position angle $241.6^\circ$. A joint fit with a single wide companion recovers a mass of $\approx$0.27\,$M_\odot$ and a semi-major axis of $\sim$110\,au, consistent with HIP 3559 B. However, \texttt{RVSearch} favors the existence of a jovian planet with an orbital period of $\approx$38 years due to additional curvature that favors a companion unrelated to HIP 3559 B, but only when the first season of HARPS data (2003) is included; without these eight measurements, a linear trend consistent with HIP 3559 B is preferred. We therefore classify HIP 3559 b as a candidate.

   \begin{figure*}
   %  \vspace{-2in}
    \setlength{\unitlength}{\textwidth}
\begin{picture}(1,0.95)   % height = corner-plot aspect ratio
  \put(0,0){\includegraphics[width=\textwidth]{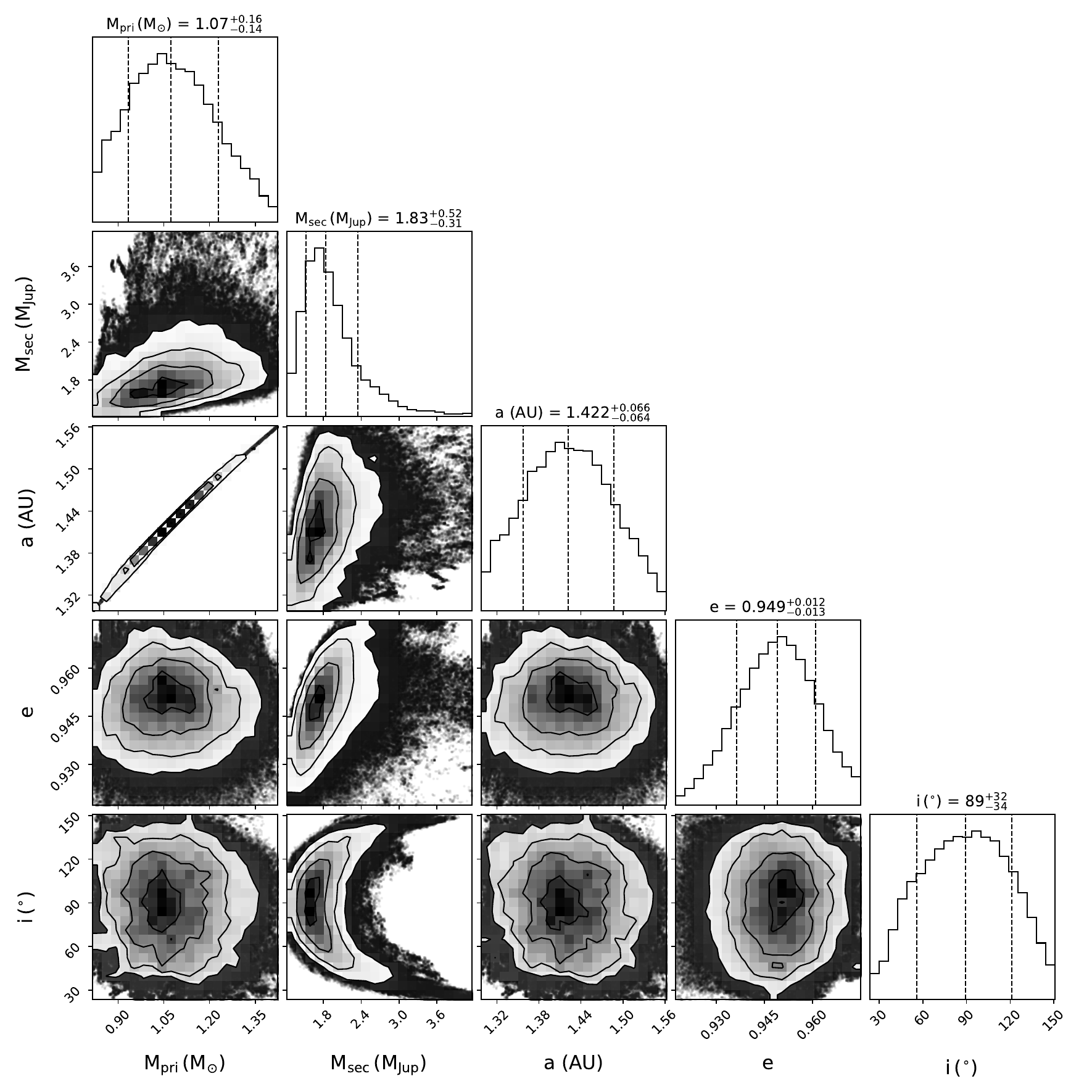}}
  \put(0.70,0.7){\includegraphics[width=0.32\textwidth]{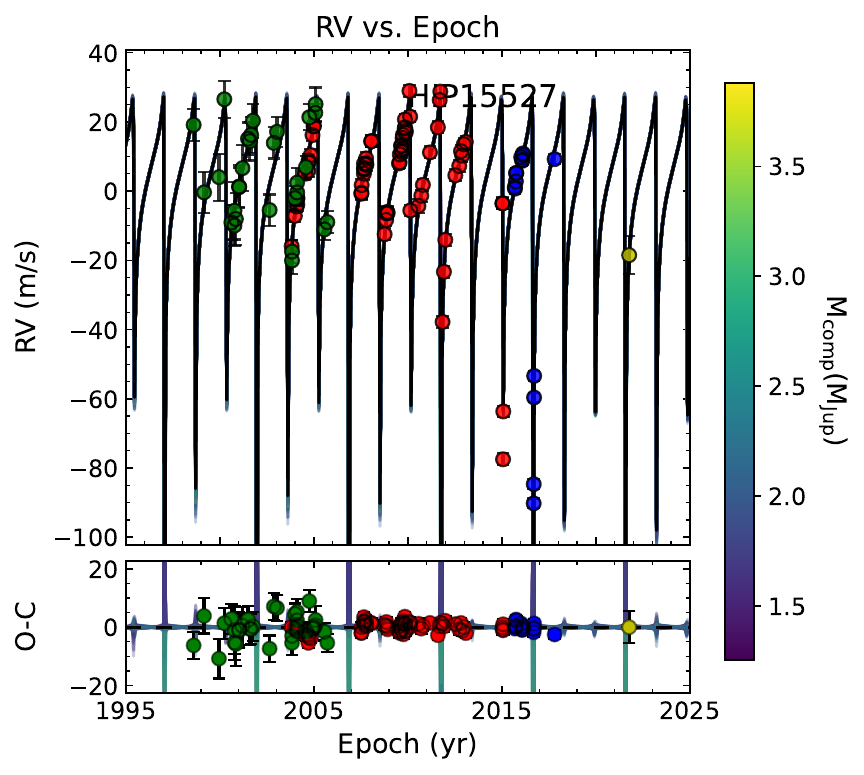}}
  \put(0.70,0.425){\includegraphics[width=0.28\textwidth]{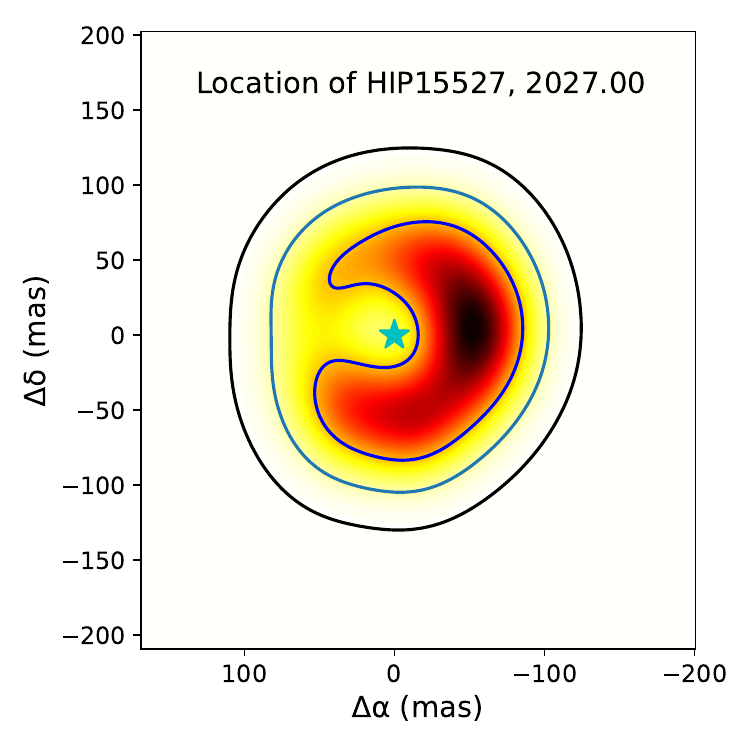}}
\end{picture}   
    \caption{Same as Figure \ref{fig:corner_plotgauss_hip669} except for HIP 15527.  }
    \vspace{-0.in}
    \label{fig:corner_plotgauss_hip15527}
\end{figure*}
\subsection{Representative Known Planets with Revised Dynamical Properties}
\label{sub: updated_mass}

\subsubsection{HIP 15527 b}
\label{subsub: HIP 15527}
HIP 15527 (HD 20782) is a G1.5 V star located at a distance of 35.87 pc (Figure \ref{fig:corner_plotgauss_hip15527}). Previous studies by \citet{Jones_2006, OToole_2009, Kane_2016, Stassun_2017, Udry_2019} reported a planet, HIP 15527 b (HD 20782 b), with an orbital period of $597.0643 \pm 0.0256$ days. The planet has a semi-major axis of $1.3649^{+0.0466}_{-0.0495}$ au, an eccentricity of $0.950 \pm 0.001$ and a minimum mass of $1.4878^{+0.1045}_{-0.1066}\,M_{\rm Jup}$.

Our RV analysis recovers this signal and finds no significant correlation between it and stellar activity indicators, while failing to find evidence for an astrometric acceleration.  The \texttt{orvara} fit yields a planet with a true mass of ${1.83}_{-0.31}^{+0.52}\,M_{\rm Jup}$, a semi-major axis of ${1.422}_{-0.064}^{+0.066}$ au, an eccentricity of ${0.949}_{-0.013}^{+0.012}$, and an inclination of ${89}_{-34}^{+32}$ deg. The orbital period is ${1.63473}_{-0.00011}^{+0.00011}$ yr, and the maximum projected angular separation is estimated to be $0.07^{+0.01}_{-0.01}\arcsec$.  Octofitter yielded consistent results.
%The orbit inferred from the MAP solution further implies a maximum projected angular separation of approximately 0\farcs{}06. 
This companion has the highest eccentricity among the planets in our sample.

\subsubsection{HIP 112414 b and c}
\label{subsub: HIP 112414}
HIP 112414 (HD 215456) is a G0.5 V star located at a distance of 39.46 pc. Previous studies by \citet{Mayor_2011}, reported two planets in the system, HIP 112414 b (HD 215456 b), and HIP 112414 c (HD 215456 c), with orbital periods of $191.99 \pm 0.73$ and $2277 \pm 67$ days, respectively. The inner planet, HIP 112414 b, has a semi-major axis of $ 0.652 \pm 0.011$ au, an eccentricity of $0.15 \pm 0.10$, and a minimum mass of $0.1013 \pm 0.0092\,M_{\rm Jup}$. The outer planet, HIP 112414 c, orbits at $3.394 \pm 0.088$ au, an eccentricity of $0.19 \pm 0.11$ and has a minimum mass of $0.2466 \pm 0.0279\,M_{\rm Jup}$.

Our RV data recovers two periodic signals with false-alarm probabilities below 0.1\%, consistent with those reported in the literature but identify no astrometric acceleration. \texttt{orvara} successfully fits a dynamical model with two planets. The inner companion, HIP 112414 b, has a true mass of ${0.128}_{-0.025}^{+0.058}\,M{\rm_{Jup}}$, a semi-major axis of ${0.703}_{-0.051}^{+0.044}$ au, an eccentricity of ${0.132}_{-0.069}^{+0.067}$, and an inclination of ${88}_{-40}^{+40}$ deg. The orbital period is ${0.52796}_{-0.00079}^{+0.00081}$ yr, and the maximum projected angular separation is estimated to be $0.019^{+0.001}_{-0.001}\arcsec$.
%The orbit inferred from the MAP solution further implies a maximum projected angular separation of approximately 0\farcs{}02. 
The outer companion, HIP 112414 c, has a true mass of ${0.342}_{-0.068}^{+0.16}\,M_{\rm Jup}$, a semi-major axis of ${3.59}_{-0.26}^{+0.23}$ au, an eccentricity of ${0.075}_{-0.051}^{+0.060}$, and an inclination of ${93}_{-43}^{+41}$ deg. The orbital period is ${6.10}_{-0.10}^{+0.11}$ yr, and the maximum projected angular separation is estimated to be $0.10^{+0.01}_{-0.01}\arcsec$.  Octofitter finds consistent solutions.
%The orbit inferred from the MAP solution further implies a maximum projected angular separation of approximately 0\farcs{}10. 
HIP 112414 b is the lowest-mass previously known planet in our sample that is fitted with \texttt{orvara}.

\subsubsection{HIP 113238 b}
\label{subsub: HIP 113238}
HIP 113238 (HD 216770) is a G9 V star located at a distance of 36.64 pc. Previous studies by \citet{Mayor_2004, Stassun_2017} reported a planet, HIP 113238 b (HD 216770 b), with an orbital period of $118.45 \pm 0.55$ days. The planet has a semi-major axis of $0.46 \pm 0.01$ au, an eccentricity of $0.37 \pm 0.06$ and a minimum mass of $0.57 \pm 0.05\, M_{\rm Jup}$.

Our RV data identifies a periodic signal consistent with this planet with false-alarm probabilities below 0.1\% and no correlation with stellar activity. The system exhibits 3.5$\sigma$-significant astrometric acceleration. The \texttt{orvara} fit yields a planet with a true mass of ${0.54}_{-0.12}^{+0.27}\,M_{\rm Jup}$, a semi-major axis of ${0.451}_{-0.039}^{+0.037}$ au, an eccentricity of ${0.106}_{-0.034}^{+0.032}$, and an inclination of ${93}_{-42}^{+40}$ deg. The orbital period is ${0.31784}_{-0.00013}^{+0.00014}$ yr, and the maximum projected angular separation is estimated to be $0.013^{+0.001}_{-0.001}\arcsec$.  Octofitter finds consistent solutions.
%The orbit inferred from the MAP solution further implies a maximum projected angular separation of approximately 0\farcs{}01.  
HIP 113238 b is the shortest-period companion in our sample for which we obtain a dynamical mass from \texttt{orvara}.

\begin{figure*}[!tp]
\includegraphics[height=0.28\textwidth]{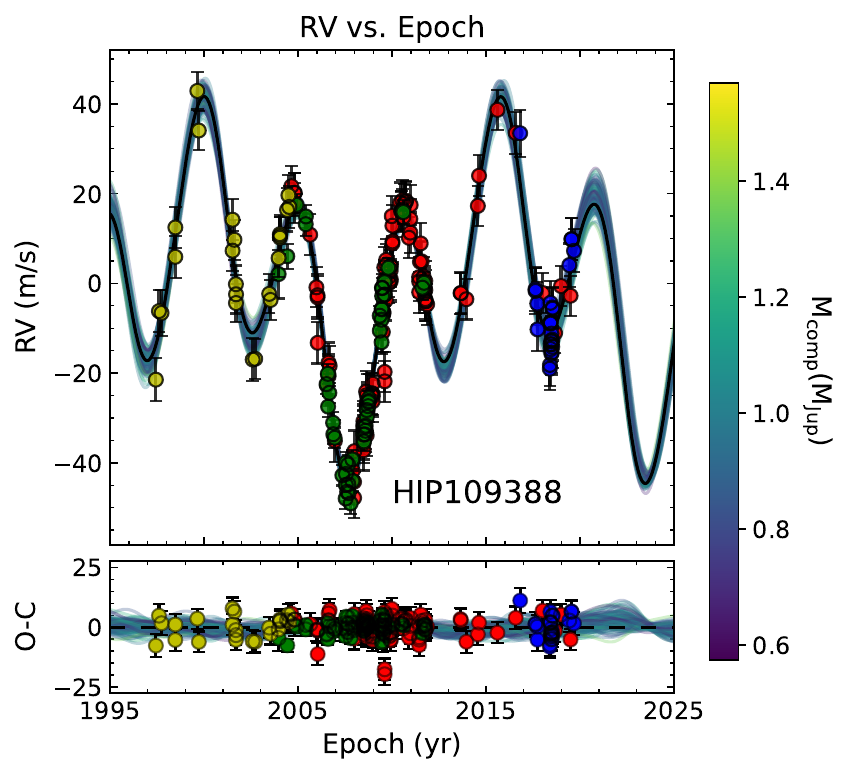}\hfill
\includegraphics[height=0.34\textwidth]{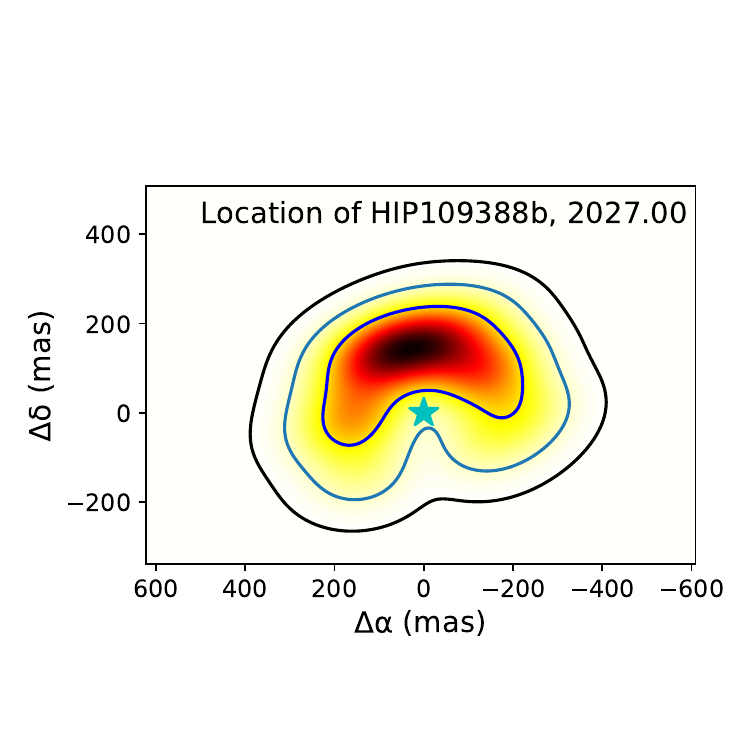}\hfill
\includegraphics[height=0.3\textwidth]{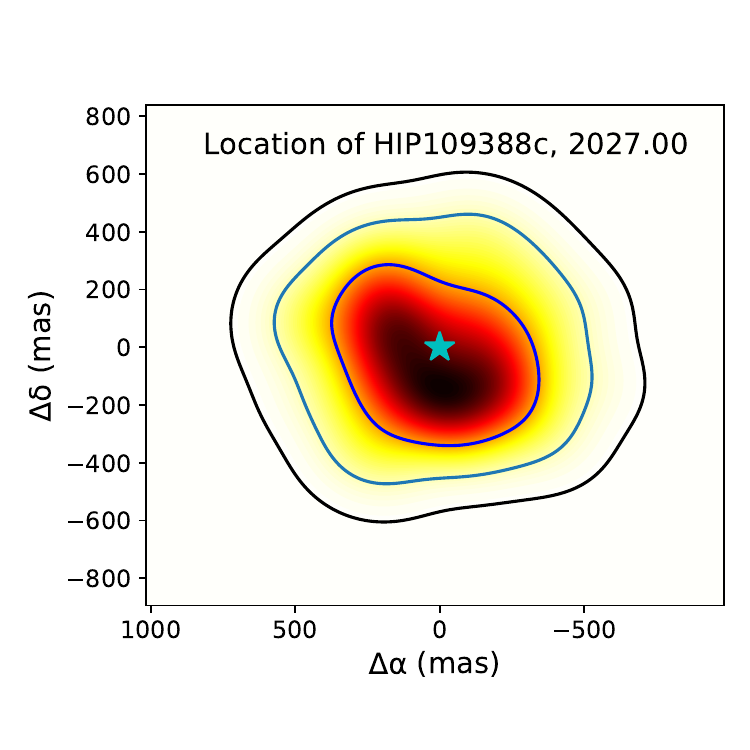}
\caption{(left) The \texttt{orvara} fit to HIP 109388 RV data, modeled as a two-planet system.  (middle, right) Posteriors distribution for the astrometric prediction of HIP 109388 b and HIP 109388 c on January 1, 2027.}
\label{fig: HIP109388}
\end{figure*}

\subsubsection{HIP 109388 b and c}
\label{subsub: HIP 109388}
HIP 109388 (GJ 849) is a M3.5 V star located at a distance of 8.81 pc (Figure \ref{fig: HIP109388}). Previous studies reported two planets in the system, HIP 109388 b (GJ 849 b), and HIP 109388 c (GJ 849 c), with orbital periods of $1925.31 \pm 6.50$ and $5990_{-100}^{+110}$ days, respectively \citep{Butler_2006, Bonfils_2013, Montet_2014, Feng_2015, Rosenthal_2021, Pinamonti_2023}. The inner planet, HIP 109388 b, has a semi-major axis of $2.32^{+0.11}_{-0.13}$ au, an eccentricity of $0.029 \pm 0.019$, and a minimum mass of $0.893^{+0.094}_{-0.097}\, M_{\rm Jup}$. The outer planet, HIP 109388 c, orbits at $4.95^{+0.25}_{-0.28}$ au, an eccentricity of $0.0920^{+0.0380}_{-0.0360}$, and has a minimum mass of $0.99\pm 0.11\,M_{\rm Jup}$.

Our RV data reveal two periodic signals with false-alarm probabilities below 0.1\% and uncorrelated with stellar activity, consistent with the companions reported in the literature. The system also exhibits 2.3-$\sigma$ significant astrometric acceleration.   The \texttt{orvara} analysis simultaneously fits two planets. The inner companion, HIP 109388 b, has a true mass of ${0.94}_{-0.16}^{+0.17}\,M_{\rm Jup}$, a semi-major axis of ${2.32}_{-0.19}^{+0.16}$ au, an eccentricity of ${0.056}_{-0.018}^{+0.017}$, and an inclination of ${95}_{-23}^{+21}$ deg. The orbital period is ${5.264}_{-0.020}^{+0.020}$ yr, and the maximum projected angular separation is estimated to be $0.27^{+0.02}_{-0.02}\arcsec$.
%The orbit inferred from the MAP solution further implies a maximum projected angular separation of approximately 0\farcs{}26. 
The outer companion, HIP 109388 c, has a true mass of ${1.04}_{-0.18}^{+0.21}\,M_{\rm Jup}$, a semi-major axis of ${4.87}_{-0.40}^{+0.35}$ au, an eccentricity of ${0.094}_{-0.036}^{+0.036}$, and an inclination of ${92}_{-23}^{+25}$ deg. The orbital period is ${16.01}_{-0.33}^{+0.37}$ yr, and the maximum projected angular separation is estimated to be $0.59^{+0.05}_{-0.05}\arcsec$.  Octofitter's parameter posterior distributions are consistent.
%The orbit inferred from the MAP solution further implies a maximum projected angular separation of approximately 0\farcs{}58. 
The host star has a mass of only ${0.45}_{-0.10}^{+0.10}~M_\odot$, making this a particularly interesting multiplanet system of giant planets orbiting a low-mass star.

%\subsection{Comparing Results from \texttt{orvara} and \texttt{Octofitter}}
%\label{sub: fitcompare}

\section{Prospects for Future High-Contrast Imaging}
\label{sec: DI}

Our companions' relatively small projected maximum angular separations and/or low masses make them challenging targets for current high-contrast imaging instruments. However, the next generation of ground- and space-based direct imaging instruments will provide substantially improved sensitivity at small angular separations.
%and may enable direct imaging and characterization of a significant fraction of these companions.
In this section, we assess prospects for directly imaging them with the upcoming Roman Coronagraph, the European Extremely Large Telescope, and HWO.
%In this section, we therefore focus on the prospects for imaging our sample with the upcoming Roman Coronagraph Instrument (Roman/CGI), future extremely large telescopes (ELTs), and the Habitable Worlds Observatory (HWO).

\subsection{Reflected Light}
\label{sub: reflect}
For simplicity, we assume that each companion has a Jupiter-like albedo of $\sim$ 0.5, yielding an expected optical reflected light contrast at maximum projected separation of:
%We first estimate the optical planet-to-star contrast of our sample using a simple reflected-light equation. Assuming an albedo of $\sim 0.5$, the expected optical contrast \citep{Currie2023b} can be approximated as:

\begin{equation}
C_{\mathrm{optical},J} \sim 1.4 \times 10^{-9} \left(\frac{r_p}{r_J}\right)^2 \left(\frac{5.2\,\mathrm{au}}{a_{proj,max}}\right)^2.
\end{equation}
%For simplicity, we assume a separation $a_{\rm p}$ derived from the semimajor axis in the \texttt{orvara} or \texttt{Octofitter} fits 
For simplicity, we derive the maximum projected separation -- $a_{proj,max}$ -- directly from the orbit posteriors for \texttt{orvara} or \texttt{Octofitter} and assume the planet is at quadrature there. For a circular orbit, the separation is identical to the semimajor axis, while it can be larger or smaller for an eccentric orbit.  For the planet radius, we assume the mean mass-radius scalings from \citet{Chen_2017} capped at 1 Jupiter radius.
%
%the 0.The radius approximation is reasonable for our sample, since \citet{Chen_2017} showed that, over the planetary-mass range in our sample, the planet mass–radius relation is relatively flat, with R $\propto$ $M^{-0.04}$: adopting a Jupiter-radius scale 
%introduces an uncertainty of approximately 5 - 10\% in radius, which has a negligible impact on the order-of-magnitude reflected-light contrast estimates considered here. 

The resulting reflected-light contrast estimates are summarized in Table \ref{tab: contrast}. The associated uncertainties were estimated by propagating the uncertainties in the maximum projected separation. Figure \ref{fig: contrast} compares these estimated contrasts with the expected high-contrast imaging capabilities of Roman/CGI and the contrast goal of HWO. 
For Roman/CGI, we show three 5$\sigma$ post-processed detection limits: 100 hr narrow-field imaging at 575 nm using the Hybrid Lyot Coronagraph, 400 hr spectroscopy at 730 nm with R=50 using the Shaped Pupil Bowtie Coronagraph, and 100 hr wide-field imaging at 825 nm using the Shaped Pupil Wide-Field Coronagraph. These contrast curves are adapted from the publicly available DI-flux-ratio-plot repository developed by Vanessa Bailey\footnote{\url{https://github.com/nasavbailey/DI-flux-ratio-plot}}.

HWO aims to directly image Earth-sized planets in the habitable zones of nearby Sun-like stars, a goal requiring a raw contrast performance of approximately $10^{-10}$ \citep{Tuchow_2025PASP}. We therefore illustrate this value as a representative HWO performance threshold. The horizontal extent of the HWO region is defined using representative coronagraph working-angle limits from recent HWO studies. \citet{Vaughan_2023} showed that an inner working angle of approximately 41 mas would significantly improve access to ocean glint and rainbow signatures, which are important indicators of surface liquid water and active water cycles on potentially habitable planets. At larger angular separations, \citet{Sagynbayeva_2025} found that an outer working angle of approximately 1440 mas is required to achieve 80–90\% visibility of cold giant planets around nearby stars. We therefore adopt 41 mas and 1440 mas as representative inner and outer boundaries for the HWO detection space shown in Figure \ref{fig: contrast}.

%Comparing our sample with these projected instrument capabilities, we find that 
Taking into account the uncertainties, five targets lie within or close to the expected Roman/CGI detection limits: HIP 17378 b, HIP 51317 c, HIP 63833 c,  HIP 109388 b, and HIP 109388 c. A sixth target -- HIP 71639 b -- nominally lies just inside the CGI inner working angle but close to the Roman/CGI limits given uncertainties in its separation and CGI's pre-flight performance predictions near 3 $\lambda$/D.  In comparison, 28 targets fall within HWO contrast goal of $10^{-10}$ and its representative working-angle range. Both planets in the HIP 109388 system lie near the Roman/CGI detection limits, making it a particularly valuable case for multi-planet reflected-light characterization. 
%The remaining Roman/CGI and HWO-accessible targets provide a broader sample for future direct-imaging studies with next-generation ground- and space-based missions.

\begin{deluxetable}{lcccccccc} 
\tablecaption{Predicted Imaging Contrast for Exoplanets\label{tab: contrast}}
\tablewidth{0pt} 
\setlength{\tabcolsep}{2pt}
\tabletypesize{\scriptsize}
\tablehead{
\colhead{HIP ID} & \colhead{(B-V)$_{star}$} & \colhead{$\log R'_{\rm HK}$} & \colhead{Age$_{star}$} & \colhead{W2$_{star}$} & \colhead{M$_{planet}$} & \colhead{Contrast$_{thermal}$} & \colhead{Contrast$_{reflected}$} & \colhead{Instrument}\\
 & (mag) & & (Gyr) & (mag) & (mag)  & \\
}
\startdata
%\textbf{blah}
HIP 669 b\tablenotemark{a} & $0.624 \pm 0.017$ & -4.67 & $2.65_{-0.99}^{+1.85}$ & $5.392 \pm 0.057$ & $18.86^{+0.64}_{-1.23}$ & $(4.11^{+8.66}_{-1.83}) \times 10^{-6}$ & $(4.05^{+1.85}_{-2.35}) \times 10^{-10}$ & METIS, HWO\\
HIP 17378 b & $0.915 \pm 0.035$ & -5.44 & 6.64\tablenotemark{g} & - & - & - & $(3.21^{+0.18}_{-0.16}) \times 10^{-9}$\tablenotemark{*} & CGI, HWO\\
HIP 18918 b & $1.190 \pm 0.012$ & -4.68 & 	4.40\tablenotemark{h} & $7.434 \pm 0.020$ & - & - & $(1.65^{+0.35}_{-0.26}) \times 10^{-7}$\tablenotemark{*} & -\\
HIP 25421 b & $0.946 \pm 0.005$ & -4.92 & 4.84\tablenotemark{g} & $5.280 \pm 0.072$ & - & - & $(1.28^{+0.91}_{-0.35}) \times 10^{-9}$ & HWO\\
HIP 31540 c & $0.714 \pm 0.025$ & -5.02 & 4.63\tablenotemark{g} & $5.902 \pm 0.036$ & $\geq 23.85$ & $\leq 6.64 \times 10^{-8}$ & $(1.74^{+2.17}_{-1.18}) \times 10^{-10}$ & HWO\\
%\textcolor{red}{HIP 31540 c} & $0.714 \pm 0.025$ & -5.02 & 4.63\tablenotemark{g} & $5.902 \pm 0.036$ & \tablenotemark{*} & \tablenotemark{*} & \tablenotemark{*} & \\
HIP 39039 b & $0.864 \pm 0.055$ & -5.11 & 5.52\tablenotemark{g} & $7.262 \pm 0.020$ & $24.36^{+0.05}_{-0.67}$ & $(1.45^{+1.25}_{-0.07}) \times 10^{-7}$ & $(9.49^{+4.18}_{-2.52}) \times 10^{-10}$ & HWO\\
%HIP 47513 b & $1.490 \pm 0.019$ & -4.77\tablenotemark{e} & - & $5.824 \pm 0.049$ & - & - & $(1.68^{+0.25}_{-0.37}) \times 10^{-6}$\tablenotemark{*} & -\\
HIP 51317 c & $1.507 \pm 0.017$ & -5.23 & - & $4.874 \pm 0.088$ & - & - & $(1.26^{+0.53}_{-0.27}) \times 10^{-9}$\tablenotemark{*} & CGI, HWO\\
HIP 64550 b & $0.643 \pm 0.024$ & -4.87 & $6.23_{-2.59}^{+3.22}$ & $5.198 \pm 0.065$ & - & - & $(3.80^{+1.26}_{-0.84}) \times 10^{-8}$\tablenotemark{*} & -\\
HIP 66487 b & $0.958 \pm 0.038$ & -5.17 & 7.58\tablenotemark{g} & $7.201 \pm 0.020$ & - & - & $(3.96^{+1.43}_{-0.93}) \times 10^{-8}$ & -\\
HIP 71639 b & $0.508 \pm 0.087$ & -4.81 & $4.97_{-2.17}^{+3.34}$ & $4.653 \pm 0.095	$ & $19.20^{+1.21}_{-1.38}$ & $(1.51^{+3.89}_{-1.02}) \times 10^{-6}$ & $(2.92^{+3.04}_{-2.02}) \times 10^{-9}$ & METIS, HWO\\
HIP 103854 b & $0.613 \pm 0.019$ & -4.93 & $7.55_{-2.98}^{+3.16}$ & $7.260 \pm 0.020$ & - & - & $(2.39^{+0.65}_{-0.68}) \times 10^{-9}$ & HWO\\
HIP 104045 b & $0.639 \pm 0.025$ & -4.88  & $6.52_{-2.68}^{+3.20}$ & $6.792 \pm 0.019$ & - & - & $(2.68^{+0.97}_{-0.63}) \times 10^{-9}$ & HWO\\
HIP 118009 b & $0.967 \pm 0.010$ & -5.07  & 4.07\tablenotemark{h} & $7.649 \pm 0.021$ & - & - & $(9.88^{+7.68}_{-3.56}) \times 10^{-9}$ & HWO\\
\hline
HIP 6511 b & $0.553 \pm 0.017$ & -4.86  & $6.07_{-2.54}^{+3.23}$ & $6.271 \pm 0.024$ & $\geq 22.54$ & $\leq 3.10 \times 10^{-7}$ & $(5.64^{+1.12}_{-0.87}) \times 10^{-9}$ & HWO\\
HIP 9094 b & $0.817 \pm 0.018$ & -5.26  & 7.60\tablenotemark{g} & $4.230 \pm 0.183$ & - & - & $(4.05^{+1.08}_{-0.77}) \times 10^{-9}$ & HWO\\
HIP 15527 b & $0.630 \pm 0.011$ & -4.85  & $5.74_{-2.43}^{+3.27}$ & $5.712 \pm 0.048$ & 20.33 & $1.42 \times 10^{-6}$ & $(5.95^{+2.15}_{-1.40}) \times 10^{-9}$ & HWO\\
%\textcolor{red}{HIP 15527 b} & $0.630 \pm 0.011$ & -4.85  & $5.74_{-2.43}^{+3.27}$ & $5.712 \pm 0.048$ & \tablenotemark{*} & \tablenotemark{*} & \tablenotemark{*} & \\
HIP 18387 b & $0.552 \pm 0.021$ & -4.87  & $6.24_{-2.59}^{+3.22}$ & $6.371 \pm 0.023$ & $\geq 22.14$ & $\leq 4.93 \times 10^{-7}$ & $(3.84^{+0.61}_{-0.50}) \times 10^{-9}$ & HWO\\
HIP 27384 b & $0.868 \pm 0.023$ & -5.03\tablenotemark{d} & 4.80\tablenotemark{i} & $6.230	\pm 0.027$ & $19.54^{+0.15}_{-0.35}$ & $(4.76^{+1.79}_{-0.62}) \times 10^{-6}$ & $(1.33^{+0.17}_{-0.14}) \times 10^{-8}$ & -\\
HIP 30579 b & $0.718 \pm 0.024$ & -4.95  & $8.16_{-3.12}^{+3.04}$ & $6.272 \pm 0.027$ & - & - & $(6.48^{+1.36}_{-0.55}) \times 10^{-8}$ & -\\
HIP 30579 c & $0.718 \pm 0.024$ & -4.95  & $8.16_{-3.12}^{+3.04}$ & $6.272 \pm 0.027$ & - & - & $(4.24^{+0.65}_{-0.53}) \times 10^{-8}$ & -\\
HIP 52409 b\tablenotemark{b} & $0.694 \pm 0.024$ & -4.99  & $11.45_{-6.14}^{+0.13}$ & $5.663 \pm 0.052$ & - & - & $(4.03^{+0.60}_{-0.70}) \times 10^{-8}$ & -\\
HIP 52409 c\tablenotemark{b} & $0.694 \pm 0.024$ & -4.99  & $11.45_{-6.14}^{+0.13}$ & $5.663 \pm 0.052$ & - & - & $(3.13^{+0.89}_{-0.98}) \times 10^{-10}$ & HWO\\
HIP 53437 b & $0.749 \pm 0.027$ & -5.17  & 5.78\tablenotemark{g} & $5.566	\pm 0.053$ & - & - & $(3.51^{+1.54}_{-0.93}) \times 10^{-9}$\tablenotemark{*} & HWO\\
HIP 54779 e & $0.617 \pm 0.033$ & -5.18\tablenotemark{e} & 7.90\tablenotemark{j} & $7.025 \pm 0.020$ & - & - & $(2.49^{+0.43}_{-0.34}) \times 10^{-8}$ & -\\
HIP 59357 b & $0.617 \pm 0.010$ & -4.83  & $5.34_{-2.29}^{+3.32}$ & $7.118 \pm 0.020$ & $\geq 23.12$ & $\leq 3.97 \times 10^{-7}$ & $(3.23^{+1.42}_{-0.86}) \times 10^{-9}$ & HWO\\
HIP 63833 b & $1.362 \pm 0.012$ & -4.89  & 4.3\tablenotemark{k} & $5.519 \pm 0.049$ & - & - & $(1.89^{+0.50}_{-0.36}) \times 10^{-8}$\tablenotemark{*} & HWO\\
HIP 63833 c & $1.362 \pm 0.012$ & -4.89  & 4.3\tablenotemark{k} & $5.519 \pm 0.049$ & $21.51^{+0.71}_{-0.04}$ & $(4.02^{+0.15}_{-1.93}) \times 10^{-7}$ & $(5.99^{+1.83}_{-1.26}) \times 10^{-9}$ & HWO, CGI\\
HIP 71481 b & $0.685 \pm 0.021$ & -4.73\tablenotemark{c}  & $3.55_{-1.64}^{+3.24}$ & $5.944 \pm 0.024$ & $18.43^{+1.00}_{-2.00}$ & $(1.01^{+5.39}_{-0.61}) \times 10^{-5}$ & $(2.25^{+1.26}_{-0.69}) \times 10^{-8}$ & METIS, HWO\\
HIP 81681 b & $0.630 \pm 0.020$ & -4.89  & $6.60_{-2.70}^{+3.20}$ & $5.511 \pm 0.047$ & - & - & $(5.38^{+1.94}_{-1.26}) \times 10^{-9}$ & HWO\\
HIP 82632 b & $0.707 \pm 0.032$ & -4.96\tablenotemark{c}  & $8.32_{-3.15}^{+2.98}$ & $7.638 \pm 0.020$ & $\geq 23.05$ & $\leq 6.84 \times 10^{-7}$ & $(1.45^{+0.64}_{-0.38}) \times 10^{-9}$ & HWO\\
HIP 83547 b & $0.565 \pm 0.025$ & -4.90  & $6.93_{-2.80}^{+3.19}$ & $5.931 \pm 0.043$ & $\geq 19.09$\tablenotemark{*} & $\leq 5.46 \times 10^{-6}$\tablenotemark{*} & $(1.37^{+0.09}_{-0.16}) \times 10^{-8}$\tablenotemark{*} & -\\
HIP 83941 b & $1.294 \pm 0.089$ & -4.67  & 5.90\tablenotemark{l} & $7.551 \pm 0.035$ & $22.26^{+0.30}_{-0.76}$ & $(1.31^{+1.32}_{-0.31}) \times 10^{-6}$ & $(2.28^{+0.65}_{-0.46}) \times 10^{-8}$ & -\\
HIP 84787 b & $0.613 \pm 0.027$ & -4.99 & $8.77_{-3.22}^{+2.78}$ & $5.026 \pm 0.066$ & - & - & $(1.08^{+0.22}_{-0.17}) \times 10^{-8}$\tablenotemark{*} & -\\
HIP 90979 b & $1.263 \pm 0.058$ & -4.91  & 4.4\tablenotemark{k} & $6.842 \pm 0.020$ & $\geq 22.51$ & $\leq 5.42 \times 10^{-7}$ & $(7.23^{+1.92}_{-1.37}) \times 10^{-9}$ & HWO\\
HIP 95262 b & $0.595 \pm 0.021$ & -4.91  & $7.10_{-2.85}^{+3.18}$ & $6.211 \pm 0.026$ & - & - & $(1.11^{+0.23}_{-0.18}) \times 10^{-8}$ & -\\
HIP 96078 b & $0.532 \pm 0.047$ & -4.72\tablenotemark{f}  & $3.36_{-1.56}^{+3.20}$ & $7.655 \pm 0.020$ & $19.49^{+0.08}_{-0.11}$ & $(1.84^{+0.20}_{-0.14}) \times 10^{-5}$ & $(1.06^{+0.17}_{-0.14}) \times 10^{-7}$ & -\\
HIP 99212 c & $0.562 \pm 0.019$ & -5.22\tablenotemark{e}  & 4.16\tablenotemark{m} & $7.590	\pm 0.020$ & $20.93^{+0.06}_{-0.11}$ & $(4.60^{+0.49}_{-0.24}) \times 10^{-6}$ & $(4.02^{+0.84}_{-0.64}) \times 10^{-8}$ & -\\
HIP 108375 b & $0.568 \pm 0.012$ & -4.81  & $4.93_{-2.15}^{+3.34}$ & $6.005 \pm 0.030$ & - & - & $(1.36^{+0.26}_{-0.20}) \times 10^{-7}$ & -\\
HIP 109378 b & $0.773 \pm 0.012$ & -5.12  & 9.76\tablenotemark{g} & $4.503 \pm 0.131$ & $19.44^{+0.07}_{-0.08}$\tablenotemark{*} & $(1.06^{+0.08}_{-0.07}) \times 10^{-6}$\tablenotemark{*} & $(1.86^{+0.12}_{-0.16}) \times 10^{-8}$\tablenotemark{*} & HWO\\
HIP 109388 b & $1.531 \pm 0.040$ & -5.27  & 4.58\tablenotemark{g} & $5.183 \pm 0.064$ & $20.98^{+0.26}_{-0.34}$ & $(4.80^{+1.80}_{-1.03}) \times 10^{-7}$ & $(6.75^{+1.12}_{-0.90}) \times 10^{-9}$ & METIS, CGI, HWO\\
HIP 109388 c & $1.531 \pm 0.040$ & -5.27  & 4.58\tablenotemark{g} & $5.183 \pm 0.064$ & $20.83^{+0.36}_{-0.39}$ & $(5.53^{+2.41}_{-1.58}) \times 10^{-7}$ & $(1.41^{+0.27}_{-0.21}) \times 10^{-9}$ & METIS, CGI, HWO\\
HIP 112414 b & $0.636 \pm 0.010$ & -4.99  & $8.82_{-3.23}^{+2.76}$ & $4.865 \pm 0.099$ & - & - & $(3.20^{+0.37}_{-0.31}) \times 10^{-8}$\tablenotemark{*} & -\\
HIP 112414 c & $0.636 \pm 0.010$ & -4.99  & $8.82_{-3.23}^{+2.76}$ & $4.865 \pm 0.099$ & - & - & $(2.54^{+0.59}_{-0.44}) \times 10^{-9}$\tablenotemark{*} & HWO\\
HIP 113238 b & $0.821 \pm 0.032$ & -5.03  & 4.3\tablenotemark{n} & $6.270 \pm 0.029$ & - & - & $(1.56^{+0.27}_{-0.21}) \times 10^{-7}$ & -\\
HIP 116906 b & $0.648 \pm 0.017$ & -4.87  & $6.18_{-2.57}^{+3.22}$ & $5.998 \pm 0.037$ & $18.80^{+0.78}_{-0.83}$ & $(7.51^{+8.62}_{-3.86}) \times 10^{-6}$ & $(9.34^{+1.29}_{-1.07}) \times 10^{-9}$ & METIS, HWO\\
\enddata
\tablecomments{The HIP 5750 B binary is also detectable with METIS but lies off the top of the plot.  \textsuperscript{a} The Li EW ($31.45 \pm 1.89$ m\AA) is adopted from \citet{Peck_2025} and used as one of the stellar age indicators.
\textsuperscript{b} The Li EW ($19$ m\AA) is adopted from \citet{White_2007} and used as one of the stellar age indicators.
\textsuperscript{c}{The $\log R'_{\rm HK}$ value is adopted from \citet{Boro_Saikia_2018} and used as one of the stellar age indicators.}
\textsuperscript{d}{The $\log R'_{\rm HK}$ value is adopted from \citet{Isaacson_2010} and used as one of the stellar age indicators.}
\textsuperscript{e}{The $\log R'_{\rm HK}$ value is adopted from \citet{Perdelwitz_2024} and used as one of the stellar age indicators.}
\textsuperscript{f}{The $\log R'_{\rm HK}$ value is adopted from \citet{Stanford-Moore_2020} and used as one of the stellar age indicators.}
\textsuperscript{g}{Age adopted from \citet{Gomes_2021}.}
\textsuperscript{h}{Age adopted from \citet{Delgado_Mena_2015}.}
\textsuperscript{i}{Age adopted from \citet{Bonfanti_2016}.}
\textsuperscript{j}{Age adopted from \citet{Casagrande_2011}.}
\textsuperscript{k}{Age adopted from \citet{Moutou_2015}.}
\textsuperscript{l}{Age adopted from \citet{Gaia_2023}.}
\textsuperscript{m}{Age adopted from \citet{Huang_2022}.}
\textsuperscript{n}{Age adopted from \citet{lorente_2021}.}
}
\tablenotetext{*}{The calculations are based on the \texttt{Octofitter} fitting results, since it provides better constraints.}
\end{deluxetable}

\subsection{Thermal Emission}
\label{sub: thermal}
The predicted detectability of a planet in thermal emission depends strongly on its mass, age, and atmospheric properties \citep[e.g.][]{Currie2023b}.  We assume planet ages equal to the ages of their host stars, since they are expected to form early in the evolution of a planetary system, typically within $\sim$1–$10$ Myr during the protoplanetary disk phase, which is negligible compared with the predominantly Gyr-scale stellar ages considered here. We primarily estimate stellar ages using the ``Bayesian Ages for Field Lower-Mass Stars'' (\texttt{BAFFLES}) code \citep{Stanford-Moore_2020}. \texttt{BAFFLES} empirically constrains stellar ages from chromospheric activity, quantified by $\log R'_{\rm HK}$, and/or the equivalent width (EW) of the Li I 6708 \AA\ absorption line, together with $B-V$ color. The chromospheric-activity calibration is applicable over $0.45 \leq B-V \leq 0.90$ and $-5.0 \leq \log R'_{\rm HK} \leq -3.7$.

We obtain $B-V$ colors from the Hipparcos Main Catalogue \citep{Perryman_1997} and correct them for interstellar reddening using $E(B-V)$ values derived from the three-dimensional dust map of \citet{Wang_2025} at the Galactic coordinates and distances of the individual targets. Most of our $\log R'_{\rm HK}$ measurements are obtained from DACE. To account for temporal variations in chromospheric activity and avoid disproportionately weighting epochs with denser sampling, we first average measurements obtained within each night, then average the nightly values within each year, and finally average the annual values to derive a representative long-term $\log R'_{\rm HK}$ for each target. For targets for which DACE does not provide $\log R'_{\rm HK}$ measurements, we adopt values from the literature. All Li EW measurements are also taken from the literature, and only two targets in our sample have available Li EW measurements. The adopted $B-V$, $\log R'_{\rm HK}$, and Li EW values, together with their references and the resulting BAFFLES age estimates, are summarized in Table \ref{tab: contrast}. This procedure yields ages for 21 targets in our sample. For targets without \texttt{BAFFLES} age estimates, we adopt previously published stellar ages from the literature (e.g., \citealt{Engle_2024, Gomes_2021}). 

%The adopted stellar ages are then used as proxies for the planet ages when estimating their thermal emission. 
As a first-order estimate of the thermal-emission detectability, we evaluate the expected planet-star contrast near $5\,\mu$m, where the contrast remains relatively favorable for cool giant planets \citep{Skemer2014}. This wavelength range corresponds to the M band of future E-ELT instruments and is close to the WISE W2 band (4.6 $\mu$m). We adopt the host-star W2 apparent magnitudes from ALLWISE \citep{Cutri_2014}.

We then use the ``spectral characterization and inference for exoplanet science'' (\texttt{species}; \citealt{Stolker_2020}) package to interpolate evolutionary model predictions based on the planet masses and ages. 
%Given the relatively old ages and low masses of our planets ($\sim0.1$--$4\ M_{\rm Jup}$), their predicted effective temperatures are generally below $\sim500$ K and thus well below the L-T transition regime. We therefore adopt the AMES-COND evolutionary models \citep{Allard_2001} to estimate the planetary intrinsic effective temperatures, surface gravities, and radii. 
We adopt the Sonora Bobcat evolutionary models \citep{Marley_2021} to estimate intrinsic temperatures and luminosities: these models are suitable for cool, cloudless substellar objects and self-luminous giant planets with temperatures down to $\sim200$ K. The evolutionary tracks span masses of approximately $0.5$–$85\,M_{\rm Jup}$ and ages up to 15 Gyr, allowing us to estimate the planetary intrinsic effective temperatures, surface gravities, and radii over a broad range of the parameter space relevant to our sample. 
Several planets in our sample have masses below the lower boundary of the available evolutionary grid. Consequently, thermal-emission properties could only be calculated for a subset of the sample, as listed in Table \ref{tab: contrast}. 
%Future extensions of evolutionary models to lower planetary masses would enable more complete predictions of the thermal emission from these companions.

\begin{figure*}[!tp]
\centering
\includegraphics[width=\textwidth]{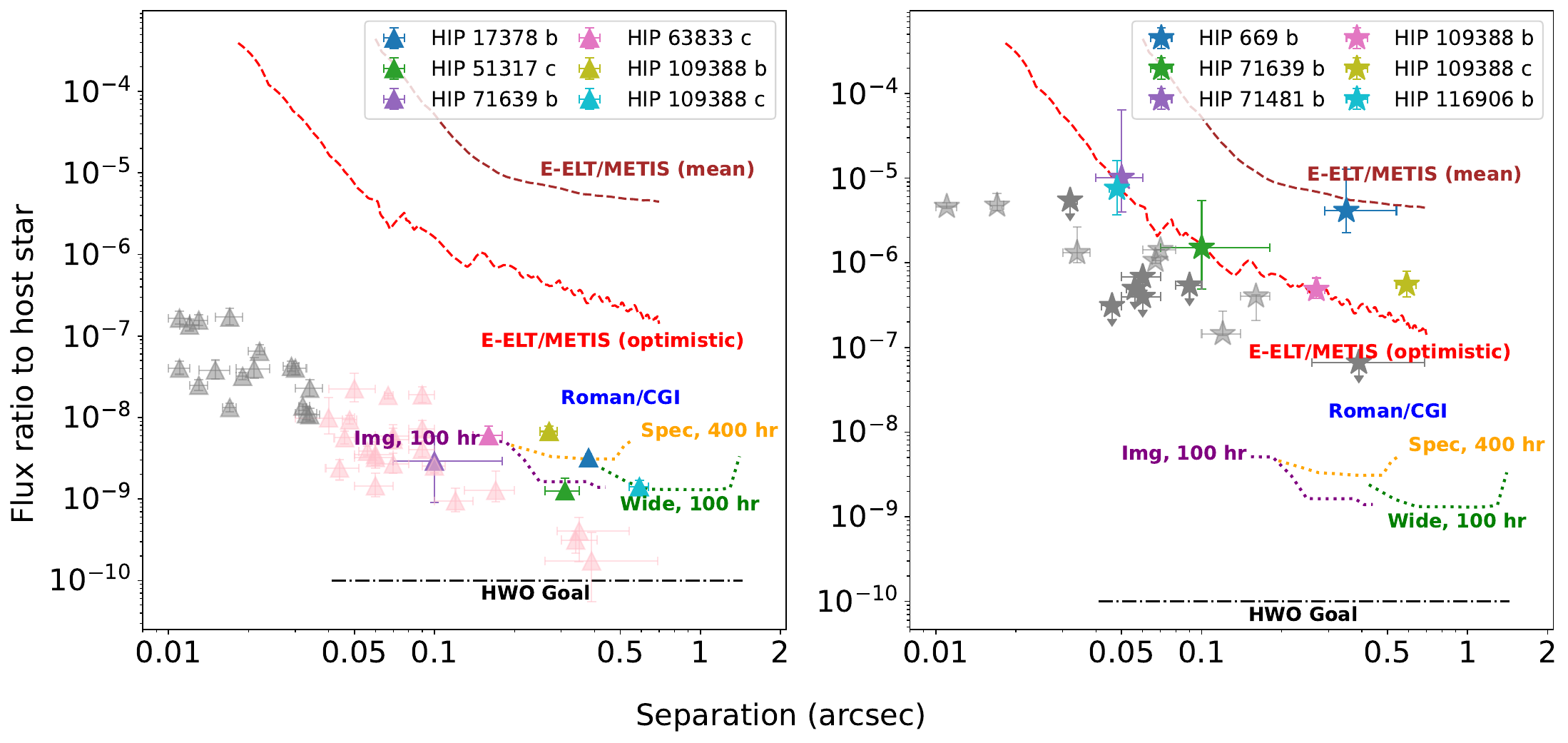}
\vspace{-0.3in}
\caption{Expected planet-to-star flux ratios as a function of angular separation for our sample. The left panel shows the reflected-light contrasts, with planets plotted as triangles. Pink triangles indicate planets that lie above the HWO contrast goal but below the Roman/CGI detection limits. Grey triangles denote planets that are not expected to be detectable in reflected light. Colored triangles shown in the legend mark planets whose predicted reflected-light contrasts are potentially accessible to Roman/CGI and also lie above the HWO contrast goal. The right panel shows the thermal-emission contrasts, with planets plotted as star symbols.
\label{fig: contrast}}
\end{figure*}

All planets in our sample are expected to have low intrinsic temperatures. Therefore, in addition to their intrinsic effective temperatures, we include a contribution from stellar irradiation, as represented by the equilibrium temperature. We assume a bond albedo of 0.5 and computed the equilibrium temperature from the semi-major axis inferred by \texttt{orvara} or from \texttt{Octofitter}. For the planetary atmosphere, we assume solar metallicity and a solar carbon-to-oxygen ratio. For the host stars, we assume solar metallicity and primarily estimate the stellar effective temperature and radius from their spectral types using the empirical relations of \citet{Pecaut_2013}. For stars without available estimates from these relations, we adopt the corresponding stellar parameters from \citet{Stassun_2019}. The stellar masses are obtained from the \texttt{orvara} or \texttt{Octofitter}. These stellar properties, together with the orbital semi-major axis and the planetary parameters derived from \texttt{species}, are then used as inputs to ``A Planetary Intensity Code for Atmospheric Spectroscopy Observations''  (\texttt{PICASO}; \citealt{Batalha_2019, Mukherjee_2023, Mang_2026}). We use the one-dimensional climate model in \texttt{PICASO}, adopting the Sonora correlated-$k$ opacity database and cloud-free atmospheres, to solve for the atmospheric structure of each planet and generate its thermal-emission spectrum. We integrate the resulting spectra over the M-bandpass to derive planetary magnitudes. The expected planet-to-star thermal-emission contrasts are then calculated by comparing the planetary M magnitudes with the stellar W2 magnitudes, where W2 is used as an approximation for the stellar M magnitude. We then conservatively estimated the upper and lower bounds of the thermal-emission contrasts using the combinations of $(M_{\rm planet} + 1\sigma_M,\ t - 1\sigma_t)$ and $(M_{\rm planet} - 1\sigma_M,\ t + 1\sigma_t)$, respectively, where the $1\sigma$ uncertainties were included when available. The predicted thermal-emission contrasts are shown in Figure \ref{fig: contrast} and summarized in Table \ref{tab: contrast}. 

We compare the sample's predicted M$_{\rm}$-band contrasts to predicted contrasts from the Mid-infrared ELT Imager and Spectrograph (METIS; \citealt{Carlomagno_2020}), a first-light instrument for the ELT providing high-contrast imaging capabilities at mid-infrared wavelengths (e.g. 3--10 $\mu m$). 
%Taking advantage of the 39-m aperture of the ELT, METIS is designed to provide diffraction-limited imaging and spectroscopy in the thermal infrared, including the atmospheric L, M, and N bands. Its high angular resolution and mid-infrared sensitivity make it especially relevant for the direct detection and characterization of cool giant planets.
We use simulated M-band contrast curves for the METIS Classical Vortex Coronagraph mode generated with the High-contrast End-to-End Performance Simulator \footnote{\url{https://github.com/vortex-exoplanet/HEEPS\#database-explorer}} (\texttt{HEEPS}; \citealt{Carlomagno_2020, Bowens_2021, Shinde_2022, Absil_2022, Delacroix_2022,OrbanDeXivry_2026}), and kindly provided by the METIS science team. The simulations cover a range of observing and instrumental conditions, including different seeing values, stellar magnitudes, and values of \texttt{ncpa\_freq}. Here, \texttt{ncpa\_freq} is treated as a simulation parameter describing the temporal behavior or update frequency of non-common path aberrations (NCPA), which can affect the residual speckle noise and thus the achievable high-contrast performance. 

In total, we consider 55 simulated contrast curves, which include the thermal background. The final contrast curves are obtained after angular differential imaging (ADI) post-processing. Each curve therefore represents the expected 5 $\sigma$ planet-to-star contrast as a function of angular separation under a specific set of observing conditions. To define a representative limiting performance for E-ELT/METIS, we take the minimum contrast value among the simulated curves at each separation. This curve is adopted as the ``optimistic" E-ELT/METIS detection limit shown in Figure \ref{fig: contrast}.  Similarly, we calculated the mean contrast amongst simulated curves at each separation, defining this as the ``mean" detection limit in Figure \ref{fig: contrast}.

Among the 21 targets for which thermal-emission properties could be calculated, 6 companions, HIP 669 b, HIP 71481 b, HIP 71639 b, HIP 109388 b, HIP 109388 c, and HIP 116906 b, have predicted M-band contrasts at or above the predicted optimistic E-ELT/METIS detection limit. These 6 companions therefore define the most favorable subset for thermal-emission follow-up in our sample. Notably, HIP 71639 b, HIP 109388 b, and HIP 109388 c are potentially accessible in both thermal emission and reflected light, making them particularly valuable targets for multi-wavelength characterization. Such observations would provide complementary constraints on the planet luminosity, atmospheric structure, and scattering properties.

\section{Summary and Discussion}
\label{sec: summary}
In this work, we analyzed archival HARPS radial-velocity data and combined them with astrometry from the HGCA to search for and characterize long-period companions. We identified 14 new planets. Joint RV–astrometry modeling allowed us to constrain their three-dimensional orbits and dynamical masses, while also improving the dynamical constraints for 32 previously known planets. We further assessed the direct imaging prospects of the planets in our sample. Five are promising targets for reflected light imaging with the Roman/CGI, with 28 targets potentially accessible to HWO and six planets detectable in thermal emission with the E-ELT/METIS.

%In addition to the specific discoveries presented here, 
Our study adds to the growing body of work demonstrating the power of combining astrometry with complementary techniques to discover new companions and refine the properties of known ones, including direct-imaging surveys such as OASIS \citep{ElMorsy2025b, Currie_2026} and previous RV–astrometry studies (e.g., \citealp{Li_2021, Philipot_2023, Xiao_2023, An_2025}). The HARPS archive represents only one of several extensive RV data sets that can be exploited in this way. Other archives, such as Keck/HIRES, together with the large collection of RV measurements available through DACE, provide substantial opportunities for additional companion discoveries and dynamical characterization. While our analysis focuses primarily on discovery and characterization, applying similar methods to well-defined samples can also enable population-level studies of companion demographics (e.g., \citealp{VanZandt_2026, Giacalone_2026}). 

Looking ahead, Gaia DR4 will greatly expand the discovery space for companions at separations of a few au \citep{Lammers_2026} and increase the number of systems showing astrometric accelerations induced by planets at wider separations, including those beyond 5 au. The number of systems with accelerations from Gaia DR4 but lacking complete orbits will greatly outnumber short-period planets with well-constrained orbits \citep{Yahalomi2026}.   Combining Gaia DR4 and prior Gaia/Hipparcos measurements with archival and continued RV monitoring and imaging will provide a powerful means of confirming such planets and constraining their dynamical masses and three-dimensional orbits.

Finally, our results have important implications for future direct imaging observations. We identify a sample of planets that may be accessible to direct imaging in reflected light with Roman/CGI and HWO or in thermal emission with the E-ELT/METIS. Nearly all of these systems are substantially older than the majority of currently known directly imaged planetary systems, while their companions overlap with the mass range of directly imaged planets. Imaging these older planets and characterizing their spectra would extend atmospheric studies to more mature planetary systems, providing an opportunity to investigate how the atmospheric properties of giant planets evolve as they cool over time. 

In addition, our analysis provides a proof of concept for vetting potential HWO targets for exo-Earth imaging using joint RV–astrometry constraints.  Identifying and characterizing long period giant planets around nearby stars can reveal those whose dynamical influence may be unfavorable to the formation or long-term stability of Earth-like planets: the newly-confirmed HIP 17378 b may be one such companion.  A systematic vetting program combining astrometry, RVs, and direct imaging is necessary to guide HWO target selection and inform mission architectures by flagging systems with exo-Earth endangering companions.  Such a program will also characterize the giant planet populations and architectures of systems that may host Earth-like planets, building on previous work considering only astrometry or RV data \citep{Painter_2025,Harada2025}.

\begin{acknowledgments}
%This research has made use of the Imaging Mission Database, which is operated by the Space Imaging and Optical Systems Lab at Cornell University. The database includes content from the NASA Exoplanet Archive, which is operated by the California Institute of Technology, under contract with the National Aeronautics and Space Administration under the Exoplanet Exploration Program, and from the SIMBAD database, operated at CDS, Strasbourg, France.
We thank the anonymous referee for a thoughtful and helpful report and Timothy Brandt for helpful comments on \texttt{orvara} coding details.
This work received computational support from ARC at UT San Antonio, operated by University Technology Solutions. 
This research has made use of the SIMBAD database, CDS, Strasbourg Astronomical Observatory, France.
This research has made use of the NASA Exoplanet Archive, which is operated by the California Institute of Technology, under contract with the National Aeronautics and Space Administration under the Exoplanet Exploration Program.
This work has made use of data from the European Space Agency (ESA) mission Gaia (\url{https://www.cosmos.esa.int/gaia}), processed by the Gaia Data Processing and Analysis Consortium (DPAC, \url{https://www.cosmos.esa.int/web/gaia/dpac/consortium}). Funding for the DPAC has been provided by national institutions, in particular the institutions participating in the Gaia Multilateral Agreement.
This work makes use of The Data \& Analysis Center for Exoplanets (DACE), which is a facility based at the University of Geneva (CH) dedicated to extrasolar planets data visualisation, exchange and analysis. DACE is a platform of the Swiss National Centre of Competence in Research (NCCR) PlanetS, federating the Swiss expertise in Exoplanet research. The DACE platform is available at \url{https://dace.unige.ch}.

We acknowledge the usage of Claude Opus 5 for grammar/proofreading and assistance with LaTeX syntax to place figure panels.

We thank the METIS science team for kindly providing the simulated E-ELT/METIS M-band contrast curves generated with HEEPS.  
Finally, we thank Paul Robertson for helpful guidance on this project in its earlier stages.
\end{acknowledgments}
\software{\texttt{astropy} \citep{2013A&A...558A..33A, 2018AJ....156..123A, 2022ApJ...935..167A},
          \texttt{corner} \citep{corner},
          \texttt{emcee} \citep{emcee},
          \texttt{matplotlib} \citep{matplotlib},
          \texttt{numpy} \citep{numpy_2011, numpy_2020},
          \texttt{pandas} \citep{pandas, mckinney-proc-scipy-2010},
          \texttt{scipy} \citep{2020SciPy-NMeth}
          \texttt{RVSearch} \citep{Rosenthal_2021}, 
          \texttt{orvara} \citep{Brandt2021},
          \texttt{BAFFLES} \citep{Stanford-Moore_2020},
          \texttt{species} \citep{Stolker_2020},
          \texttt{PICASO} \citep{Batalha_2019, Mukherjee_2023, Mang_2026},
          \texttt{HEEPS} \citep{Carlomagno_2020, Bowens_2021, Shinde_2022, Absil_2022, Delacroix_2022}
          }
%\clearpage

\begin{appendix}
\restartappendixnumbering
   \section{Tables of New Companions Detected and Previously Identified Companions Recovered}
   \startlongtable
   \begin{deluxetable*}{lccccccccc} 
\tabletypesize{\scriptsize}
\tablecaption{Newly Identified Planetary and Substellar Systems \label{tab: new_planetnew_harpshgca}}
\tablewidth{0pt} 
\tablehead{
\colhead{HIP ID} & \colhead{SpT} & \colhead{M$_{star}$} & \colhead{d} & \colhead{$\chi^{2}_{\rm HGCA}$} & \colhead{Instrument(s)} & \colhead{N$_{\rm RV}$} & \colhead{Pearson r (p)\tablenotemark{*}} & \colhead{FAP$_{\rm RVSearch}$} & \colhead{P$_{\rm RVSearch}$}
\\ & & (M$_{\odot}$) & (pc) & & & & & & (days)
}
\startdata
%\textbf{blah}
HIP 669 & G1V & 1.03 & 27.31 & 82.78 & HARPS & 55 & 0.23 (0.09)  & $1.20 \times 10^{-4}$ & 9890.27 \\
HIP 3559 & G0V & 1.06  & 32.38  & 12.54   & HARPS, HIRES  & 172  & -0.17 (0.13)  & $6.50 \times 10^{-1}$ & 13698.34 \\
HIP 5750 & F5V & 1.33  & 111.83  & 388.77  & HARPS & 38  & 0.10 (0.58)  &  $4.72 \times 10^{-18}$ & 546.29\\
HIP 17378 & K0IV  & 1.35\tablenotemark{a}  & 9.09  & 2.76  & HARPS, HIRES, APF & 1835  & -0.08 (0.006)  & $6.45 \times 10^{-5}$ & 601.46 \\
HIP 18918 & K5V  & 0.70  & 38.77  & 1.90  & HARPS & 47  & -0.09 (0.55)\tablenotemark{b} & $1.70 \times 10^{-4}$ & 108.13 \\
HIP 23575 & G6V  & 0.97  & 48.47  & 5.78  & HARPS & 207  & 0.01 (0.90)\tablenotemark{b}  & $4.63 \times 10^{-4}$ & 11.06 \\
HIP 25421 & K2/3V  & 0.79\tablenotemark{a} & 17.86  & 6.76  & HARPS & 80  & -0.17 (0.12)  & $1.41 \times 10^{-4}$ & 1871.06 \\
%HIP 25878 & M1.5V  &  0.47 & 5.70  & 13.01  & HARPS, HIRES & 160  & -0.05 (0.61)\tablenotemark{b}  & $9.63 \times 10^{-4}$ & 32.89 \\
HIP 39039 & K0V  & 0.88  & 54.20  & 7.53  & HARPS & 67  & 0.04 (0.75)  & $3.90 \times 10^{-1}$ & 3485.92 \\
HIP 47513 & M2.5V  & 0.40  & 11.54  & 0.87  & HARPS, HIRES & 156  & -0.15 (0.13)\tablenotemark{b}  & $2.24 \times 10^{-4}$ & 28.93 \\
HIP 64550 & G1.5V  & 0.94\tablenotemark{a}  & 25.01  & 0.41  & HARPS & 147  & 0.15 (0.07)\tablenotemark{b}  & $4.29 \times 10^{-5}$ & 74.07 \\
HIP 66487 & K2V  & 0.82  &  47.87 & 1.38  & HARPS & 35  & 0.23 (0.19)  & $7.00 \times 10^{-6}$ & 369.10 \\
HIP 71639 & F8V  & 1.18  &  34.00 & 0.58  & HARPS & 23  & -0.05 (0.82)  & $4.29 \times 10^{-11}$ & 1527.00 \\
HIP 103854 & G2V & 1.00 & 89.52 & 1.89 & HARPS & 178 & 0.02 (0.75) & $1.65 \times 10^{-4}$ & 2375.65\\
HIP 104045 & G5V  & 0.98  & 53.79  & 0.17  & HARPS & 54 & 0.20 (0.15)  & $1.58 \times 10^{-5}$  & 2462.23  \\
HIP 118009 & K3V  &  0.78 & 49.63  & 0.69  & HARPS & 22  & -0.31 (0.16)  & $3.73 \times 10^{-6}$ & 1166.11\\
\enddata
\tablecomments{Spectral types are given from \textit{SIMBAD}, while masses are inferred from \citet{Pecaut_2013} given the star's spectral type.}
\vspace{-5pt}
\tablenotetext{*}{The Pearson correlation coefficient and corresponding p-value are calculated using the RVs and $S_{\rm MW}$.}
\vspace{-5pt}
\tablenotetext{a}{For spectral types not tabulated in \citet{Pecaut_2013}, stellar masses are adopted from \citet{Anders_2019}.}
\vspace{-5pt}
\tablenotetext{b}{For signals with periods shorter than 180 days, BIS is used as the stellar activity indicator instead of $S_{\rm MW}$, and the Pearson correlation coefficient and corresponding p-value are calculated using the RVs and BIS.}
\vspace{-5pt}
\end{deluxetable*}

\startlongtable
\begin{deluxetable*}{lccccccccccc} 
%\tabletypesize{\scriptsize}
\tabletypesize{\tiny}
\setlength{\tabcolsep}{1pt}
\tablecaption{Exoplanetary Systems with Previously Published Detections \label{tab: recovered_planetnew_harpshgca}}
\tablewidth{0pt} 
\tablehead{
\colhead{HIP ID} & \colhead{SpT} & \colhead{M$_{star}$} & \colhead{d} & \colhead{$\chi^{2}_{\rm HGCA}$} & \colhead{Instrument(s)} & \colhead{N$_{\rm RV}$} & \colhead{Pearson r (p)\tablenotemark{*}} & \colhead{FAP$_{\rm RVSearch}$} & \colhead{P$_{\rm RVSearch}$} & \colhead{P$_{\rm Published}$\tablenotemark{$\dagger$}} & \colhead{Reference} 
\\ & & (M$_{\odot}$) & (pc) & & & & (days) & (days)
} 
\startdata
%\textbf{blah}
HIP 6511 & G0V  & 1.06  & 55.35 & 3.05  & HARPS  & 62  & 0.01 (0.91)  & $5.65 \times 10^{-25}$  & 1287.33 & 1313 & 1\\
HIP 9094 & G9V  & 0.90  & 33.57  & 7.30  & HARPS, HIRES, APF  & 522  & 0.13 (0.09)\tablenotemark{b}  & $2.21 \times 10^{-34}$  & 37.93 & 37.9196 & 2, 3 \\
  &    &    &    &    &    &    & 0.04 (0.55)  & $5.24 \times 10^{-61}$  & 1930.89 & 1935.0 & 2, 3, 4, 5\\
HIP 15527 & G1.5V  & 0.99\tablenotemark{a}  & 35.87  & 1.08  & HARPS, UCLES, ESPRESSO  & 115  &  -0.06 (0.57) & $7.89 \times 10^{-11}$  & 584.81 & 597.0643 & 6, 7, 8, 9, 10\\
HIP 18387 & F8V  & 1.18  & 55.78  & 6.43  & HARPS  & 63  & 0.08 (0.53)  & $2.43 \times 10^{-20}$  & 1806.46 & 1802.29 & 11, 12\\
HIP 27384 & G8IV  & 1.07\tablenotemark{a}  & 90.67  & 3.94  & HARPS, HIRES, HDS  & 49  & -0.08 (0.72)  & $1.38 \times 10^{-18}$  & 685.71 & 685.25 & 7, 11, 13 ,14, 15\\
HIP 30579 & G8V  & 0.94  & 34.32  & 0.15 & HARPS  & 124  & -0.00 (0.98)  & $2.34 \times 10^{-19}$  & 227.90 & 225.79 & 16, 17\\
  &    &    &    &   &    &    &    &  $3.38 \times 10^{-17}$  & 344.00 & 345.43 & 16, 17 \\
HIP 31540 & G6V  &  0.97 & 37.39  & 2.50  & HARPS, HIRES & 239  & 0.08 (0.31)\tablenotemark{b}  & $1.89 \times 10^{-21}$ & 4.08 & 4.0845 & 18\\
  &    &    &    &   &    &    &  -0.11 (0.14)  &  $6.18 \times 10^{-30}$  & 6252.95 & - & - \\
%HIP 36208 &   &   &   &   &   &   &   &   &  \\
HIP 51317 & M2V  & 0.44  & 7.04  & 1.71  & HARPS, HIRES, ESPRESSO  & 266  & -0.07 (0.38)  & $1.68 \times 10^{-4}$  & 731.51 & - & -\\
HIP 52409 & G6V  & 0.97  & 34.55  & 20.36  & HARPS, HIRES, CORALIE  & 314  & -0.03 (0.81)  & $3.64 \times 10^{-11}$  & 325.69 & 325.705 & 3, 7, 19, 20, 21, 22\\
  &    &    &    &   &    &    &    &  $7.92 \times 10^{-8}$  & 7320.39 & 8360 & 3, 19\\
HIP 53437 & G3/5V  & 1.13\tablenotemark{a}  & 57.62  & 1.44  & HARPS  & 131  & 0.12 (0.17)  &  $6.28 \times 10^{-21}$ & 2107.26 & 2164 & 23\\
HIP 54779 & G3V  & 0.99  & 64.96  & 0.08  & HARPS, ESPRESSO  & 196  & 0.12 (0.12)\tablenotemark{b}  & $1.66 \times 10^{-4}$  & 13.09 & 13.0766 & 24\\
  &    &    &    &   &    &    &   &  $7.99 \times 10^{-5}$  & 25.46 & 25.502672 & 24, 25 \\
  &    &    &    &   &    &    &  -0.14 (0.06)  &  $4.43 \times 10^{-13}$  & 203.34 & 204.59 & 24 \\
HIP 56572 & G8V  & 0.94  & 49.61  & 173.36  & HARPS  & 37  & -0.07 (0.70)  & $4.76 \times 10^{-7}$  & 195.51\tablenotemark{e} & - & -\\
  &    &    &    &   &    &    &   &  $8.11 \times 10^{-10}$  & 390.18 & 383 & 7, 26, 27\\
HIP 59357 & G3V  & 0.99  & 55.10  & 5.06  & HARPS  & 59  & -0.06 (0.63)  & $4.20 \times 10^{-15}$  & 2338.31 & 2412 & 28\\
HIP 63833 & K9V  & 0.59  & 16.31  & 7.76  & HARPS  & 101  & 0.19 (0.06)  & $2.33 \times 10^{-13}$  & 650.47 & 663.2 & 12, 29\\
  &    &    &    &   &    &    &   &  $4.59 \times 10^{-20}$  & 1833.82 & 1818 & 12, 29 \\
HIP 71481 & K2.5IV  & 0.85\tablenotemark{a}  & 26.05  & 17.23  & HARPS  & 30  & -0.09 (0.64)  &  $1.12 \times 10^{-4}$ & 298.14 & 298.2 & 30\\
HIP 81681 & G5V  & 0.98  & 30.38  & 0.37  & HARPS, HIRES  & 174  & -0.18 (0.05)  & $4.66 \times 10^{-14}$  & 1011.10 & 1096.2 & 31\\
HIP 82632 & G6V  & 0.97  & 88.07  & 0.64  & HARPS, MIKE  & 68  & 0.12 (0.38)  & $2.09 \times 10^{-6}$  & 2978.77 & 2918.92 & 11, 30, 32\\
HIP 83547 & F8V  & 1.18 & 48.41  & 0.48  & HARPS, ESPRESSO  & 67  & 0.12 (0.32)  & $8.60 \times 10^{-27}$  & 498.68 & 499.39999 & 7, 33\\
HIP 83941 & M0V & 0.65\tablenotemark{c} & 38.33 & 1.31  & HARPS  & 114  & -0.06 (0.52)  & $1.68 \times 10^{-9}$  & 272.60 & 272.7 & 34\\
HIP 84787 & G1V  & 1.03  & 55.66  & 3.63  & HARPS  & 72  & -0.21 (0.08)  & $1.88 \times 10^{-12}$  & 854.73 & 842.2 & 1\\
HIP 90979 & K7V  & 0.64  & 27.21  & 7.26  & HARPS  & 48  & 0.24 (0.10)  & $2.70 \times 10^{-12}$  & 1636.38 & 1634 & 29, 35\\
HIP 95262 & G1V  & 1.03  & 59.96  & 0.04  &  HARPS & 118  & -0.03 (0.79)  & $1.02 \times 10^{-26}$  & 984.00 & 956 & 7, 36\\
HIP 96078 & F8V  & 1.18  & 108.47  & 0.73  & HARPS, HIRES  & 116  & -0.01 (0.92)  & $2.23 \times 10^{-7}$  & 141.61 & 141.630 & 7, 37, 38\\
HIP 99212 & G0V  & 1.06  & 89.04  & 1.09  & HARPS  & 118  & 0.17 (0.07) & $1.57 \times 10^{-35}$  & 320.06 & 315.7 & 39\\
%HIP 104045 & G5V  & 0.98  & 53.79  & 0.17  & HARPS & 54 & 0.19 (0.16)  & $1.58 \times 10^{-5}$  & 2462.23  \\ (Paper was withdrawn by Thiago Ferreira Dos Santos)
HIP 108375 & G1/3V  & 1.08\tablenotemark{a}  & 44.91  & 4.60  & HARPS, UCLES  & 190  & 0.11 (0.23)\tablenotemark{b}  & $5.33 \times 10^{-21}$  & 129.31 & 129.36 & 5, 40, 41, 42\\
HIP 109378 & G8V  & 0.94  & 21.34  & 1.11  & HARPS, HIRES, CORALIE, ESPRESSO & 558  & -0.15 (0.22)  & $4.81 \times 10^{-112}$  & 443.00 & 442.943 & 3, 5, 7, 43, 44\\
HIP 109388 & M3.5V  & 0.45\tablenotemark{d}  &  8.81 & 7.82  & HARPS, HIRES  & 214  & -0.01 (0.93)  & $1.23 \times 10^{-55}$  & 1910.95 & 1925.31 & 3, 5, 7, 45, 46, 47, 48\\
  &    &    &    &   &    &    &   &  $1.22 \times 10^{-40}$  & 5829.23 & 5990 & 3, 45, 47 \\
HIP 112414 & G0.5V  & 1.25\tablenotemark{a}  & 39.46  & 3.23  & HAPRS  & 269  & 0.04 (0.50)  & $1.92 \times 10^{-12}$  & 192.78 & 191.99 & 31\\
  &    &    &    &   &    &    &   &  $7.53 \times 10^{-8}$  & 2224.30 & 2277.0 & 31 \\
HIP 113238 & G9V  & 0.90  & 36.64  & 15.03  & HARPS, CORALIE  & 150  & -0.18 (0.04)\tablenotemark{b}  & $8.31 \times 10^{-15}$  & 116.09 & 118.45 & 7, 21\\
HIP 116906 & G5V  & 0.98  & 42.02  & 5.10  & HARPS, HIRES  & 122  & 0.05 (0.73)  & $4.35 \times 10^{-10}$  & 572.54 & 572.537 & 3, 5, 7, 49\\
\enddata
\tablecomments{References - 1) \citet{Naef_2010}, 2) \citet{Wright_2009}, 3) \citet{Rosenthal_2021}, 4) \citet{Bryan_2016}}, 5) \citet{Butler_2006}, 6) \citet{Udry_2019}, 7) \citet{Stassun_2017}, 8) \citet{Kane_2016}, 9) \citet{OToole_2009}, 10) \citet{Jones_2006}, 11) \citet{Barbato_2018}, 12) \citet{Moutou_2011}, 13) \citet{Luhn_2019}, 14) \citet{Ment_2018}, 15) \citet{Harakawa_2010}, 16) \citet{Li_2022}, 17) \citet{Correia_2009}, 18) \citet{Bouchy_2009}, 19) \citet{Rickman_2019}, 20) \cite{Wittenmyer_2019}, 21) \citet{Mayor_2004}, 22) \citet{Fischer_2001}, 23) \citet{Frensch_2023}, 24) \citet{Gandolfi_2025}, 25) \citet{Giacalone_2021}, 26) \citet{Philipot_2023}, 27) \citet{Naef_2007}, 28) \citet{Sreenivas_2022}, 29) \citet{Moutou_2015}, 30) \citet{Jenkins_2017}, 31) \citet{Mayor_2011}, 32) \citet{Arriagada_2010}, 33) \citet{Moutou_2009}, 34) \citet{Schorr_2025}, 35) \citet{Barbato_2020}, 36) \citet{Santos_2010}, 37) \citet{Ment_2018}, 38) \citet{Fischer_2007}, 39) \citet{Cabrera_2023}, 40) \citet{Rubenstein_2025}, 41) \citet{Gregory_2007}, 42) \citet{Tinney_2005}, 43) \citet{Wittenmyer_2007}, 44) \citet{Marcy_1999}, 45) \citet{Pinamonti_2023}, 46) \citet{Feng_2015}, 47) \citet{Montet_2014}, 48) \citet{Bonfils_2013}, 49) \citet{Vogt_2000}
\vspace{-5pt}
\tablenotetext{*}{The Pearson correlation coefficient and corresponding p-value are calculated using the RVs and $S_{\rm MW}$.}
\vspace{-5pt}
\tablenotetext{\dagger}{For systems with multiple published measurements of a consistent orbital period, only the value from the most recent publication are listed.}
\vspace{-5pt}
\tablenotetext{a}{For spectral types not tabulated in \citet{Pecaut_2013}, stellar masses are adopted from \citet{Anders_2019}.}
\vspace{-5pt}
\tablenotetext{b}{For signals with periods shorter than 180 days, BIS is used as the stellar activity indicator instead of $S_{\rm MW}$, and the Pearson correlation coefficient and corresponding p-value are calculated using the RVs and BIS.}
\vspace{-5pt}
\tablenotetext{c}{We adopt the stellar mass reported by \citet{Schorr_2025} rather than an estimate based on spectral type, as its unusually high metallicity may cause it to deviate from the typical mass expected for its spectral type.}
\vspace{-5pt}
\tablenotetext{d}{We adopt the stellar mass reported by \citet{Pinamonti_2023} rather than an estimate based on spectral type, as its unusually high metallicity may cause it to deviate from the typical mass expected for its spectral type.  }
\tablenotetext{e}{This signal appears to be an alias of the 390-day-period HIP 56572 b.}
\vspace{-5pt}

\end{deluxetable*}

 \section{Tables of Derived Companion Properties} 

 \startlongtable
\begin{deluxetable*}{lcccccccc} 
\tablecaption{Dynamical Properties of New Exoplanets Successfully Fit with \texttt{orvara}\label{tab: new_planet}}
\tablewidth{\textwidth}
\setlength{\tabcolsep}{2pt}
\tabletypesize{\small}
\tablehead{
\colhead{HIP ID} & \colhead{M$_{star}$} & \colhead{M$_{comp}$} & \colhead{a} & \colhead{$e$} & \colhead{$i$} &P & \colhead{Sep$_{max}$}
\\ & (M$_{\odot}$) & (M$_{\rm Jup}$) & (au) & & (deg) & (yrs) &  (\arcsec{})
} 
\startdata
\textbf{orvara Solutions}\\
HIP 669 b & ${1.03}_{-0.25}^{+0.25}$ & ${4.37}_{-0.91}^{+1.23}$ & ${8.61}_{-1.22}^{+3.62}$ & ${0.17}_{-0.12}^{+0.24}$ & ${61}_{-20}^{+20}$ or ${116}_{-18}^{+20}$ & ${24.4}_{-3.7}^{+18}$ & ${0.35}_{-0.06}^{+0.19}$\\
%HIP 3559 b & ${1.06}_{-0.25}^{+0.25}$ & ${2.3}_{-1.0}^{+1.7}$ & ${14.4}_{-4.0}^{+11}$ & ${0.32}_{-0.20}^{+0.27}$ & ${39}_{-19}^{+34}$ or ${147}_{-34}^{+15}$ & ${53}_{-20}^{+73}$ & ${0.53}_{-0.17}^{+0.55}$\\
HIP 5750 B & ${1.33}_{-0.25}^{+0.25}$ & ${101}_{-23}^{+30}$ & ${2.55}_{-0.49}^{+0.62}$ & ${0.813}_{-0.16}^{+0.078}$ & ${41}_{-17}^{+55}$ & ${3.44}_{-0.88}^{+1.4}$ & ${0.04}_{-0.01}^{+0.01}$ \\
HIP 25421 b & ${0.79}_{-0.24}^{+0.25}$ & ${0.102}_{-0.037}^{+0.088}$ & ${1.36}_{-0.15}^{+0.12}$ or ${2.81}_{-0.37}^{+0.34}$ & ${0.18}_{-0.13}^{+0.22}$ & ${92}_{-50}^{+50}$\tablenotemark{a} & ${1.78}_{-0.48}^{+0.64}$ or ${5.30}_{-1.56}^{+2.24}$ & ${0.17}_{-0.04}^{+0.03}$\\
HIP 31540 c & ${0.97}_{-0.25}^{+0.25}$ & ${0.75}_{-0.17}^{+0.30}$ & ${12.6}_{-4.1}^{+10}$ & ${0.78}_{-0.13}^{+0.10}$ & ${89}_{-39}^{+39}$\tablenotemark{a} & ${46}_{-20}^{+61}$ & ${0.39}_{-0.13}^{+0.30}$\\
HIP 39039 b & ${0.88}_{-0.25}^{+0.25}$ & ${1.37}_{-0.36}^{+0.52}$ & ${4.51}_{-0.52}^{+0.54}$ & ${0.66}_{-0.10}^{+0.11}$ & ${59}_{-21}^{+22}$ or ${120}_{-22}^{+22}$ & ${10.27}_{-0.82}^{+0.53}$ & ${0.12}_{-0.02}^{+0.02}$\\
HIP 66487 b & ${0.82}_{-0.25}^{+0.25}$ & ${0.66}_{-0.19}^{+0.40}$ & ${0.945}_{-0.11}^{+0.087}$ & ${0.14}_{-0.10}^{+0.14}$ & ${89}_{-43}^{+43}$\tablenotemark{a} & ${1.0129}_{-0.0034}^{+0.0045}$ & ${0.021}_{-0.003}^{+0.003}$\\
HIP 71639 b & ${1.18}_{-0.25}^{+0.25}$ & ${6.1}_{-1.3}^{+2.1}$ & ${3.15}_{-0.87}^{+1.3}$ & ${0.34}_{-0.22}^{+0.23}$ & ${89}_{-12}^{+12}$ &  ${5.6}_{-2.2}^{+2.9}$ & ${0.10}_{-0.03}^{+0.08}$\\
HIP 103854 b & ${1.00}_{-0.25}^{+0.25}$ & ${0.271}_{-0.073}^{+0.17}$ & ${3.62}_{-0.36}^{+0.44}$ & ${0.127}_{-0.089}^{+0.15}$ & ${89}_{-43}^{+43}$\tablenotemark{a} & ${6.76}_{-0.24}^{+0.48}$ & ${0.044}_{-0.005}^{+0.008}$\\
HIP 104045 b & ${0.98}_{-0.25}^{+0.25}$ & ${0.62}_{-0.14}^{+0.21}$ & ${3.52}_{-0.33}^{+0.29}$ & ${0.142}_{-0.082}^{+0.079}$ & ${91}_{-35}^{+35}$\tablenotemark{a} & ${6.69}_{-0.18}^{+0.18}$ & ${0.07}_{-0.01}^{+0.01}$\\
HIP 118009 b & ${0.78}_{-0.25}^{+0.25}$ & ${0.59}_{-0.15}^{+0.24}$ & ${2.00}_{-0.24}^{+0.20}$ &  ${0.096}_{-0.066}^{+0.093}$ & ${91}_{-38}^{+37}$\tablenotemark{a} & ${3.192}_{-0.043}^{+0.045}$ & ${0.04}_{-0.01}^{+0.01}$\\
\hline
\textbf{Octofitter Solutions}\\
HIP 669 b & $1.05^{+0.24}_{-0.24}$ & $4.50^{+1.01}_{-0.88}$ & $9.04^{+4.15}_{-1.45}$ & $0.19^{+0.21}_{-0.14}$ & $62^{+20}_{-17}$ or $116^{+18}_{-17}$ & $25.90^{+20.79}_{-4.69}$ & ${0.37}_{-0.07}^{+0.18}$\\
%HIP 3559 b & $1.04^{+0.28}_{-0.24}$ & $3.56^{+1.17}_{-0.87}$ & $16.23^{+16.46}_{-4.49}$ & $0.37^{+0.26}_{-0.24}$ & $19^{+11}_{-5}$ or $157^{+6}_{-10}$ & $61.92^{+118.55}_{-21.88}$ & ${0.63}_{-0.21}^{+0.92}$\\
HIP 5750 B & $1.31^{+0.28}_{-0.19}$ & $92.97^{+15.11}_{-9.54}$ & $2.24^{+0.15}_{-0.11}$ & $0.69^{+0.04}_{-0.04}$ & $29^{+2}_{-2}$ & $2.93^{+0.06}_{-0.26}$ & ${0.034}_{-0.002}^{+0.003}$\\
HIP 25421 b & $0.79^{+0.25}_{-0.25}$ & $0.11^{+0.18}_{-0.06}$ & ${1.31}_{-0.21}^{+0.14}$ or ${2.82}_{-0.36}^{+0.34}$ & $0.21^{+0.31}_{-0.14}$ & $88^{+62}_{-62}$\tablenotemark{a} & $1.69^{+0.69}_{-0.56}$ or $5.33^{+2.32}_{-1.54}$ & ${0.16}_{-0.05}^{+0.03}$\\
HIP 31540 c & $1.12^{+0.09}_{-0.14}$\tablenotemark{b} & $0.88^{+0.44}_{-0.11}$ & $10.17^{+6.05}_{-1.33}$\tablenotemark{b} & $0.64^{+0.14}_{-0.06}$ & $90^{+43}_{-40}$ & $31.19^{+32.49}_{-5.45}$ & ${0.34}_{-0.06}^{+0.15}$\\
HIP 39039 b & $0.85^{+0.25}_{-0.23}$ & $1.40^{+0.54}_{-0.37}$ & $4.54^{+5.18}_{-0.52}$ & $0.68^{+0.13}_{-0.12}$ & $61^{+20}_{-19}$ or $121^{+22}_{-21}$ & $10.36^{+22.35}_{-0.92}$ & ${0.12}_{-0.02}^{+0.18}$\\
HIP 66487 b & $0.88^{+0.23}_{-0.28}$ & $0.68^{+0.36}_{-0.18}$ & $0.97^{+0.08}_{-0.12}$ & $0.14^{+0.14}_{-0.09}$ & $89^{+42}_{-40}$\tablenotemark{a} & $1.0130^{+0.0043}_{-0.0035}$ & ${0.022}_{-0.002}^{+0.003}$\\
HIP 71639 b & $1.12^{+0.30}_{-0.25}$ & $5.47^{+1.21}_{-1.07}$ & $2.65^{+0.73}_{-0.58}$ & $0.24^{+0.22}_{-0.15}$ & $93^{+8}_{-19}$ & $3.80^{+2.08}_{-1.14}$ & ${0.09}_{-0.02}^{+0.02}$\\
HIP 103854 b & $1.02^{+0.24}_{-0.24}$ & $0.28^{+0.17}_{-0.08}$ & $3.64^{+0.43}_{-0.33}$ & $0.13^{+0.16}_{-0.09}$ & $89^{+43}_{-44}$\tablenotemark{a} & $6.76^{+0.47}_{-0.25}$ & ${0.044}_{-0.005}^{+0.008}$\\
HIP 104045 b & $0.98^{+0.26}_{-0.25}$ & $0.62^{+0.23}_{-0.14}$ & $3.52^{+0.30}_{-0.35}$ & $0.14^{+0.08}_{-0.08}$ & $90^{+36}_{-34}$\tablenotemark{a} & $6.69^{+0.18}_{-0.18}$ & ${0.07}_{-0.01}^{+0.01}$\\
HIP 118009 b & $0.77^{+0.25}_{-0.24}$ & $0.59^{+0.26}_{-0.15}$ & $1.99^{+0.19}_{-0.24}$ & $0.10^{+0.09}_{-0.07}$ & $89^{+39}_{-36}$\tablenotemark{a} & $3.19^{+0.05}_{-0.04}$ & ${0.043}_{-0.005}^{+0.005}$\\
\enddata
\tablenotetext{a}{Although the inclination posterior is largely prior-dominated, we can rule out nearly face-on solutions that would imply massive companions.
\tablenotetext{b}{The multimodal posterior may render the parameter estimates unreliable.}}
\end{deluxetable*}

\startlongtable
\begin{deluxetable*}{lccccccccc} 
\tablecaption{\texttt{RVSearch}-derived Orbital Properties and Minimum Masses for Short-Period Companions\label{tab:new_planet_low_mass}}
\tablewidth{0pt}
\tabletypesize{\small}
\tablehead{
\colhead{HIP ID} & \colhead{M$_{\rm comp}\sin i^{*}$} & \colhead{a} & \colhead{$e$} & \colhead{$K$} &P
\\ & (M$_{\rm Jup}$) & (au) & & (m/s) & (days)
} 
\startdata
HIP 17378 b & $0.103 \pm 0.006$ & $1.542 \pm 0.004$ & $0.185 \pm 0.092$ & $2.07 \pm 0.11$ & $601.6 \pm 2.6$\\
HIP 18918 b & $0.184 \pm 0.022$  & $0.3946 \pm 0.0003$ & $0.085 \pm 0.078$  & $10.0 \pm 1.2$ & $108.12 \pm 0.11$ \\
HIP 23575 b & $0.0393 \pm 0.0050$  & $0.09621 \pm 0.00003$  & $0.18 \pm 0.12$  & $3.72 \pm 0.47$  & $11.0598 \pm 0.0048$\\
%HIP 25878 b & $0.0570 \pm 0.0083$  & $0.15624 \pm 0.00007$  & $0.445 \pm 0.092$  & $6.68 \pm 0.91$  & $32.881 \pm 0.021$ \\
HIP 47513 b$^{a}$ & $0.0364 \pm 0.0051$  & $0.13595 \pm 0.00005$  & $0.437 \pm 0.097$  & $4.93 \pm 0.64$   & $28.929 \pm 0.017$ \\
HIP 51317 c & $0.052 \pm 0.008$  & $1.209 \pm 0.001$ & $0.853 \pm 0.036$  &  $3.86  \pm 0.45$  &  $731.6 \pm 0.8$ \\
%HIP 56572 c & $0.651 \pm 0.035$  & $0.6377 \pm 0.0015$  & $0.152 \pm 0.026$  & $24.2 \pm 1.3$  & $191.85 \pm 0.66$ \\
HIP 64550 b & $0.043 \pm 0.005$  & $0.3382 \pm 0.0003$  & $0.14 \pm 0.09$  & $2.21 \pm 0.23$  & $74.092 \pm 0.092$ \\
\enddata
\tablenotetext{*}{The minimum companion masses were calculated using host-star masses estimated from their spectral types.}
\tablenotetext{a}{We identify this signal as a compelling \textit{candidate} planet requiring further confirmation. See main text.}
\end{deluxetable*}

\startlongtable
\begin{deluxetable*}{lcccccccc} 
\tablecaption{Dynamical Properties of New Short-Period Exoplanets Successfully Fit by \texttt{Octofitter}\label{tab: new_planet_octofitter}}
\tablewidth{\textwidth}
\setlength{\tabcolsep}{2pt}
\tabletypesize{\small}
\tablehead{
\colhead{HIP ID} & \colhead{M$_{star}$} & \colhead{M$_{comp}$} & \colhead{a} & \colhead{$e$} & \colhead{$i$} &P & \colhead{Sep$_{max}$}
\\ & (M$_{\odot}$) & (M$_{\rm Jup}$) & (au) & & (deg) & (yrs) & (\arcsec{})
} 
\startdata
HIP 17378 b & $1.34^{+0.03}_{-0.04}$ & $5.31^{+1.70}_{-1.22}$ & $1.83^{+0.01}_{-0.02}$ & $0.92^{+0.01}_{-0.01}$ & $1.2^{+0.4}_{-0.3}$ or $178.8^{+0.3}_{-0.4}$ & $2.127^{+0.001}_{-0.001}$ & ${0.385}_{-0.005}^{+0.004}$\\
HIP 18918 b & $0.87^{+0.08}_{-0.17}$ & $0.24^{+0.13}_{-0.05}$ & $0.42^{+0.01}_{-0.03}$ & $0.08^{+0.10}_{-0.06}$ & $90^{+42}_{-42}$\tablenotemark{b} & $0.2960^{+0.0003}_{-0.0003}$ & ${0.011}_{-0.001}^{+0.001}$\\
%HIP 25878 b \\
HIP 47513 b$^{a}$ & $0.59^{+0.19}_{-0.14}$ & $0.06^{+0.03}_{-0.01}$ & $0.15^{+0.02}_{-0.01}$ & $0.42^{+0.09}_{-0.10}$ & $90^{+44}_{-42}$\tablenotemark{b} & $0.07920^{+0.00004}_{-0.00004}$ & ${0.017}_{-0.002}^{+0.002}$\\
HIP 51317 c & $0.45^{+0.18}_{-0.18}$ & $0.06^{+0.03}_{-0.02}$ & $1.22^{+0.14}_{-0.20}$ & $0.84^{+0.03}_{-0.05}$ & $90^{+41}_{-41}$\tablenotemark{b} & $2.003^{+0.004}_{-0.004}$ & ${0.31}_{-0.05}^{+0.04}$\\
HIP 64550 b & $0.98^{+0.15}_{-0.25}$ & $0.05^{+0.03}_{-0.01}$ & $0.34^{+0.02}_{-0.03}$ & $0.14^{+0.10}_{-0.08}$ & $90^{+44}_{-44}$\tablenotemark{b} & $0.2028^{+0.0003}_{-0.0002}$ & ${0.015}_{-0.002}^{+0.002}$\\
\enddata
\tablenotetext{a}{We identify this signal as a compelling \textit{candidate} planet requiring further confirmation. See main text.}
\tablenotetext{b}{Although the inclination posterior is largely prior-dominated, we can rule out nearly face-on solutions that would imply massive companions.}
\end{deluxetable*}

\startlongtable
\begin{deluxetable*}{lccccccccc} 
\tablecaption{Updated Dynamical Properties of Exoplanets in Previously Identified Systems\label{tab: update}}
\tablewidth{0pt} 
\tabletypesize{\small}
\setlength{\tabcolsep}{3pt}
\tablehead{
\colhead{HIP ID} & \colhead{M$_{star}$} & \colhead{M$_{comp}$} & \colhead{a} & \colhead{$e$} & \colhead{$i$} & P & {Sep$_{max}$}
\\ & (M$_{\odot}$) & (M$_{\rm Jup}$) & (au) & & (deg) & (yrs) & (\arcsec{})
} 
\startdata
\textbf{orvara Solutions}\\
HIP 6511 b & ${1.06}_{-0.25}^{+0.25}$ & ${0.79}_{-0.16}^{+0.30}$ & ${2.37}_{-0.20}^{+0.17}$ & ${0.122}_{-0.040}^{+0.040}$ & ${86}_{-38}^{+41}$\tablenotemark{a} & ${3.533}_{-0.038}^{+0.044}$ & ${0.046}_{-0.004}^{+0.004}$\\
HIP 9094 b & ${0.90}_{-0.24}^{+0.25}$ & ${0.72}_{-0.19}^{+0.46}$ & ${2.93}_{-0.29}^{+0.25}$ &  ${0.040}_{-0.027}^{+0.032}$ & ${51}_{-23}^{+27}$ or ${124}_{-25}^{+28}$ & ${5.282}_{-0.029}^{+0.030}$ & ${0.09}_{-0.01}^{+0.01}$\\
HIP 15527 b & ${1.07}_{-0.14}^{+0.16}$ & ${1.83}_{-0.31}^{+0.52}$ & ${1.422}_{-0.064}^{+0.066}$ & ${0.949}_{-0.013}^{+0.012}$ & ${89}_{-34}^{+32}$\tablenotemark{a} & ${1.63473}_{-0.00011}^{+0.00011}$ & ${0.07}_{-0.01}^{+0.01}$ \\
HIP 18387 b & ${1.18}_{-0.25}^{+0.25}$ & ${1.22}_{-0.31}^{+0.70}$ & ${3.07}_{-0.23}^{+0.20}$ & ${0.032}_{-0.022}^{+0.035}$ & ${49}_{-22}^{+29}$ or ${132}_{-28}^{+20}$ & ${4.953}_{-0.047}^{+0.050}$ & ${0.056}_{-0.004}^{+0.004}$\\
HIP 27384 b & ${1.07}_{-0.25}^{+0.25}$ & ${10.6}_{-2.1}^{+3.9}$ & ${1.56}_{-0.13}^{+0.11}$ & ${0.020}_{-0.013}^{+0.017}$ & ${109}_{-39}^{+29}$ & ${1.8768}_{-0.0027}^{+0.0028}$ & ${0.017}_{-0.001}^{+0.001}$\\
HIP 30579 b & ${1.03}_{-0.20}^{+0.20}$ & ${0.245}_{-0.041}^{+0.077}$ & ${0.738}_{-0.051}^{+0.044}$ & ${0.029}_{-0.020}^{+0.032}$ & ${90}_{-35}^{+36}$\tablenotemark{a} & ${0.62392}_{-0.00053}^{+0.00052}$ & ${0.022}_{-0.002}^{+0.001}$\\
HIP 30579 c & ${1.03}_{-0.20}^{+0.20}$ & ${0.75}_{-0.13}^{+0.22}$ & ${0.972}_{-0.068}^{+0.058}$ & ${0.023}_{-0.014}^{+0.014}$ & ${85}_{-33}^{+37}$\tablenotemark{a} & ${0.94186}_{-0.00050}^{+0.00050}$ & ${0.029}_{-0.002}^{+0.002}$\\
HIP 52409 b & ${1.005}_{-0.079}^{+0.091}$ & ${5.7}_{-1.3}^{+2.0}$ & ${0.930}_{-0.025}^{+0.027}$ & ${0.3636}_{-0.0038}^{+0.0039}$ &  ${36}_{-10}^{+15}$ & ${0.891659}_{-0.000039}^{+0.000038}$ & ${0.030}_{-0.002}^{+0.003}$\\
HIP 52409 c & ${1.005}_{-0.079}^{+0.091}$ & ${3.22}_{-0.39}^{+0.66}$ & ${9.02}_{-0.75}^{+1.2}$ & ${0.426}_{-0.069}^{+0.083}$ & ${62}_{-16}^{+19}$ or ${117}_{-19}^{+18}$ & ${26.9}_{-3.1}^{+5.6}$ & ${0.34}_{-0.04}^{+0.07}$ \\
HIP 53437 b & ${1.13}_{-0.25}^{+0.25}$ & ${0.55}_{-0.11}^{+0.20}$ & ${3.36}_{-0.27}^{+0.23}$ & ${0.412}_{-0.043}^{+0.044}$ & ${89}_{-37}^{+37}$\tablenotemark{a} & ${5.782}_{-0.046}^{+0.055}$ & ${0.06}_{-0.01}^{+0.01}$\\
HIP 54779 e & ${0.99}_{-0.25}^{+0.25}$ & ${0.141}_{-0.032}^{+0.078}$ & ${0.675}_{-0.062}^{+0.052}$ & ${0.223}_{-0.070}^{+0.069}$ & ${90}_{-43}^{+43}$\tablenotemark{a} & ${0.5573}_{-0.0044}^{+0.0044}$ & ${0.013}_{-0.001}^{+0.001}$\\
HIP 59357 b & ${0.99}_{-0.25}^{+0.25}$ & ${0.85}_{-0.20}^{+0.45}$ & ${3.43}_{-0.32}^{+0.27}$ & ${0.062}_{-0.043}^{+0.068}$ & ${80}_{-41}^{+47}$ & ${6.40}_{-0.10}^{+0.11}$ & ${0.06}_{-0.01}^{+0.01}$\\
HIP 63833 b & ${0.60}_{-0.23}^{+0.24}$ & ${0.46}_{-0.15}^{+0.35}$ & ${1.24}_{-0.19}^{+0.15}$ & ${0.304}_{-0.060}^{+0.060}$ & ${89}_{-48}^{+46}$\tablenotemark{a} & ${1.7810}_{-0.0089}^{+0.0090}$ & ${0.09}_{-0.01}^{+0.01}$\\
HIP 63833 c & ${0.60}_{-0.23}^{+0.24}$ &  ${1.00}_{-0.28}^{+0.26}$ & ${2.49}_{-0.38}^{+0.30}$ & ${0.186}_{-0.038}^{+0.039}$ & ${92}_{-16}^{+16}$ & ${5.056}_{-0.057}^{+0.057}$ & ${0.16}_{-0.02}^{+0.02}$\\
HIP 71481 b & ${0.87}_{-0.22}^{+0.24}$ & ${5.2}_{-3.0}^{+4.9}$ & ${0.835}_{-0.079}^{+0.070}$ & ${0.772}_{-0.093}^{+0.12}$ & ${164}_{-32}^{+8}$ & ${0.8163}_{-0.0012}^{+0.0011}$ & ${0.05}_{-0.01}^{+0.01}$\\
HIP 81681 b & ${0.98}_{-0.25}^{+0.25}$ & ${0.179}_{-0.040}^{+0.078}$ & ${1.96}_{-0.18}^{+0.15}$ & ${0.082}_{-0.056}^{+0.076}$ & ${91}_{-40}^{+40}$\tablenotemark{a} & ${2.764}_{-0.023}^{+0.022}$ & ${0.07}_{-0.01}^{+0.01}$\\
HIP 82632 b & ${0.97}_{-0.25}^{+0.25}$ & ${2.65}_{-0.53}^{+0.71}$ & ${4.02}_{-0.37}^{+0.32}$ &  ${0.509}_{-0.015}^{+0.014}$ & ${91}_{-32}^{+32}$\tablenotemark{a} & ${8.155}_{-0.074}^{+0.074}$ & ${0.06}_{-0.01}^{+0.01}$\\
HIP 83547 b & ${1.18}_{-0.25}^{+0.25}$ & ${3.23}_{-0.58}^{+1.1}$ & ${1.300}_{-0.10}^{+0.086}$ & ${0.356}_{-0.016}^{+0.016}$ & ${95}_{-36}^{+37}$\tablenotemark{a} & ${1.3635}_{-0.0056}^{+0.0049}$ & ${0.032}_{-0.003}^{+0.002}$ \\
HIP 83941 b & ${0.68}_{-0.20}^{+0.23}$ & ${1.34}_{-0.45}^{+1.2}$ & ${0.725}_{-0.078}^{+0.072}$ & ${0.799}_{-0.057}^{+0.087}$ & ${90}_{-41}^{+42}$\tablenotemark{a} & ${0.74634}_{-0.00031}^{+0.00031}$ & ${0.034}_{-0.004}^{+0.004}$\\
HIP 84787 b & ${1.03}_{-0.25}^{+0.25}$ & ${0.76}_{-0.16}^{+0.38}$ & ${1.78}_{-0.16}^{+0.13}$ & ${0.072}_{-0.050}^{+0.067}$ & ${91}_{-42}^{+41}$\tablenotemark{a} & ${2.344}_{-0.014}^{+0.025}$ & ${0.034}_{-0.003}^{+0.003}$\\
HIP 90979 b & ${0.64}_{-0.24}^{+0.25}$ & ${0.78}_{-0.26}^{+0.49}$ & ${2.35}_{-0.34}^{+0.27}$ & ${0.065}_{-0.044}^{+0.054}$ & ${52}_{-29}^{+29}$ or ${135}_{-28}^{+18}$ & ${4.483}_{-0.041}^{+0.046}$ & ${0.09}_{-0.01}^{+0.01}$\\
HIP 95262 b & ${1.03}_{-0.25}^{+0.25}$ & ${0.51}_{-0.11}^{+0.26}$ & ${1.96}_{-0.17}^{+0.15}$ & ${0.029}_{-0.020}^{+0.031}$ & ${91}_{-42}^{+41}$\tablenotemark{a} & ${2.697}_{-0.018}^{+0.018}$ & ${0.033}_{-0.003}^{+0.003}$\\
HIP 96078 b & ${1.19}_{-0.22}^{+0.24}$ & ${1.19}_{-0.20}^{+0.49}$ & ${0.564}_{-0.036}^{+0.036}$ & ${0.193}_{-0.020}^{+0.020}$ & ${93}_{-39}^{+38}$\tablenotemark{a} & ${0.387713}_{-0.000084}^{+0.000085}$ & ${0.0055}_{-0.0004}^{+0.0004}$\\
HIP 99212 c & ${1.07}_{-0.22}^{+0.24}$ & ${0.86}_{-0.17}^{+0.36}$ & ${0.935}_{-0.069}^{+0.066}$ & ${0.092}_{-0.027}^{+0.027}$ & ${88}_{-38}^{+40}$\tablenotemark{a} & ${0.8756}_{-0.0036}^{+0.0035}$ & ${0.011}_{-0.001}^{+0.001}$\\
%HIP 104045 b & ${0.98}_{-0.25}^{+0.25}$ & ${0.62}_{-0.14}^{+0.21}$ & ${3.52}_{-0.33}^{+0.29}$ & ${0.142}_{-0.082}^{+0.079}$ & ${90.53}_{-37.71}^{+37.12}$ & ${6.69}_{-0.18}^{+0.18}$ & 0.08 \\
HIP 108375 b & ${1.11}_{-0.22}^{+0.24}$ & ${0.55}_{-0.11}^{+0.25}$ & ${0.518}_{-0.038}^{+0.035}$ & ${0.377}_{-0.024}^{+0.024}$ & ${90}_{-42}^{+40}$\tablenotemark{a} & ${0.35408}_{-0.00016}^{+0.00018}$ & ${0.012}_{-0.001}^{+0.001}$\\
HIP 109378 b & ${1.06}_{-0.15}^{+0.19}$ & ${1.45}_{-0.20}^{+0.34}$ & ${1.159}_{-0.058}^{+0.067}$ & ${0.4686}_{-0.0036}^{+0.0035}$ & ${87}_{-30}^{+30}$\tablenotemark{a} & ${1.21286}_{-0.00011}^{+0.00011}$ & ${0.066}_{-0.004}^{+0.005}$\\
HIP 109388 b & ${0.45}_{-0.10}^{+0.10}$ & ${0.94}_{-0.16}^{+0.17}$ & ${2.32}_{-0.19}^{+0.16}$ & ${0.056}_{-0.018}^{+0.017}$ & ${95}_{-23}^{+21}$ & ${5.264}_{-0.020}^{+0.020}$ & ${0.27}_{-0.02}^{+0.02}$\\
HIP 109388 c & ${0.45}_{-0.10}^{+0.10}$ & ${1.04}_{-0.18}^{+0.21}$ & ${4.87}_{-0.40}^{+0.35}$ & ${0.094}_{-0.036}^{+0.036}$ & ${92}_{-23}^{+25}$ & ${16.01}_{-0.33}^{+0.37}$ & ${0.59}_{-0.05}^{+0.05}$\\
HIP 112414 b & ${1.25}_{-0.25}^{+0.25}$ & ${0.128}_{-0.025}^{+0.058}$ & ${0.703}_{-0.051}^{+0.044}$ & ${0.132}_{-0.069}^{+0.067}$ & ${88}_{-40}^{+40}$\tablenotemark{a} & ${0.52796}_{-0.00079}^{+0.00081}$ & ${0.019}_{-0.001}^{+0.001}$\\
HIP 112414 c & ${1.25}_{-0.25}^{+0.25}$ & ${0.342}_{-0.068}^{+0.16}$ & ${3.59}_{-0.26}^{+0.23}$ & ${0.075}_{-0.051}^{+0.060}$ & ${93}_{-43}^{+41}$\tablenotemark{a} & ${6.10}_{-0.10}^{+0.11}$ & ${0.10}_{-0.01}^{+0.01}$\\
HIP 113238 b & ${0.91}_{-0.22}^{+0.24}$ & ${0.54}_{-0.12}^{+0.27}$ & ${0.451}_{-0.039}^{+0.037}$ & ${0.106}_{-0.034}^{+0.032}$ & ${93}_{-42}^{+40}$\tablenotemark{a} & ${0.31784}_{-0.00013}^{+0.00014}$ & ${0.013}_{-0.001}^{+0.001}$\\
HIP 116906 b & ${1.01}_{-0.21}^{+0.20}$ & ${9.0}_{-1.4}^{+1.8}$ & ${1.360}_{-0.10}^{+0.082}$ & ${0.7608}_{-0.0038}^{+0.0038}$ & ${103}_{-27}^{+24}$ & ${1.56755}_{-0.00015}^{+0.00015}$ & ${0.048}_{-0.003}^{+0.003}$\\
\hline
\textbf{Octofitter Solutions}\\
HIP 6511 b & $1.07^{+0.26}_{-0.27}$ & $0.78^{+0.30}_{-0.16}$ & $2.38^{+0.18}_{-0.22}$ & $0.13^{+0.04}_{-0.04}$ & $92^{+37}_{-40}$\tablenotemark{a} & $3.54^{+0.05}_{-0.04}$ & ${0.046}_{-0.004}^{+0.004}$\\
HIP 9094 b & $0.95^{+0.24}_{-0.22}$ & $0.68^{+0.24}_{-0.15}$ & $2.99^{+0.23}_{-0.25}$ & $0.04^{+0.03}_{-0.03}$ & $90^{+39}_{-37}$\tablenotemark{a} & $5.28^{+0.03}_{-0.03}$ & ${0.09}_{-0.01}^{+0.01}$\\
HIP 15527 b & $1.05^{+0.07}_{-0.07}$\tablenotemark{b} & $1.71^{+0.57}_{-0.14}$ & $1.41^{+0.03}_{-0.03}$\tablenotemark{b} & $0.948^{+0.001}_{-0.001}$ & $82^{+35}_{-33}$ & $1.6344^{+0.0001}_{-0.0001}$ & ${0.065}_{-0.002}^{+0.005}$\\
HIP 18387 b & $1.17^{+0.27}_{-0.23}$ & $1.54^{+0.76}_{-0.51}$ & $3.07^{+0.22}_{-0.22}$ & $0.03^{+0.04}_{-0.02}$ & $37^{+30}_{-12}$ or $145^{+12}_{-25}$ & $4.96^{+0.05}_{-0.05}$ & ${0.056}_{-0.004}^{+0.004}$\\
HIP 27384 b & $1.17^{+0.25}_{-0.16}$ & $10.36^{+1.50}_{-1.27}$ & $1.61^{+0.11}_{-0.07}$ & $0.02^{+0.02}_{-0.01}$ & $83^{+20}_{-20}$ & $1.877^{+0.003}_{-0.003}$ & ${0.018}_{-0.001}^{+0.001}$\\
%HIP 30579 b &  \\
HIP 30579 c & $1.04^{+0.04}_{-0.05}$ & $0.77^{+0.30}_{-0.08}$ & $0.97^{+0.01}_{-0.02}$ & $0.03^{+0.03}_{-0.02}$ & $90^{+36}_{-37}$\tablenotemark{a} & $0.938^{+0.001}_{-0.001}$ & ${0.029}_{-0.001}^{+0.001}$\\
%HIP 52409 b &  \\
%HIP 52409 c & \\
HIP 53437 b & $1.12^{+0.27}_{-0.25}$ & $0.58^{+0.27}_{-0.13}$ & $3.34^{+0.25}_{-0.28}$ & $0.41^{+0.04}_{-0.04}$ & $99^{+39}_{-46}$ & $5.78^{+0.05}_{-0.05}$ & ${0.06}_{-0.01}^{+0.01}$\\
HIP 54779 e & $1.02^{+0.23}_{-0.30}$ & $0.14^{+0.09}_{-0.03}$ & $0.68^{+0.05}_{-0.07}$ & $0.22^{+0.07}_{-0.07}$ & $90^{+45}_{-43}$\tablenotemark{a} & $0.557^{+0.005}_{-0.004}$ & ${0.013}_{-0.002}^{+0.001}$\\
HIP 59357 b & $0.99^{+0.24}_{-0.24}$ & $1.09^{+0.63}_{-0.34}$ & $3.44^{+0.27}_{-0.30}$ & $0.06^{+0.07}_{-0.04}$ & $38^{+26}_{-14}$ or $139^{+19}_{-32}$ & $6.40^{+0.11}_{-0.10}$ & ${0.07}_{-0.01}^{+0.01}$\\
HIP 63833 b & $0.58^{+0.22}_{-0.13}$ & $0.36^{+0.10}_{-0.06}$ & $1.23^{+0.15}_{-0.10}$ & $0.31^{+0.06}_{-0.06}$ & $92^{+14}_{-14}$ & $1.78^{+0.01}_{-0.01}$ & ${0.09}_{-0.01}^{+0.01}$\\
HIP 63833 c & $0.58^{+0.22}_{-0.13}$ & $0.96^{+0.24}_{-0.15}$ & $2.46^{+0.297}_{-0.203}$ & $0.18^{+0.04}_{-0.04}$ & $92^{+14}_{-14}$ & $5.05^{+0.06}_{-0.06}$ & ${0.15}_{-0.01}^{+0.02}$\\
HIP 71481 b & $1.10^{+0.21}_{-0.22}$ & $8.19^{+2.48}_{-2.58}$ & $0.90^{+0.05}_{-0.07}$ & $0.78^{+0.11}_{-0.10}$ & $12^{+7}_{-3}$ & $0.816^{+0.001}_{-0.001}$ & ${0.06}_{-0.01}^{+0.01}$\\
HIP 81681 b & $0.98^{+0.25}_{-0.24}$ & $0.18^{+0.09}_{-0.04}$ & $1.96^{+0.15}_{-0.18}$ & $0.08^{+0.08}_{-0.06}$ & $89^{+39}_{-41}$\tablenotemark{a} & $2.76^{+0.02}_{-0.02}$ & ${0.07}_{-0.01}^{+0.01}$\\
HIP 82632 b & $0.96^{+0.25}_{-0.24}$ & $2.56^{+0.67}_{-0.49}$ & $4.00^{+0.32}_{-0.36}$ & $0.51^{+0.01}_{-0.01}$ & $90^{+29}_{-28}$ & $8.16^{+0.07}_{-0.07}$ & ${0.06}_{-0.01}^{+0.01}$\\
HIP 83547 b & $1.21^{+0.14}_{-0.11}$ & $3.22^{+1.24}_{-0.35}$ & $1.31^{+0.05}_{-0.04}$ & $0.35^{+0.02}_{-0.02}$ & $108^{+30}_{-31}$ & $1.36^{+0.01}_{-0.01}$ & ${0.032}_{-0.001}^{+0.002}$\\
HIP 83941 b & $0.69^{+0.03}_{-0.03}$ & $1.72^{+0.95}_{-0.32}$ & $0.73^{+0.01}_{-0.01}$ & $0.87^{+0.03}_{-0.02}$ & $89^{+38}_{-37}$\tablenotemark{a} & $0.7462^{+0.0003}_{-0.0003}$ & ${0.035}_{-0.001}^{+0.001}$\\
HIP 84787 b & $0.98^{+0.30}_{-0.24}$ & $1.04^{+0.84}_{-0.36}$ & $1.76^{+0.16}_{-0.16}$ & $0.07^{+0.07}_{-0.05}$ & $52^{+28}_{-23}$ or $148^{+14}_{-25}$ & $2.35^{+0.02}_{-0.02}$  & ${0.034}_{-0.003}^{+0.004}$\\
HIP 90979 b & $0.63^{+0.33}_{-0.29}$ & $0.67^{+0.32}_{-0.23}$ & $2.33^{+0.35}_{-0.44}$ & $0.06^{+0.05}_{-0.04}$ & $67^{+42}_{-27}$ & $4.49^{+0.05}_{-0.05}$ & ${0.09}_{-0.02}^{+0.01}$\\
HIP 95262 b & $0.99^{+0.28}_{-0.25}$ & $0.49^{+0.23}_{-0.11}$ & $1.93^{+0.17}_{-0.18}$ & $0.03^{+0.03}_{-0.02}$ & $90^{+40}_{-38}$\tablenotemark{a} & $2.70^{+0.02}_{-0.02}$ & ${0.033}_{-0.003}^{+0.003}$\\
HIP 96078 b & $1.18^{+0.05}_{-0.04}$ & $1.16^{+0.62}_{-0.13}$ & $0.56^{+0.01}_{-0.01}$ & $0.19^{+0.02}_{-0.02}$ & $89^{+40}_{-40}$\tablenotemark{a} & $0.3878^{+0.0001}_{-0.0001}$ & ${0.0055}_{-0.0001}^{+0.0002}$\\
HIP 99212 c & $0.97^{+0.24}_{-0.19}$ & $0.82^{+0.40}_{-0.16}$ & $0.91^{+0.07}_{-0.06}$ & $0.09^{+0.03}_{-0.03}$ & $90^{+41}_{-42}$\tablenotemark{a} & $0.875^{+0.003}_{-0.003}$ & ${0.011}_{-0.001}^{+0.001}$\\
HIP 108375 b & $1.12^{+0.04}_{-0.04}$ & $0.53^{+0.28}_{-0.06}$ & $0.52^{+0.01}_{-0.01}$ & $0.38^{+0.03}_{-0.02}$ & $89^{+43}_{-41}$\tablenotemark{a} & $0.3541^{+0.0002}_{-0.0002}$ & ${0.012}_{-0.001}^{+0.001}$\\
HIP 109378 b & $1.09^{+0.04}_{-0.04}$ & $1.56^{+0.39}_{-0.20}$ & $1.17^{+0.01}_{-0.02}$ & $0.47^{+0.00}_{-0.00}$ & $102^{+30}_{-46}$ & $1.2129^{+0.0001}_{-0.0001}$ & ${0.067}_{-0.002}^{+0.003}$\\
HIP 109388 b & $0.57^{+0.14}_{-0.20}$ & $1.02^{+0.18}_{-0.25}$ & $2.52^{+0.19}_{-0.34}$ & $0.05^{+0.02}_{-0.02}$ & $93^{+7}_{-7}$ & $5.27^{+0.02}_{-0.02}$ & ${0.29}_{-0.04}^{+0.02}$\\
HIP 109388 c & $0.57^{+0.14}_{-0.20}$ & $1.13^{+0.21}_{-0.27}$ & $5.32^{+0.42}_{-0.70}$ & $0.12^{+0.04}_{-0.03}$ & $93^{+7}_{-7}$ & $16.23^{+0.40}_{-0.34}$ & ${0.65}_{-0.09}^{+0.06}$\\
HIP 112414 b & $1.24^{+0.05}_{-0.06}$ & $0.14^{+0.09}_{-0.02}$ & $0.70^{+0.01}_{-0.01}$ & $0.13^{+0.07}_{-0.07}$ & $101^{+43}_{-53}$ & $0.528^{+0.001}_{-0.001}$ & ${0.019}_{-0.001}^{+0.001}$\\
HIP 112414 c & $1.24^{+0.05}_{-0.06}$ & $0.36^{+0.23}_{-0.07}$ & $3.59^{+0.06}_{-0.07}$ & $0.08^{+0.06}_{-0.05}$ & $101^{+43}_{-53}$ & $6.09^{+0.10}_{-0.10}$ & ${0.096}_{-0.004}^{+0.006}$\\
HIP 113238 b & $0.90^{+0.30}_{-0.17}$ & $0.57^{+0.52}_{-0.14}$ & $0.45^{+0.05}_{-0.03}$ & $0.11^{+0.03}_{-0.03}$ & $93^{+49}_{-49}$\tablenotemark{a} & $0.3178^{+0.0001}_{-0.0001}$ & ${0.013}_{-0.001}^{+0.001}$\\
HIP 116906 b & $1.02^{+0.07}_{-0.09}$\tablenotemark{b} & $9.91^{+1.51}_{-1.25}$ & $1.36^{+0.03}_{-0.04}$\tablenotemark{b} & $0.76^{+0.00}_{-0.00}$ & $58^{+15}_{-10}$ or $126^{+8}_{-15}$ & $1.5675^{+0.0002}_{-0.0001}$ & ${0.050}_{-0.002}^{+0.002}$\\
\enddata
\tablenotetext{a}{Although the inclination posterior is largely prior-dominated, we can rule out nearly face-on solutions that would imply massive companions.
\tablenotetext{b}{The multimodal posterior may render the parameter estimates unreliable.}}
\end{deluxetable*}
 
\end{appendix}

%\clearpage

\bibliography{bibliography}{}
\bibliographystyle{aasjournal}

\end{document}